\documentclass[sigconf]{acmart}

\usepackage{acmart-taps}

\usepackage{multirow} 
\usepackage{array} 
\newcolumntype{C}[1]{>{\centering\arraybackslash}p{#1}} 

\usepackage{float} 

\usepackage{siunitx} 
\usepackage{etoolbox} 
\usepackage{dcolumn} 
\usepackage{booktabs} 

\usepackage{fontawesome5}

\usepackage{xcolor} 
\usepackage{soul} 
\soulregister{\cite}7 
\soulregister{\citep}7
\soulregister{\citet}7
\soulregister{\ref}7
\soulregister{\pageref}7

\usepackage{framed}

\newenvironment{myfancybox}
  {\begin{framed}\noindent}
  {\end{framed}}
  
\usepackage{xspace} 

\definecolor{oxfordblue}{rgb}{0.0, 0.13, 0.28}
\definecolor{harvardcrimson}{rgb}{0.79, 0.0, 0.09}
\definecolor{dartmouthgreen}{rgb}{0.05, 0.5, 0.06}
\definecolor{princetonorange}{rgb}{1.0, 0.56, 0.0}
\definecolor{yaleblue}{rgb}{0.06, 0.3, 0.57}
\definecolor{usccardinal}{rgb}{0.6, 0.0, 0.0}
\definecolor{uclablue}{rgb}{0.33, 0.41, 0.58}
\definecolor{msugreen}{rgb}{0.09, 0.27, 0.23}
\definecolor{cornellred}{rgb}{0.7, 0.11, 0.11}
\definecolor{pomegranate}{RGB}{192, 57, 43}
\definecolor{anti-pomegranate}{RGB}{43,178,192}
\definecolor{alizarin}{RGB}{231, 76, 60}
\definecolor{anti-belize}{RGB}{185, 41, 56}
\definecolor{belize}{RGB}{41, 128, 185}
\definecolor{sky}{RGB}{52, 152, 219}
\definecolor{green}{RGB}{22, 160, 133}
\definecolor{anti-green}{RGB}{160,22,118}
\definecolor{turquoise}{RGB}{26, 188, 156}
\definecolor{pumpkin}{RGB}{211, 84, 0}
\definecolor{anti-pumpkin}{RGB}{0,22,211}
\definecolor{carrot}{RGB}{230, 126, 34}
\definecolor{wisteria}{RGB}{142, 68, 173}
\definecolor{anti-wisteria}{RGB}{99,173,68}
\definecolor{amethyst}{RGB}{155, 89, 182}
\definecolor{nephritis}{RGB}{39, 174, 96}
\definecolor{anti-nephritis}{RGB}{174,39,117}
\definecolor{gray-bg}{RGB}{242,242,235}
\definecolor{light-bg}{RGB}{249,249,249}
\definecolor{extended-blue}{RGB}{59,130,246}
\definecolor{extended-red}{RGB}{239,68,68}
\definecolor{extended-orange}{RGB}{249,115,22}
\definecolor{extended-violet}{RGB}{99,102,241}
\definecolor{extended-green}{RGB}{16,185,129}

\newcommand{\remove}[1]{}
\newcommand{\revision}[1]{{#1}}
\newcommand{\correct}[1]{{#1}}

\newcommand{\eg}{e.g.,\ }

\newcommand{\ie}{i.e.,\ }

\newcommand{\cf}{cf.\ }

\newcommand{\tool}{\textsc{Narrix}\xspace}

\AtBeginDocument{%
  }

\newcommand{\namedparagraph}[1]{\vspace{0.2cm}\noindent\textbf{#1:}}

\newcommand{\fig}[1]{Fig. {#1}}
\newcommand{\figref}[1]{\fig{\ref{#1}}}
\newcommand{\tab}[1]{Table {#1}}
\newcommand{\tabref}[1]{\tab{\ref{#1}}}
\newcommand{\secref}[1]{§\ref{#1}}

\usepackage{pifont}  

\usepackage[most]{tcolorbox}
\tcbuselibrary{skins, breakable}

\newtcbox{\buttonlabel}{on line,
  tcbox width=auto limited, tcbox raise base, enhanced,
  colback=gray!5, colframe=black!60,
  fontupper=\sffamily\footnotesize\bfseries,
  boxrule=0.5pt, arc=1pt,
  left=2pt, right=2pt, top=0.5pt, bottom=0.5pt,
  boxsep=0pt, before upper={\strut}
}

\newtcbox{\strategybutton}{on line,
  tcbox width=auto limited, tcbox raise base, enhanced,
  colback=white, colframe=cyan!60!blue,
  fontupper=\sffamily\footnotesize\bfseries,
  boxrule=0.5pt, arc=1pt,
  left=2pt, right=2pt, top=0.5pt, bottom=0.5pt,
  boxsep=0pt, before upper={\strut}
}

\newtcbox{\lexicalcuebutton}{on line,
  tcbox width=auto limited, tcbox raise base, enhanced,
  colback=yellow!50, colframe=yellow!80!orange,
  fontupper=\ttfamily\footnotesize\bfseries,
  boxrule=0.5pt, arc=1pt,
  left=2pt, right=2pt, top=0.5pt, bottom=0.5pt,
  boxsep=0pt, before upper={\strut}
}

\newtcbox{\prompt}{on line,
  tcbox width=auto limited, tcbox raise base, enhanced,
  colback=gray!5,
  fontupper=\ttfamily\footnotesize\bfseries,
  boxrule=0pt, arc=1pt,
  left=2pt, right=2pt, top=0.5pt, bottom=0.5pt, boxsep=0pt,
  before upper={\strut}
}

\NewDocumentCommand{\promptsplit}{m m}{%
  \prompt{#1}\allowbreak\ \prompt{#2}%
}

\definecolor{dimension-plot}{RGB}{59,130,246}       
\definecolor{dimension-character}{RGB}{139,92,246}  
\definecolor{dimension-information}{RGB}{16,185,129}
\definecolor{dimension-emotional}{RGB}{239,68,68}   
\definecolor{dimension-linguistic}{RGB}{250,204,21} 
\definecolor{dimension-pacing}{gray}{0.5}       
\definecolor{dimension-thematic}{RGB}{99,102,241}   
\definecolor{dimension-engagement}{RGB}{236,72,153} 

\newcommand{\cdplot}{{\small\ttfamily \textbf{Plot}}}
\newcommand{\cdcharacter}{{\small\ttfamily \textbf{Character}}}
\newcommand{\cdinformation}{{\small\ttfamily \textbf{Information}}}
\newcommand{\cdemotional}{{\small\ttfamily \textbf{Emotional}}}
\newcommand{\cdlinguistic}{{\small\ttfamily \textbf{Linguistic}}}
\newcommand{\cdpacing}{{\small\ttfamily \textbf{Pacing}}}
\newcommand{\cdthematic}{{\small\ttfamily \textbf{Thematic}}}
\newcommand{\cdengagement}{{\small\ttfamily \textbf{Engagement}}}

\definecolor{tp-opportunity}{RGB}{16,185,129}      
\definecolor{tp-changeofplans}{RGB}{245,158,11}    
\definecolor{tp-pointofnoreturn}{RGB}{239,68,68}   
\definecolor{tp-majorsetback}{RGB}{139,92,246}     
\definecolor{tp-climax}{RGB}{236,72,153}           

\newcommand{\tpOpportunity}[1]{{\small\ttfamily \textbf{#1}}}
\newcommand{\tpChangeOfPlans}[1]{{\small\ttfamily \textbf{#1}}}
\newcommand{\tpPointOfNoReturn}[1]{{\small\ttfamily \textbf{#1}}}
\newcommand{\tpMajorSetback}[1]{{\small\ttfamily \textbf{#1}}}
\newcommand{\tpClimax}[1]{{\small\ttfamily \textbf{#1}}}

\newcommand{\icon}[2][gray]{{\small\textcolor{#1}{#2}}}

\acmSubmissionID{6700}

\copyrightyear{2026}
\acmYear{2026}
\setcopyright{cc}
\setcctype{by-nc-nd}
\acmConference[CHI '26]{Proceedings of the 2026 CHI Conference on Human Factors in Computing Systems}{April 13--17, 2026}{Barcelona, Spain}
\acmBooktitle{Proceedings of the 2026 CHI Conference on Human Factors in Computing Systems (CHI '26), April 13--17, 2026, Barcelona, Spain}
\acmDOI{10.1145/3772318.3790813}
\acmISBN{979-8-4007-2278-3/2026/04}

\begin{document}

\title[Narrix: Remixing Narrative Strategies from Examples for Story Writing]{Narrix: Remixing Narrative Strategies from Examples for Story Writing}

\author{Chao Zhang}
\email{cz468@cornell.edu}
\orcid{0000-0003-4286-8468}
\authornote{Work done during an internship at Adobe.}
\affiliation{ %
 \institution{Cornell University}
 \city{Ithaca, NY}
 \country{USA}}

\author{Shunan Guo}
\email{sguo@adobe.com}
\orcid{0000-0001-5355-8399}
\affiliation{ %
 \institution{Adobe Research}
 \city{San Jose, CA}
 \country{USA}}

\author{Abe Davis}
\email{abedavis@cornell.edu}
\orcid{0000-0003-1469-2696}
\affiliation{ %
 \institution{Cornell University}
 \city{Ithaca, NY}
 \country{USA}}

\author{Eunyee Koh}
\email{eunyee@adobe.com}
\orcid{0000-0003-2091-5972}
\affiliation{ %
 \institution{Adobe Research}
 \city{San Jose, CA}
 \country{USA}}

\renewcommand{\shortauthors}{Zhang et al.}

\begin{abstract}
%


Experienced storytellers decompose stories into local narrative strategies and how these strategies shape higher-level arcs. 
This decomposition helps writers recognize patterns in others' work and adapt those patterns to tell new stories.
Novices, however, struggle to identify these strategies or to reuse them effectively.
We present \tool, a novel writing tool that helps novice writers recognize narrative strategies in example stories and repurpose these strategies in their own writing. 
\tool analyzes strategies in example stories, highlights them with color-coded lexical cues and explanations, and situates them on an interactive story arc for exploration by emotional shifts and turning points. 
Writers then drag strategies onto multi-dimensional tracks and apply block-scoped edits to revise or continue their drafts through controlled generation steered by specified strategies. 
Through a within-subjects study (\(N=12\)), \tool showed improved participants' retention, confidence, and creative adaptation of narrative strategies compared to a baseline chat-based writing interface.
\end{abstract}



\begin{teaserfigure}
 \includegraphics[width=\linewidth]{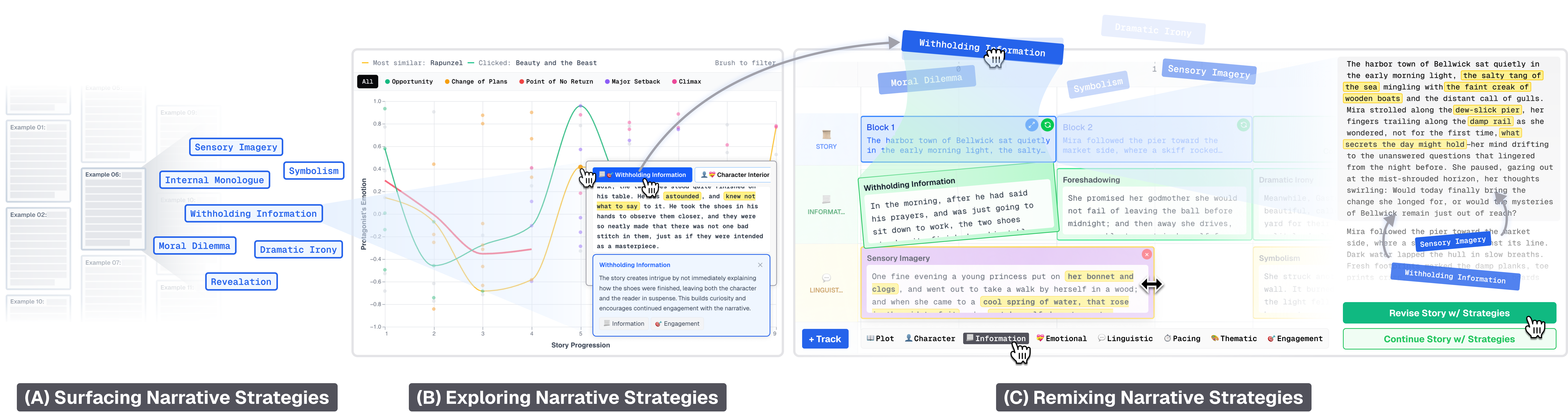}
 \caption{Three main features of \tool: (A) Surfacing Narrative Strategies: \tool automatically extracts narrative strategies from example stories, highlighting relevant lexical cues in color and explaining their function in the story. (B) Exploring Narrative Strategies: Users can retrieve example story content and their strategies based on evolving higher-level storytelling intents (\eg emotional shifts, turning points) via an interactive story arc. (C) Remixing Narrative Strategies: Users can drag and drop selected strategies onto multi-dimensional tracks (\eg character, plot, linguistic) to steer the AI in revising or continuing their story by integrating those strategies.\looseness=-1}
 \Description{Three main features of Narrix: (A) Surfacing Narrative Strategies: Narrix automatically extracts narrative strategies from example stories, highlighting relevant lexical cues in color and explaining their function in the story. (B) Exploring Narrative Strategies: Users can retrieve example story content and their strategies based on evolving higher-level storytelling intents (e.g., emotional shifts, turning points) via an interactive story arc. (C) Remixing Narrative Strategies: Users can drag and drop selected strategies onto multi-dimensional tracks (e.g., character, plot, linguistic) to steer the AI in revising or continuing their story by integrating those strategies.}
 \label{fig:teaser_figure}
\end{teaserfigure}

\maketitle

\section{Introduction}


The writing of a good story often hinges on the execution of key narrative moments---an unexpected twist, a moment of emotional tension, or a sudden shift in tone---the impact of each paragraph depends on not just what the author says, but also how they choose to say it. 
For example, a suspenseful scene might build tension through sensory imagery and delayed revelation, or build toward  a dramatic twist with misdirection and a sudden shift in point of view.
However, novice writers often lack the creative repertoire needed to shape such moments in compelling ways.
To build this repertoire, both cognitive apprenticeship~\cite{collinsCognitiveApprenticeship2006,collinsCognitiveApprenticeshipMaking1991} and writing pedagogy~\cite{grahamBestPracticesWriting2007,grahamWritingNextEffective2007,murrayWriterTeachesWriting2004} emphasize the value of learning from examples---not by simply lifting surface-level content, but by dissecting, internalizing, and remixing the \textit{strategies} that underlie impactful storytelling.
This principle is echoed across writing communities. 
As one writer in r/writing\footnote{The subreddit r/writing (\url{https://www.reddit.com/r/writing/}), with over 3.2 million members as of 2025, is one of the largest and most active online communities for writers, offering a representative hub for sharing advice, experiences, and resources across genres and skill levels.} puts it:\looseness=-1

\begin{quote}
   \textit{``\textit{Take their ideas, how they write dialogue, develop characters, conduct prose, etc. and turn it into your beautiful Frankenstein monster... amalgamate the things you love from your favorite stories and combine them into something that is truly yours, that entertains you, and inspires you, and that you are proud of.}''}
\end{quote}

However, narrative strategies are often tacit, subtly woven into the fabric of the text.
Novices may sense the impact of a compelling scene, but struggle to articulate how it works or to recreate its effect in their own writing.
Exposure to too many examples can further compound the challenge, overwhelming rather than guiding writers as they attempt to locate, compare, and adapt useful techniques.
Prior systems such as IntroAssist~\cite{huiIntroAssistToolSupport2018,huiLettersmithScaffoldingWritten2023} and CorpusStudio~\cite{dangCorpusStudioSurfacingEmergent2025} surface genre norms and conventional structures in domains like email and academic writing.  
Yet, recognizing and reusing narrative strategies for story writing remains comparatively underexplored. 
Addressing this challenge requires more than surfacing patterns: it calls for scaffolds that helps novice writers interpret, experiment with, and internalize strategies within their own drafts.\looseness=-1


To this end, we present \tool (\figref{fig:teaser_figure}), an AI-assisted tool for narrative writing. Grounded in the theory of cognitive apprenticeship~\cite{collinsCognitiveApprenticeship2006,collinsCognitiveApprenticeshipMaking1991}, \tool helps users recognize narrative strategies in example stories and repurpose these strategies in their own writing.
To do so, it starts by decomposing example stories into local pieces---which we call \emph{blocks}---and the narrative strategies those pieces use. 
It then helps the users recognize and understand narrative strategies by naming and labeling them in the interface, highlighting relevant lexical cues with color-coded, in-place annotations, and explaining the function of each strategy within the story.
An interactive story-arc view supports users in exploring example blocks and their strategies according to evolving, higher-level storytelling intents (\eg emotional shifts, turning points). 
Drawing on the metaphor of a digital audio workstation (DAW)---where musicians remix a song by arranging sound clips on a multi-track timeline and layering audio filters---\tool enables users to drag and drop strategies onto multi-dimensional tracks representing distinct story elements, such as character, plot, and linguistic, to guide the AI in revising or continuing their story while applying those strategies as ``filters.''\looseness=-1

We evaluated \tool in a within-subjects study (\(N=12\)) against a chat-based writing interface. 
Participants using \tool recalled and understood more narrative strategies, applied and remixed a greater number of strategies in their own writing, and reported higher confidence, satisfaction, and perceived creativity support than with a baseline chat-based writing interface. 
Their interactions reflected a cyclical exploration-learning-remix workflow, with remix functioning as the backbone and deliberate learning interwoven throughout. 
Strategy use followed both goal-driven and exploratory paths, supported by track-based layout and block-scoped editing that facilitated controlled, strategy-steered revisions.
Overall, our work demonstrates how reifying tacit narrative strategies into visible, manipulatable units and supporting their adaptation can transform AI-assisted writing: from passive content generation into a process of deliberate learning and creative control.\looseness=-1

\section{Related Work}

\subsection{Examples in Creative Support Tools}
CSTs in HCI help users discover, understand, and reuse examples across domains like design, programming, and writing. 
In this section, we review how CSTs (including writing tools) support creative work through three core practices: (1) finding relevant examples, (2) learning from examples, and (3) reusing examples in new contexts.

Examples can inspire and guide creativity, but locating the \emph{right} example at the \emph{right} moment remains challenging~\cite{ngoonShownAdaptiveConceptual2021}.
HCI systems address this with a range of retrieval methods~\cite{ritchieDtourStylebasedExploration2011,kangParagonOnlineGallery2018,xuIdeateRelateExamplesGallery2021,leeDesigningInteractiveExample2010}.
In design, tools like d.tour~\cite{ritchieDtourStylebasedExploration2011} support stylistic querying via descriptors such as ``colorful'' or ``image-heavy''), while IdeateRelate~\cite{xuIdeateRelateExamplesGallery2021} provides more abstract, concept-level matching.
In writing, CorpusStudio~\cite{dangCorpusStudioSurfacingEmergent2025} retrieves examples via textual (sentence-level) and structural (section-title) similarity; TaleStream~\cite{chouTaleStreamSupportingStory2023} suggests tropes (\ie storytelling conventions) during early ideation; and ScriptViz~\cite{raoScriptVizVisualizationTool2024a} automatically retrieves movie screenshots that satisfy user-defined SQL constraints to aid scriptwriting.
However, these systems seldom model a creator's evolving \emph{intent} during the process. 
In storytelling, authors often need examples keyed to current narrative purpose (\eg executing a turning point or shifting emotional trajectory) rather than only topic, convention, or lexical similarity.
Our system supports such intent-driven retrieval by externalizing the writer's goals through an evolving story arc and aligning examples to narrative position and affect.\looseness=-1

The pedagogical value of examples is well established in HCI and language education~\cite{charneyLearningWriteGenre1995,driscollGenreKnowledgeWriting2020,tardyBuildingGenreKnowledge2009,tardyTeachingResearchingGenre2020}.
In design, novices benefit disproportionately from galleries because multiple examples help them infer underlying principles~\cite{leeDesigningInteractiveExample2010}.
In writing, tools improve lexical and syntactic fluency by surfacing patterns from corpora: WriteAhead mines academic collocations~\cite{changWriteAhead2MiningLexical2015}; Corpus of Contemporary American English (COCA)-based systems support ESL learners' usage~\cite{mansourUsingCOCAFoster2017}; Langsmith suggests fluent academic sentences~\cite{itoLangsmithInteractiveAcademic2020}; and Lettersmith scaffolds professional email writing with annotated checklists and aligned exemplars~\cite{huiLettersmithScaffoldingWritten2023}.
Yet these systems primarily focus on linguistic patterns or genre convention, while deeper creative strategies still remain implicit.
We address this gap by helping users explicitly discover, interpret, and repurpose narrative strategies instantiated in examples, moving beyond linguistic conventions toward strategy-level learning.\looseness=-1


Many CSTs support direct reuse of examples.
In visual domains, style-transfer and component-level remix support rapid adaptation~\cite{gatysImageStyleTransfer2016a,leeDesigningInteractiveExample2010,luMistyUIPrototyping2025}.
For example, Misty~\cite{luMistyUIPrototyping2025} helps users blend specific aspects (\eg color, layout, content) from one UI example into their work-in-progress designs.
In writing, systems surface example text for composition and reuse: prior work supports recycling past replies in email~\cite{naeemSmartEmailClient2018}, checklists with expert-annotated examples for help-seeking~\cite{huiIntroAssistToolSupport2018}, and corpus-driven retrieval and highlights for academic writing~\cite{dangCorpusStudioSurfacingEmergent2025}.
However, these tools largely emphasize content-level borrowing (\eg snippets, structural cues, stylistic conventions) rather than identifying and repurposing underlying creative strategies.
Our approach builds on this practice but shifts the focus to strategy-level reuse, enabling writers to insert, adapt, and combine named strategies in their own drafts.\looseness=-1

\subsection{Cognitive Apprenticeship Support}
\label{sec:apprenticeship}

Cognitive apprenticeship supports learning complex skills by making expert strategies visible and guiding learners through \textit{modeling}, \textit{coaching}, \textit{scaffolding}, \textit{reflection}, and \textit{exploration}~\cite{collinsCognitiveApprenticeship2006,collinsCognitiveApprenticeshipMaking1991}.
Modeling, coaching, and scaffolding help students acquire integrated skills through observation and guided practice; 
reflection help students focus their observations of expert problem solving and gain conscious access to (and control of) their own problem-solving strategies; 
and exploration promotes autonomy by encouraging the application of strategies in new, self-defined contexts.\looseness=-1

Several prior tools provide partial support for this pedagogy. 
For example, IntroAssist~\cite{huiIntroAssistToolSupport2018} and Lettersmith~\cite{huiLettersmithScaffoldingWritten2023} surface expert-annotated texts and prompt self-reflection, thereby supporting modeling, coaching, and reflection.
However, they rely heavily on manual annotations and still require users to perform much of the interpretive work themselves (particularly challenging for novices facing complex narrative strategies) and offer limited scaffolding for deeper reasoning or creative experimentation. 
\revision{Schemex~\cite{wangSchemexInteractiveStructural2025}, though designed for text analysis rather than writing, introduced an AI-powered workflow that enables users to extract patterns from examples through clustering, abstraction, and refinement using contrasting examples. More relatedly,}
CorpusStudio~\cite{dangCorpusStudioSurfacingEmergent2025} models corpus-level norms (\eg section headings, common sentence patterns) and supports reflection, yet leaves underlying narrative strategies relatively underexplored.\looseness=-1

In contrast, \tool extends cognitive apprenticeship support by adding active scaffolding and exploration. 
It leverages AI to automatically surface tacit narrative strategies from examples, situate them by narrative position and affect, and guide writers as they experiment with and adapt these strategies within their own drafts. 
This approach helps novices grasp not only \textit{what} makes a narrative moment compelling, but also \textit{how} to repurpose strategies in context.\looseness=-1

\begin{figure*}
\centering
\includegraphics[width=\linewidth]{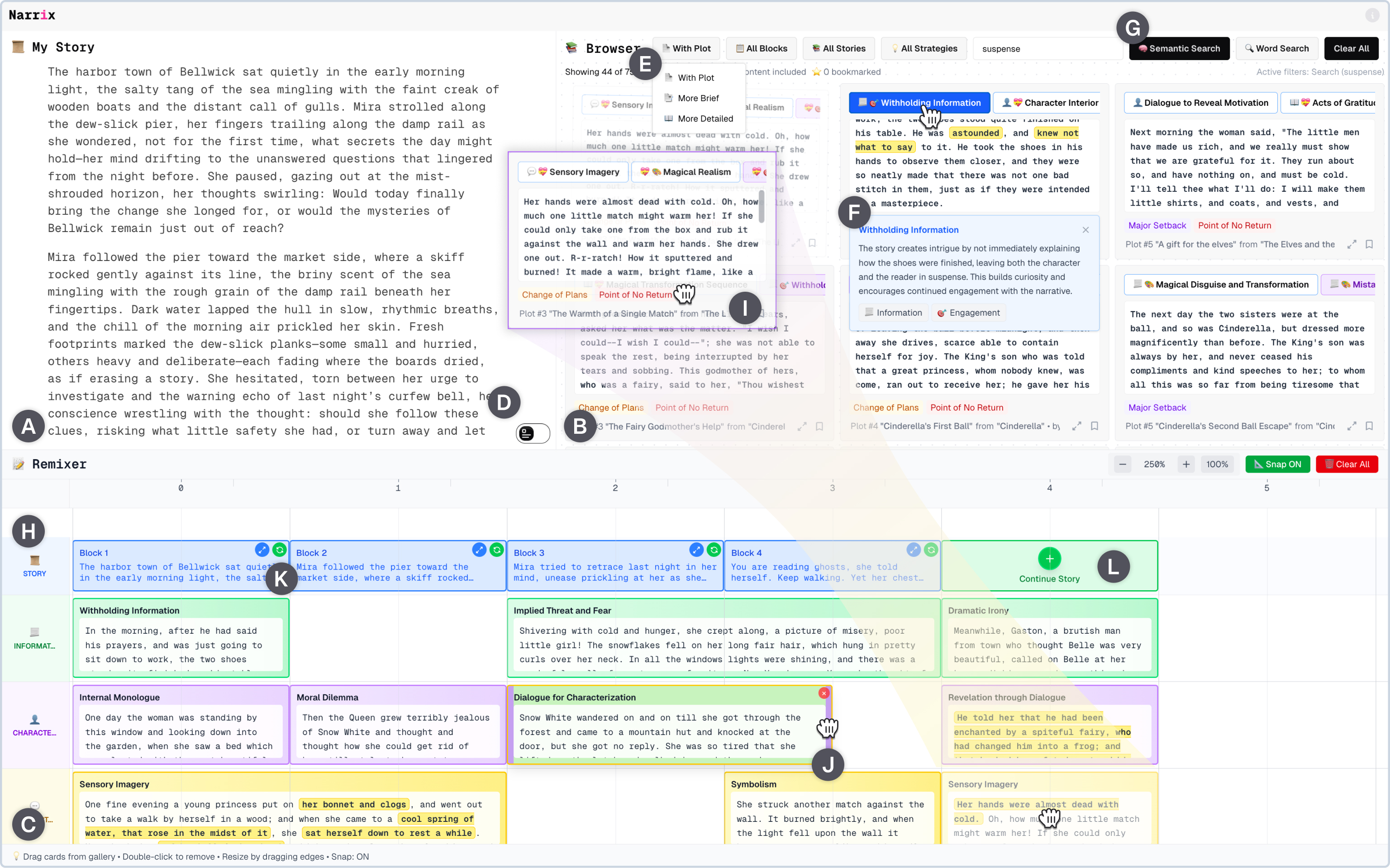}
\caption{\tool's \textbf{user interface}, organized into three views: (A) \textbf{Markdown Editor} for drafting the user’s story, with a mode switch (D) to toggle to the \textbf{Story-Arc Inspector} (see \figref{fig:story_arc}); (B) \textbf{Browser} for exploring example content blocks and their narrative strategies; and (C) \textbf{Remix Tracks} for layering strategies across creative dimensions. In the Browser, users can choose different card views (E), open strategy explanation panels (F), and perform semantic or word searches for strategies (G). Users can drag strategies from example cards onto relevant creative-dimension tracks (I) below the story track (H). Strategy tiles in the Remix Tracks can be resized to span multiple blocks (J). Users can steer AI to revise blocks (K) or continue the story (L) using the selected strategies.}
\Description{Narrix's user interface, organized into three views: (A) Markdown Editor for drafting the user’s story, with a mode switch (D) to toggle to the Story-Arc Inspector; (B) Browser for exploring example content blocks and their narrative strategies; and (C) Remix Tracks for layering strategies across creative dimensions. In the Browser, users can choose different card views (E), open strategy explanation panels (F), and perform semantic or word searches for strategies (G). Users can drag strategies from example cards onto relevant creative-dimension tracks (I) below the story track (H). Strategy tiles in the Remix Tracks can be resized to span multiple blocks (J). Users can steer AI to revise blocks (K) or continue the story (L) using the selected strategies.}
\label{fig:user_interface}
\end{figure*}

\subsection{\revision{Intelligent Writing Tools}}
\revision{The HCI community has a long-standing interest in intelligent writing tools~\cite{leeDesignSpaceIntelligent2024}, supporting writers across various stages, such as brainstorming ideas~\cite{geroSparksInspirationScience2022,schmittCharacterChatSupportingCreation2021,chouTaleStreamSupportingStory2023}, planning outlines~\cite{wanPolymindParallelVisual2025a,riedlVignettebasedStoryPlanning2008}, drafting content~\cite{dhillonShapingHumanAICollaboration2024,hoqueHaLLMarkEffectSupporting2024,jakeschCoWritingOpinionatedLanguage2023,kimAuthorsValuesAttitudes2024,kimMechanicalNovelCrowdsourcing2017}, and refining text~\cite{afrinEffectiveInterfacesStudentDriven2021a,itoUseAIpoweredRewriting2023,leeInteractiveChildrenStory2022,rezaABScribeRapidExploration2023,turkayIteroRevisionHistory2018}.
These tools span diverse genres, including argumentative writing~\cite{zhangVISARHumanAIArgumentative2023,zhangFrictionDecipheringWriting2025}, story writing~\cite{chungTaleBrushSketchingStories2022a,yuanWordcraftStoryWriting2022,huangHeteroglossiaInSituStory2020}, and scientific writing~\cite{shenConvXAIDeliveringHeterogeneous2023a,sunMetaWriterExploringPotential2024}.}\looseness=-1

\revision{In story writing, early} systems emphasized autonomous generation via computational planning with rules~\cite{lebowitzCreatingCharactersStorytelling1984,meehanTALESPINInteractiveProgram1977,riedlNarrativePlanningBalancing2010}, character-based simulation~\cite{cavazzaCharactersSearchAuthor2001}, case-based reasoning~\cite{gervasStoryPlotGeneration2005,perezMEXICAComputerModel2001,riedlVignettebasedStoryPlanning2008,turnerMinstrelComputerModel1993}, and, increasingly, language models for sentence infilling or interpolation~\cite{ammanabroluAutomatedStorytellingCausal2020,huangINSETSentenceInfilling2020,ippolitoUnsupervisedHierarchicalStory2019,wangNarrativeInterpolationGenerating2020}.
\revision{Nowadays, large language models (LLMs) enable writers to steer generation text through prompting~\cite{duvalBreakingWritersBlock2021,fanHierarchicalNeuralStory2018,ippolitoUnsupervisedHierarchicalStory2019a,sunIGAIntentGuidedAuthoring2021,xuMEGATRONCNTRLControllableStory2020,zhangMathemythsLeveragingLarge2024}.
While prompt-based interactions offer flexibility, they also introduce several challenges for our target user and scenario.
First, prior research has shown that users without AI expertise may struggle to design effective prompts~\cite{zamfirescu-pereiraWhyJohnnyCant2023}.
This challenge is amplified when our target users are also novices in writing: they may lack narratological knowledge (\eg narrative strategies), which hinders their ability to craft prompts that analyze example stories, extract strategies, and transfer them into their own writing.
Second, conversational interfaces typically constrain users to a linear flow.
Writers have limited visibility into global story progression, find it hard to compare their drafts with example stories at the level of story arcs, and struggle to coordinate strategies across broader story structure.
Third, prompt-based interactions offers limited pedagogical transparency.
LLM outputs rarely reveal \textit{which} narrative techniques are employed, \textit{why} they work, and \textit{how} they can transfer to new context, making it harder for novice users to learn story-writing skills.\looseness=-1

To address these limitations, recent work has explored interfaces that enable direct manipulation or embed structured interactions into human-AI co-writing process.
For example, TaleBrush allows authors to sketch story arcs to shape narratives~\cite{chungTaleBrushSketchingStories2022}; PatchView supports worldbuilding through dust-and-magnet metaphors~\cite{chungPatchviewLLMpoweredWorldbuilding2024a}; VISAR~\cite{zhangVISARHumanAIArgumentative2023} and Polymind~\cite{wanPolymindParallelVisual2025a} provides node-based visual programming for interactive co-writing;
Dramatron~\cite{mirowskiCoWritingScreenplaysTheatre2023} offers hierarchical story generation, enabling users to control screenplay generation across different narrative elements (\eg characters, plots, locations, and dialogue); and Friction~\cite{zhangFrictionDecipheringWriting2025} and Synthia~\cite{zhangSynthiaVisuallyInterpreting2025} integrates a feedback-based revision workflow into AI-assisted writing to help users reflect on comments and improve their drafts.
While these tools provide richer structural guidance and afford more control than prompt-only interfaces, their assistance tend to focus on shaping story structure or organizing content, and the narrative techniques invoked by the model remain implicit within generated text. In contrast, \tool surfaces narrative strategies extracted from exemplar stories, visualizes how they unfold across an arc, and enables users to directly manipulate and combine them. This explicit focus on strategy visibility, explanation, and transfer complements existing structural and generative approaches.

To sum up, \tool advances beyond prompting-based assistants and structured co-writing tools by centering both controllability and learning around example-based story writing.
Our visual interfaces allow users to inspect how strategies unfold across a story arc and orchestrate them when shaping their own drafts with LLM support.
Grounded in cognitive apprenticeship principles, this workflow encourages learning by exposing narrative strategies from examples, explaining when and why a strategy works, and scaffolding their application in new contexts.}
 


\section{System Design}

We present \tool, an interactive system that supports novice writers in discovering, interpreting, and remixing narrative strategies from examples into their own stories. 
In this section, we first introduce the five design goals, grounded in the principles of cognitive apprenticeship from our related work \secref{sec:apprenticeship}, that guided the development of \tool.
Then, we present the main interfaces and features of \tool and illustrate their use through a usage scenario featuring Zoey, a novice writer, leverages \tool in daily writing practice.
Lastly, we provide the technical details on how narrative examples are processed and modeled to enable these interactions, along with implementation details of the system architecture.\looseness=-1


\subsection{Design Goals}
\label{sec:design_goals}


Ward~\cite{wardStructuredImaginationRole1994} defines example-based creativity as a form of ``structured imagination''--modifying an existing solution and applying it to a new context.
We aim to support novice writers by making tacit narrative strategies visible and guiding them through applying, interpreting, and experimenting with those strategies in their own writing. 
To this end, our system design is grounded in the five components of cognitive apprenticeship~\cite{collinsCognitiveApprenticeship2006}, operationalized through the following design goals.
First, \tool aims to \textbf{model} (\textbf{DG-M}) narrative techniques by surfacing them in example texts, highlighting how they are realized in compelling storytelling moments, Second, it \textbf{coaches} (\textbf{DG-C}) users by directing their attention to specific narrative moves and explaining their contextual effectiveness. Third, it \textbf{scaffolds} (\textbf{DG-S}) the intepretation and application of these strategies by bridging the gap between abstract understanding and concrete writing actions. Fourth, it encourages \textbf{reflection} (\textbf{DG-R}) by helping users compare their own writing with example texts to identify similarities, differences, and areas for improvement. Finally, it enables \textbf{exploration} (\textbf{DG-E}) by assisting users to experiment with combining different narrative strategies and observe how these choices shape storytelling in creative and varied ways.\looseness=-1

\subsection{Interface \& Features} 

Informed by the design goals, we design and develop \tool.
The interface of \tool consists of three coordinated views (\figref{fig:user_interface}), each supporting a key aspect of example-based story writing: a Markdown Editor (\figref{fig:user_interface}A) / Story-Arc Inspector (\figref{fig:story_arc}) (with a mode switch for drafting or visualizing story arcs, \figref{fig:user_interface}D), a Browser (\figref{fig:user_interface}B) for exploring example story blocks and their narrative strategies, and Remixer (\figref{fig:user_interface}C) for remixing and layering strategies across different creative dimensions.
In this section, we will introduce the key features in these three views that help users (1) surface, (2) explore, and (3) remix narrative strategies.

\subsubsection{Surfacing Narrative Strategies}
\label{sec:surfacing_strats}

In Browser (\figref{fig:user_interface}B), each uploaded story is segmented into coherent blocks, each comprising several sentences that represent a distinct narrative beat.
To make abstract strategies visible and understandable to writers, \tool surfaces strategies, explanations, and concrete textual evidence directly within each block.\looseness=-1

\namedparagraph{Strategy Browser}
Every block appears as a card. 
Users can switch between different card views (\figref{fig:user_interface}E): with-plot (a detailed card with the story content of the block, as shown in \figref{fig:user_interface}F), more-brief (a brief summary card without any story content), or more-detailed (a context-rich card showing more story context). 
Users can expand \icon{\faExpand*} cards to view the full story or bookmark \icon{\faBookmark} them for easy access.\looseness=-1

\namedparagraph{Strategy Annotation}
For each card, the system infers one or more narrative strategies, such as
\strategybutton{Sensory Imagery},
\strategybutton{Internal Monologue},
\strategybutton{Personification}~\textbf{[DG-M]}.
To help writers understand how abstract strategies function within authentic story contexts, \tool incorporates an explanation panel for each narrative strategy (\figref{fig:user_interface}F) which activates upon user click. This panel (i) describes how the strategy operates in the story, (ii) lists its dimension tags (\eg character, information, linguistic), and (iii) highlights lexical cues (\eg words, phrases) such as \lexicalcuebutton{glimmering}, \lexicalcuebutton{whispered to herself}, or \lexicalcuebutton{as if the trees could talk} that signal the strategy in context~\textbf{[DG-C]}.
To further support exploration, the system also provides semantic search (\figref{fig:user_interface}G) that allows users to find instances of specific strategies across all uploaded examples based on their names, explanations, and lexical features.\looseness=-1

\subsubsection{Exploring Narrative Strategies}
\label{sec:exploring_strats}

As the segmented story blocks and their strategies accumulate, even a modest set of example stories can quickly balloons into a large and complex information space. Dozens of narratives may expand into hundreds of blocks, each associated with multiple candidate strategies. While this richness offers learning potential, it can also overwhelm writers, especially those seeking targeted inspiration or tactical guidance. 
To help writers find strategies most relevant to their current creative goals, \tool provides interactive visualizations that organize the example space by storytelling intent, both structurally through narrative arcs and functionally through turning points and strategy roles.\looseness=-1

\begin{figure}
\centering
\includegraphics[width=\linewidth]{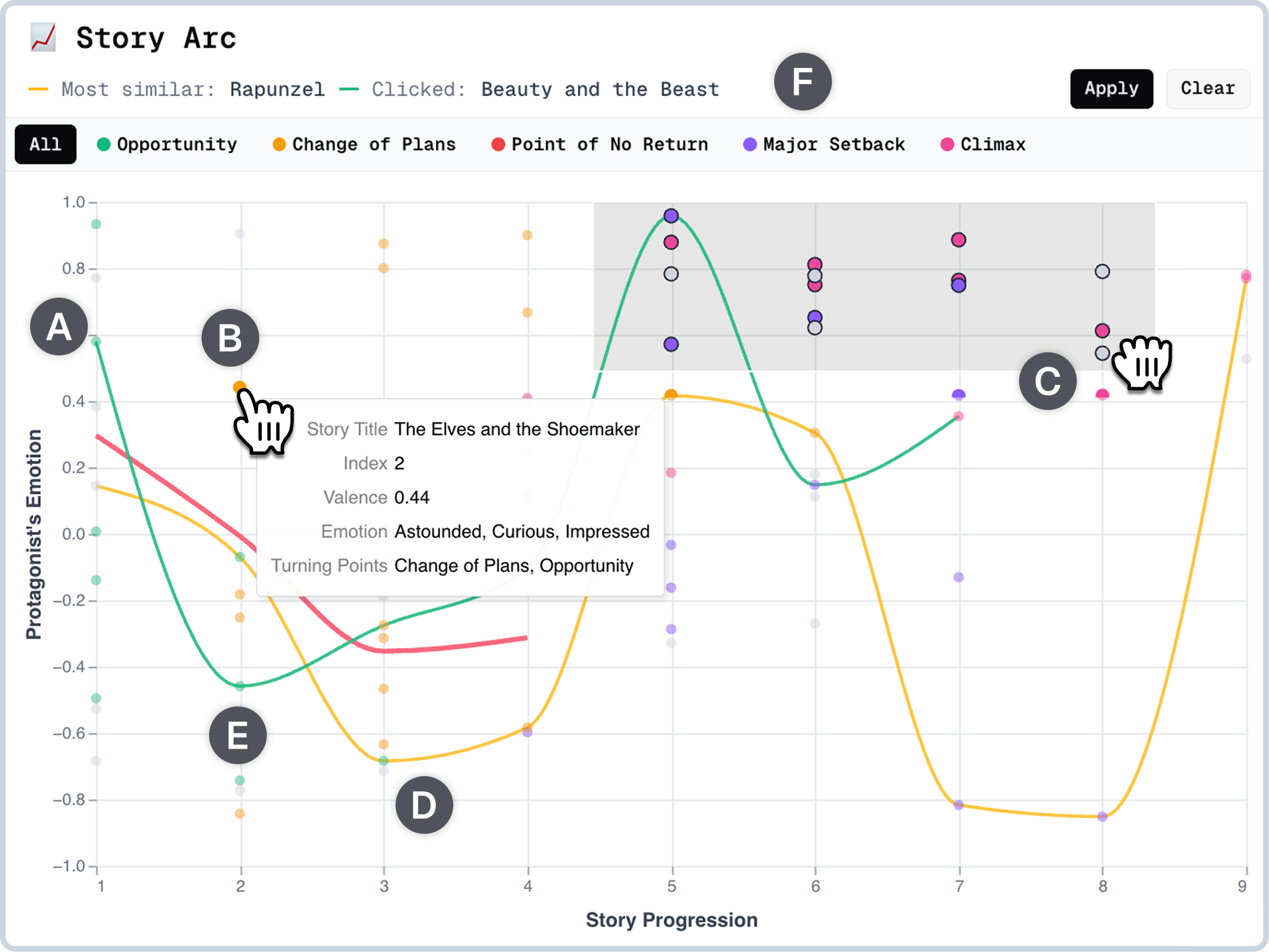}
  \caption{\tool's \textbf{Story-Arc Inspector}, which juxtaposes the user's evolving story arc (red line) with arcs from example stories. Each point represents a story block, positioned by story progression (x-axis) and the protagonist's emotional valence (y-axis). Hovering over a point (B) reveals its details and highlights the corresponding card in the Browser; brushing the points (C) selects a range of story progression and emotion. The system automatically overlays the most similar example arc in yellow (D) and optionally additional arcs in green (E) toggled by user click for comparison. Users can filter story blocks by turning-point type (F) to focus on strategies serving specific narrative purposes.}
  \Description{Narrix's Story-Arc Inspector, which juxtaposes the user's evolving story arc (red line) with arcs from example stories. Each point represents a story block, positioned by story progression (x-axis) and the protagonist's emotional valence (y-axis). Hovering over a point (B) reveals its details and highlights the corresponding card in the Browser; brushing the points (C) selects a range of story progression and emotion. The system automatically overlays the most similar example arc in yellow (D) and optionally additional arcs in green (E) toggled by user click for comparison. Users can filter story blocks by turning-point type (F) to focus on strategies serving specific narrative purposes.}
  \label{fig:story_arc}
\end{figure}

\namedparagraph{Story-Arc Visualization}
A story arc charts the protagonist's emotional journey of a narrative and is widely used to convey storytelling intent~\cite{tianAreLargeLanguage2024}. 
\tool juxtaposes the user's evolving arc with those derived from the example corpus, displayed within a shared coordinate space: story block order appears on the \(x\)-axis and affective valence on the \(y\)-axis. 
The user draft is rendered as a red line (\figref{fig:story_arc}A), while each example block is shown as a scatterplot point \icon[gray]{\faCircle} \textbf{[DG-M]}. 
Selecting or hovering over a point (\figref{fig:story_arc}A) highlights the corresponding card in the Browser (\figref{fig:story_arc}B), while brushing the points (\figref{fig:story_arc}C) allows users to select ranges of story progression and emotion valence \textbf{[DG-S]}.\looseness=-1 

To help writers discover examples with comparable emotional trajectories and draw attention to structurally relevant strategies, \tool automatically identifies and overlays the most similar example arc in yellow (\figref{fig:story_arc}D) (\secref{sec:similarity_story_arc}). 
Optionally, users can overlay additional arcs in green by clicking other points, enabling side-by-side comparison of multiple narrative trajectories \textbf{[DG-R]}.\looseness=-1

\namedparagraph{Turning-Point Filtering}
Turning points (\tpOpportunity{Opportunity}, \tpChangeOfPlans{Change of Plans}, \tpPointOfNoReturn{Point of No Return}, \tpMajorSetback{Major Setback}, and \tpClimax{Climax}) represent the underlying goal or function of individual story blocks and offer a complementary lens for exploring narrative strategies~\cite{tianAreLargeLanguage2024,papalampidiMoviePlotAnalysis2019}.
Each story block in the example corpus is automatically categorized by its turning point type, if any (\secref{sec:turning_points})~\textbf{[DG-M]}.
Users can filter the scatterplot points and Browser cards simultaneously by turning points (\figref{fig:story_arc}F), enabling focused exploration of strategies that serve specific narrative purposes within their story.\looseness=-1

\subsubsection{Remixing Narrative Strategies}
Once they identify preferred strategies, users can incorporate and experiment with them in their own drafts using Remixer interface. 
This view, designed based on the metaphor of a digital audio workstation (DAW), enables writers to apply narrative strategies with fine-grained, track-based control.\looseness=-1

In a DAW, musicians remix a song by arranging sound clips on a multi-track timeline and layering audio filters to achieve the desired effect. 
Analogously, the Remixer interface in \tool arranges story blocks (\cf sound clips) in draft order and allows writers to layer narrative strategies (\cf audio filters) across multiple creative tracks, such as characterization, linguistic style, and information delivery. 
This metaphor provides a familiar mental model of ``drag-and-drop, layer, tweak, and audition'' interactions, enabling users to engage with narrative constructions through what Ward~\cite{wardStructuredImaginationRole1994,reaganEmotionalArcsStories2016} describes as ``structured imagination.'' 
\looseness=-1

\begin{figure}
  \centering
  \includegraphics[width=\linewidth]{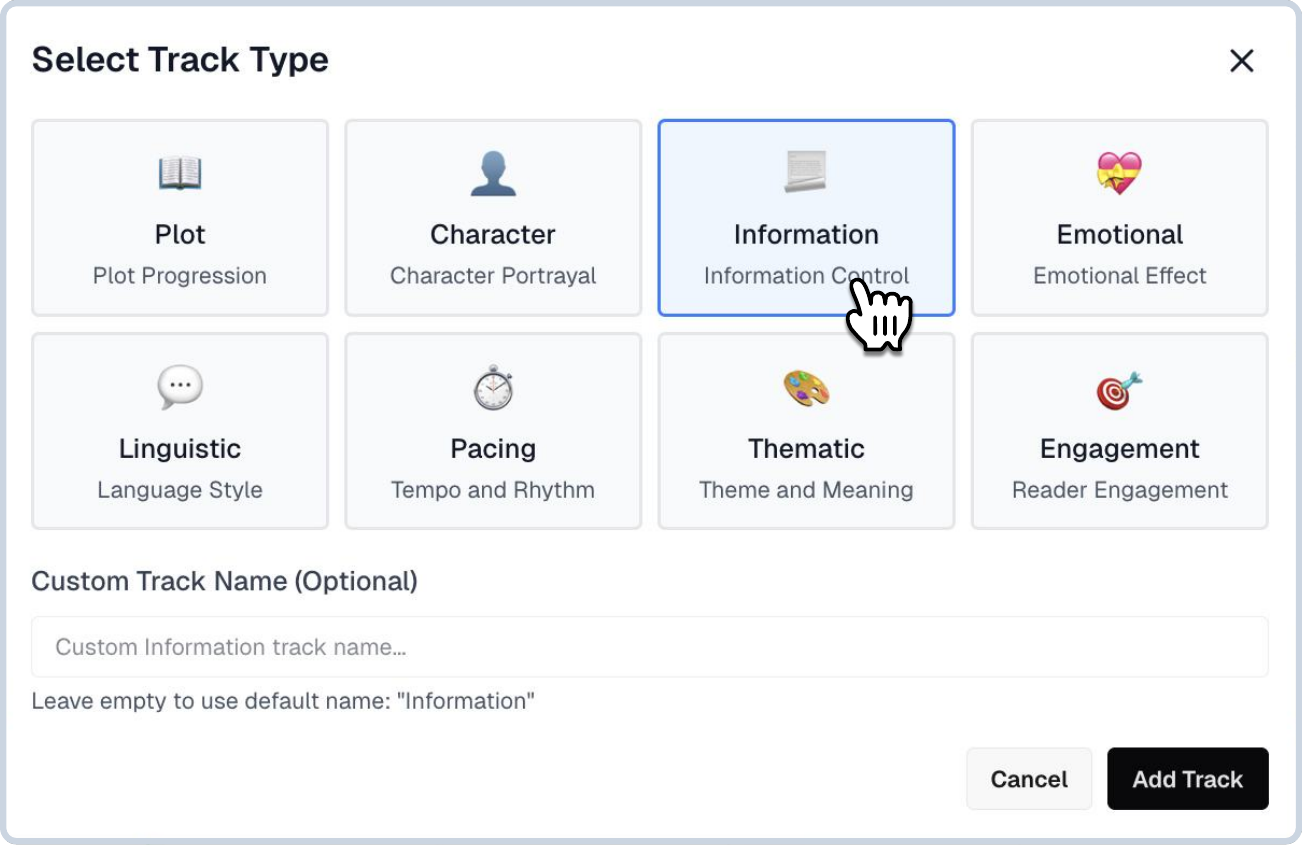}
  \caption{\tool's \textbf{Track Selection} panel for adding tracks of different creative dimensions to the Remix workspace. Users can choose from eight predefined dimensions, \eg Plot, Character, Information, Emotional, Linguistic, Pacing, Thematic, and Engagement. 
  }
  \Description{Narrix's Track Selection panel for adding tracks of different creative dimensions to the Remix workspace. Users can choose from eight predefined dimensions, e.g., Plot, Character, Information, Emotional, Linguistic, Pacing, Thematic, and Engagement.}
  \label{fig:creative_track}
\end{figure}

\namedparagraph{Track Setup} 
The top track (\figref{fig:user_interface}H) is the story track, displaying user-authored blocks in sequence. The system creates new block whenever the user presses Enter in Editor.
Below the story track, users can add tracks that represent different dimensions of storytelling by clicking \buttonlabel{+Track} and selecting from (\figref{fig:creative_track}): 
\cdplot{}, 
\cdcharacter{}, 
\cdinformation{}, 
\cdemotional{}, 
\cdlinguistic{}, 
\cdpacing{}, 
\cdthematic{}, and 
\cdengagement{}.
These dimensions are grounded in foundational theories from narratology and writing studies~\cite{balNarratologyIntroductionTheory2004,kennedyLiteratureIntroductionFiction2016,mckeeStoryStyleStructure1997,princeNarratologyFormFunctioning2012}, capturing key creative aspects emphasized in story writing.
Appendix~\ref{appendix:creative_dimensions} provides detailed definitions of each dimension.

Multiple tracks (of the same or different dimensions) can be added as needed~\textbf{[DG-E]}. As users interact with the Remixer, the Browser remains synchronized with the currently focused track dimension, automatically updating the displayed information in each card with the relevant strategies.\looseness=-1

\begin{figure}
  \centering
  \includegraphics[width=\linewidth]{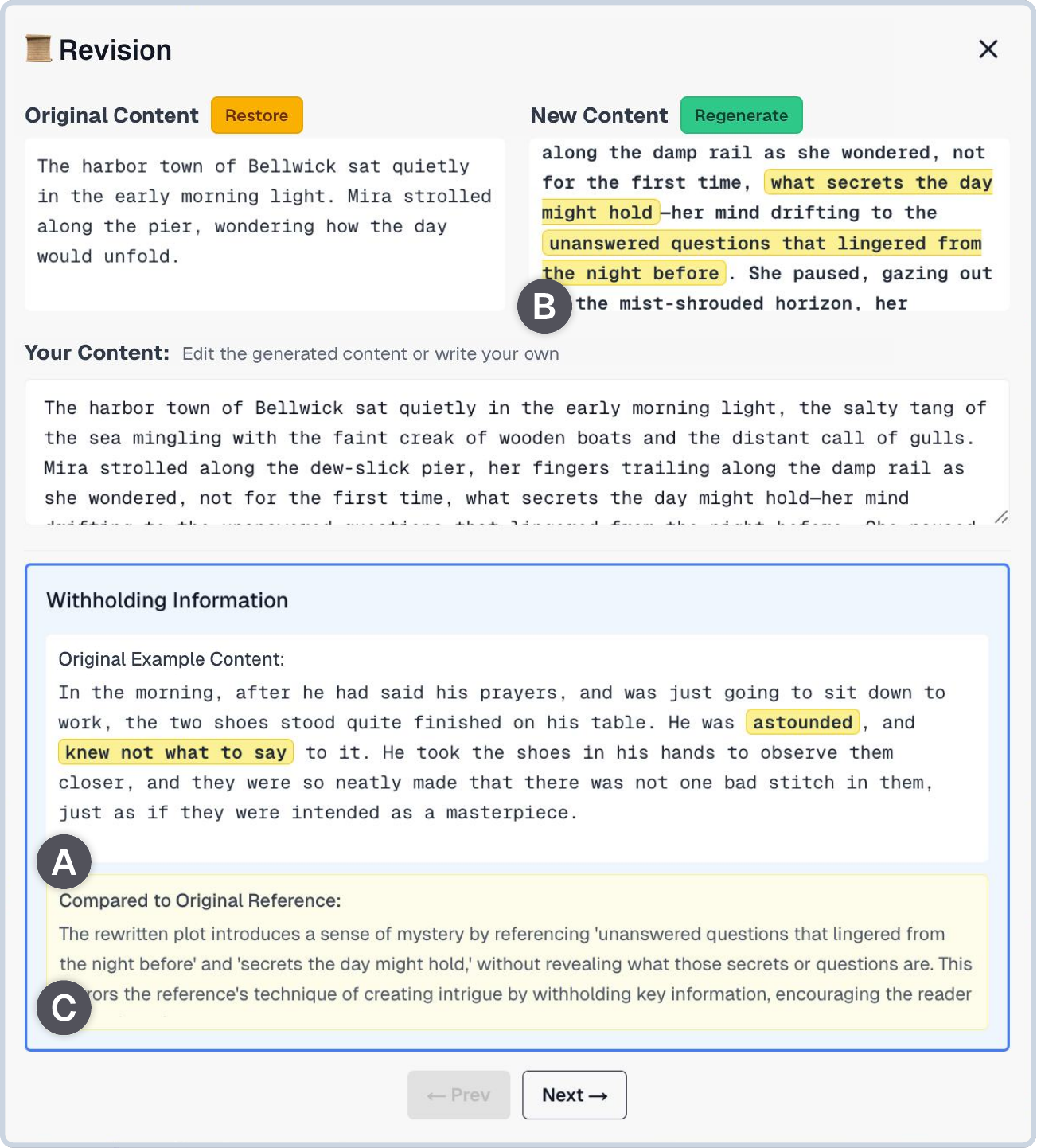}
  \caption{\tool's \textbf{Reflective Comparison} view, which contrasts: (A) the original example content and its highlighted lexical cues for the selected strategy, (B) the revised user content with corresponding cues, and (C) a generated comparison explaining how the strategy is realized in both texts and what differences or similarities exist. 
}
  \Description{Narrix's Reflective Comparison view, which contrasts: (A) the original example content and its highlighted lexical cues for the selected strategy, (B) the revised user content with corresponding cues, and (C) a generated comparison explaining how the strategy is realized in both texts and what differences or similarities exist.}
  \label{fig:ai_output}
\end{figure}

\namedparagraph{Drag-and-Drop Remixing}
Users can drag cards from the Browser onto any track (\figref{fig:user_interface}I).
If multiple strategies in a card are associated with the target dimension, the system prompts the user to select which to apply, based on its explanations and relevant lexical features highlighted in the example.
Selected strategies are displayed as tiles beneath the corresponding block, and users can resize these tiles to span multiple blocks by dragging handles on either side (\figref{fig:user_interface}J).
This drag-and-drop interaction enables users to experiment with different combinations of narrative strategies, supporting planning and creative experimentation~\textbf{[DG-E]}.

\namedparagraph{Revision and Continuation}
When strategy tiles are present under a story block tile, users can click a revise \icon[gray]{\faRedo*} icon (\figref{fig:user_interface}K) to have the system revise that block by applying the selected strategies, while maintaining narrative coherence \textbf{[DG-S]}.
Similarly, if strategies are present beneath the last placeholder tile in the story track, users can click \buttonlabel{Continue Story} (\figref{fig:user_interface}L), optionally provide a short description of the desired next story block, and have the AI continue the story by applying the specified strategies.
Users can further revise, regenerate, or add the generated content to their story. 
An expand \icon[gray]{\faExpand*} icon (\figref{fig:user_interface}K) allows users to review the generation history and restore previous versions as needed.

\namedparagraph{Reflective Comparison}
After each revision or continuation, \tool generates a side-by-side comparison view (\figref{fig:ai_output}) to support learning through contrast and reflection \textbf{[DG-R]}. The view shows: (i) how the strategy operates in the original example, including its lexical features (\figref{fig:ai_output}A); (ii) how it is realized in the revised user content, with corresponding lexical features (\figref{fig:ai_output}B); and (iii) the key differences and similarities between them (\figref{fig:ai_output}C). 
By explicitly connecting the strategies intent with textual outcome, this view helps writers internalize how strategies operate and facilitates the transfer of learned strategies to new contexts.\looseness=-1

\subsection{Example Usage Scenario}

Zoey, a novice fiction writer, is drafting a 1{,}000-word micro story for a community challenge. 
After completing an opening scene, Zoey finds the scene lacking in tension and vivid imagery. To strengthen the draft, they upload ten classic fairy-tale retellings (\eg \textit{Cinderella}, \textit{The Frog Prince}, \textit{Rapunzel}) to \tool to surface narrative strategies and might by repurposed.\looseness=-1

\subsubsection{Browsing and Searching for Narrative Strategies}  
Zoey begins by exploring the Browser, which displays block cards segmented from the uploaded stories, each annotated with candidate strategies.  
Seeking to add suspense to the opening, Zoey types ``suspense'' into the search bar. 
Cards tagged with strategies such as \strategybutton{Withholding Information} rise to the top (\figref{fig:user_interface}F).  
Clicking on this strategy opens a panel (\figref{fig:user_interface}F) to read a detailed explanation panel where Zoey read how it works and study highlighted lexical cues such as \lexicalcuebutton{astounded} and \lexicalcuebutton{knew not what to say}.
Before moving on, Zoey skims several other cards and bookmarks the ones that seem most promising.\looseness=-1

\subsubsection{Remixing Strategies with Creative Tracks}  
To weave these tactics into the opening, Zoey clicks \buttonlabel{+Track} to first add a \cdinformation{} track, then a \cdcharacter{} track, and finally a \cdlinguistic{} track (\figref{fig:user_interface}C).  
Zoey drags \strategybutton{Withholding Information} onto the \cdinformation{} track to heighten narrative stakes; \strategybutton{Sensory Imagery} from \textit{The Frog Prince} goes onto the \cdlinguistic{} track, enriching descriptive detail; and  \strategybutton{Internal Monologue} from \textit{Rapunzel} drops onto the \cdcharacter{} track, revealing the protagonist's unease.  
\tool rewrites the opening paragraphs with these layered strategies after Zoey clicks the revise button (\figref{fig:user_interface}K). 
A comparative panel (\figref{fig:ai_output}) contrasts their use in the exemplars with the revised draft, spotlighting phrases such as \lexicalcuebutton{secrets the day might hold} and \lexicalcuebutton{lingered from the night before}.

\subsubsection{Exploring Strategies via Interactive Story Arcs}  
As the draft develops, Zoey switches to Story-Arc Inspector to reflect on the story's evolving emotional structure.
The protagonist's emotional trajectory appears as a red arc that drops sharply and then plateaus (\figref{fig:story_arc}A). 
\tool automatically highlights the most similar example arc---in this case, from \textit{Rapunzel}---in yellow (\figref{fig:story_arc}D).
Zoey clicks the \tpClimax{Climax} filter to isolate example story blocks that serve as climactic moments (\figref{fig:story_arc}F); Browser cards and scatter points update to show only the narrative endgame.  
Next, Zoey brushes a band along the \(y\)-axis, focusing on points with high emotional intensity, to surface strategies that resolve conflict through dramatic payoff and to refine the set of candidate endings.\looseness=-1

\subsubsection{Guided Continuation and Iterative Refinement}  
To complete the story, Zoey drags \strategybutton{Dramatic Irony} onto the \cdinformation{} track beneath the last placeholder tile and clicks \buttonlabel{Continue Story}.  
After typing \promptsplit{Let readers grasp the hidden truth while the protagonist remains}{unaware.}, \tool generates an ending applying the selected strategy.  
Satisfied with this ending, Zoey revisits the middle paragraphs. By layering additional strategies---\strategybutton{Symbolism} onto the \cdlinguistic{} track and \strategybutton{Moral Dilemma} and \strategybutton{Dialogue for Characterization} onto the \cdcharacter{} track---Zoey deepens tension, voice, and thematic texture, leaving the draft poised for further iterative polish with \tool's strategy-driven support.


\subsection{Technical Details of Processing Narrative Examples}
\label{sec:technial_pipeline}

Our system employs a modular processing pipeline that transforms raw narrative examples into structured, analyzable components that support storytelling.
In this section, we describe the technical pipelines (Fig~\ref{fig:technical_pipeline}) and evaluations for (1) inferring narrative strategies, (2) detecting key turning points from narrative examples, and (3) modeling story arcs.\looseness=-1

\subsubsection{Inferring Narrative Strategies}
\label{sec:narrative_strategies}
\begin{figure*}
\centering
\includegraphics[width=\linewidth]{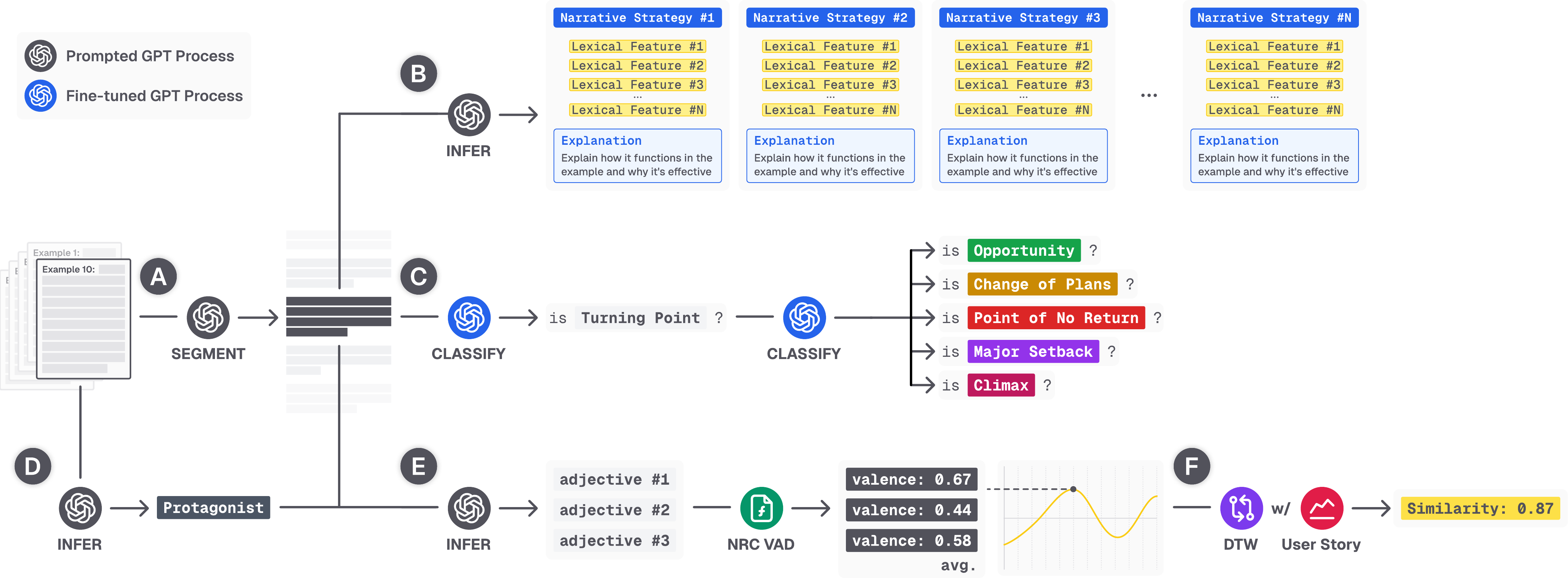}
\caption{\tool's \textbf{technical pipeline} for processing narrative examples. (A) Segment each example story into content blocks. (B) Infer narrative strategies for each block, including a concise label, explanatory description, and highlighted lexical features. (C) Classify each block for the presence of turning points using fine-tuned models. (D) Identify the story's protagonist. (E) Infer the protagonist's emotions per block, map them to valence scores using the NRC VAD Lexicon, and sketch a story arc. (F) Compare the user's arc to example arcs using DTW to retrieve the most similar narrative trajectory.}
\Description{Narrix's technical pipeline for processing narrative examples. (A) Segment each example story into content blocks. (B) Infer narrative strategies for each block, including a concise label, explanatory description, and highlighted lexical features. (C) Classify each block for the presence of turning points using fine-tuned models. (D) Identify the story's protagonist. (E) Infer the protagonist's emotions per block, map them to valence scores using the NRC VAD Lexicon, and sketch a story arc. (F) Compare the user's arc to example arcs using DTW to retrieve the most similar narrative trajectory.}
\label{fig:technical_pipeline}
\end{figure*}

This component is responsible for identifying and explaining abstract narrative strategies within story context. 
The pipeline begins by prompting GPT-4.1 to segment each example story into story blocks, each comprising several sentences that represent distinct narrative beats (Fig~\ref{fig:technical_pipeline}A). 
We then prompt GPT-4.1 to infer narrative strategies (as short labels) from each block, along with explanations and relevant lexical cues (i.e., words or phrases that signal the strategy) (\figref{fig:technical_pipeline}B).
\revision{Given known LLM failure modes, such as hallucination and bias~\cite{zhangNavigatingFogHow2025,jiSurveyHallucinationNatural2023}, we describe our iterative prompt design process, discuss risks and mitigations, and report human evaluation of the final prompts, including representative failure cases.
}

\namedparagraph{Prompting Techniques}
Following established prompt design patterns~\cite{whitePromptPatternCatalog2023b}, we used a two-part prompt comprising a structured system instruction and a specific context. The system prompt positioned the model as an ``expert literary analyst,'' provided clear definitions and scope for ``narrative strategies,'' and specified the required outputs: (1) a concise name, (2) a brief explanation, and (3) verbatim lexical features from the example. The context prompt embedded the story block and reinforced evidence extraction requirements. 
A fixed JSON schema ensured consistent formatting and grounded each strategy in explicit textual evidence.

\revision{We iterated the prompts through short, repeated cycles of diagnostic review, in which we performed systematic inspection of sampled model failures to identify recurring error patterns.
We addressed the error patterns identified in each iteration by adding constraints in the prompt: (a) \textbf{evidence grounding} via mandatory verbatim lexical cues that ties each claim to observable text; (b) a \textbf{constrained schema} reduces free-form generation and simplifies downstream verification; and (c) \textbf{definition priming} that narrows scope to functional narratology rather than plot summary.
These constraints specifically target common failure modes (e.g., over-generalization, fabricated evidence, and hallucinations) and are supported by NLP findings that extractive, source-attributed justifications improve factual faithfulness and auditability~\cite{leiRationalizingNeuralPredictions2016a}, while verification-style prompting has been shown to reduce hallucinations by forcing consistency checks against evidence~\cite{dhuliawalaChainofVerificationReducesHallucination2024}. 
\correct{However, we acknowledge that constraints in a prompt encourage grounding but do not technically guarantee or ``tie'' it in a deterministic sense.
Although such techniques do not eliminate the risk of ungrounded output, they can increase transparency and make failures easier to detect and adjudicate.}
The final prompts are provided in Appendix~\ref{appendix:technical_details} for replication.\looseness=-1
}

\namedparagraph{Expert Evaluation}
Because inferring narrative strategies from text is inherently subjective---with no single correct answer or established ground truth---we conducted a human-centered evaluation with ten creative writing experts, each with over ten years of writing experience.
We processed the dataset from Tian et al.~\cite{tianAreLargeLanguage2024}, and then sampled 100 candidates from output, each consisting of a story block, an inferred strategy, its explanation, and highlighted lexical features. 
Each expert rated 20 unique examples, with two experts per example to support inter-rater reliability analysis.
\looseness=-1

We asked experts to assess two aspects: (1) whether the inferred strategy accurately captured the functional narratology of the story block and whether the accompanying explanation was coherent and insightful; (2) whether the lexical cues reflected the identifies strategy.
Ratings were recorded on a 3-point scale (1 = incorrect, 2 = correct but partially informative, 3 = correct and fully informative).
The intra-class correlation coefficient (\textit{ICC}) was 0.62, indicating good agreement among raters~\cite{cicchettiGuidelinesCriteriaRules1994}.

The average score for strategy quality was 2.54 (\(SD = 0.62\)), with 89\% of examples received a score of 2 or above from both experts, suggesting that hallucinations or major interpretive failures were rare. 
\revision{Among the failure cases, raters noted several recurring patterns: plausible labels but were not sufficiently supported from the local block (e.g., naming a MacGuffin without showing how it drives decisions), explanations that summarized a ``chain reaction'' instead of isolating the causal hinge, or references to a line of dialogue or a setting detail without clarifying their role in the scene.}
Lexical cues received an average rating of 2.48 (\(SD = 0.58\)), with only 8\% of examples rated as incorrect. 
\revision{Errors reported by raters mainly involved generic or misaligned cues, such as selecting long spans instead of pinpointing the words that realize the claimed strategy.}\looseness=-1

Overall, the results suggest our methods lead to model's outputs that were perceived as largely accurate and interpretable.
\revision{Nonetheless, LLM errors cannot be fully eliminated. For users, these errors may surface as overconfident labels, incomplete explanations, or overgeneral cues. 
\tool buffers these risks by (i) surfacing evidence highlights, (ii) giving users control over strategy selection, and (iii) requiring confirmation for every AI change to the text.
We provide a detailed error analysis in Appendix~\ref{appendix:failure_case}, and encourage future work to extend our methods.\looseness=-1
}

\subsubsection{Detecting Turning Points}
\label{sec:turning_points}

To detect turning points in narrative flow, we modeled the task as five independent binary classification problems, one per turning point type $T = \{t_1, t_2, t_3, t_4, t_5\}$, as each story block $b_i$ may contain zero or more turning points($|T_i| \geq 0$), where $T_i \subseteq T$ (\figref{fig:technical_pipeline}C).
For each turning point type $t_k \in T$, we constructed a dataset comprising 300 positive story blocks $P_k = \{b_i \mid b_i \text{ contains } t_k\}$ and 300 negative story blocks $N_k = \{b_j \mid b_j \text{ does not contain } t_k\}$ based on the labels in the dataset of Tian et al.~\cite{tianAreLargeLanguage2024}. Each dataset was split into training (60\%), validation (20\%), and test (20\%) sets.
We then fine-tuned a GPT-4o-based binary classifier $f_k$ for each type of turning point using the corresponding training and validation sets., where $f_k(b) = 1$ if the story block $b$ contains turning point type $t_k$, and $f_k(b) = 0$ otherwise.
The fine-tuned models achieved consistently strong performance across all types on the test set, with average accuracy (0.88), precision (0.87), recall (0.92), and F1 scores (0.89), outperforming a prompting-only GPT-4.1 baseline in terms of accuracy: Opportunity (\textit{ours}: 0.91 vs. \textit{baseline}: 0.78), Change of Plans (\textit{ours}: 0.88 vs. \textit{baseline}: 0.61), Point of No Return (\textit{ours}: 0.85 vs. \textit{baseline}: 0.69), Major Setback (\textit{ours}: 0.84 vs. \textit{baseline}: 0.66), and Climax (\textit{ours}: 0.91 vs. \textit{baseline}: 0.87).\looseness=-1

\subsubsection{Modeling Story Arcs}
This component constructs quantitative representations of narrative structures and enables comparison between stories based on their affective trajectories. 
Here, we first describe how we sketch story arcs and then outline our approach for measuring the similarity between story arcs.\looseness=-1

\namedparagraph{Sketching Story Arcs}
\label{sec:affective_valence}
We followed the method from Tian et al.~\cite{tianAreLargeLanguage2024} to derive emotional arcs from stories.
We first instruct GPT-4o to identify the main character of the story (\figref{fig:technical_pipeline}D). 
Then, for each story block $b_i$ in a narrative, we ask the same LLM to infer three adjectives, $W_i = \{w_{i1}, w_{i2}, w_{i3}\}$, that describe the protagonist's emotions as the story progresses (\eg \textit{amused}, \textit{relaxed}, \textit{anxious}) (\figref{fig:technical_pipeline}E). 
To quantify these emotions, we utilize the NRC VAD Lexicon~\cite{mohammadNRCVADLexicon2025}
to obtain the valence scores of $w_{ij}$, ranging from 0 to 1, where $j \in \{1, 2, 3\}$. 
NRC VAD Lexicon provides human ratings of valence for more than 55,000 English words and phrases.
For each story block, we use the average scores of $w_{ij}$ to represent the valence of $b_i$ obtaining $V(b_i)$, creating a quantitative emotional trajectory for the entire narrative.




\namedparagraph{Measuring Similarity of Story Arcs}
\label{sec:similarity_story_arc}
We compare emotional trajectories (valence curves) using Dynamic Time Warping (DTW)~\cite{berndtUsingDynamicTime1994} (Fig.~\ref{fig:technical_pipeline}F).
Let $A=[a_1,\ldots,a_m]$ and $B=[b_1,\ldots,b_n]$ denote two valence sequences (bounded in $[-1,1]$).
The dynamic programming (DP) recurrence is
\begin{equation}
\begin{aligned}
D_{i,j} &= \lvert a_i - b_j \rvert
+ \min\{ D_{i-1,j},\, D_{i,j-1},\, D_{i-1,j-1} \},\\
&\qquad 1\le i\le m,\; 1\le j\le n.
\end{aligned}
\end{equation}
with boundary conditions enforcing a shared start: \(D_{0,0}=0\), \(D_{i,0}=+\infty\) for \(1\le i\le m\), and \(D_{0,j}=+\infty\) for \(1\le j\le n\).
Given the user story can be incomplete, we allow the alignment to end anywhere in the reference, so the DTW distance is \(D^\star=\min_{1\le j\le n} D_{m,j}\).
Lastly, we convert $D^\star$ to a similarity score $S\in[0,1]$ using an upper-bound normalization:
\begin{equation}
S = \max\!\left(0,\; 1 - \frac{D^\star}{2\cdot\max(m,n)}\right).
\end{equation}
We retrieve the most similar arc by ranking stories in the corpus by $S$ and visualize it alongside the user's arc.

\subsection{Implementation Notes}
\tool is built with the Next.js\footnote{\url{https://nextjs.org/}} framework, which supports server-side rendering for API calls, including those to the OpenAI APIs\footnote{\url{https://openai.com/api/}} for instructing pre-trained GPT models, and the Firebase\footnote{\url{https://firebase.google.com/}} APIs for logging user events.
We use Vega-Lite\footnote{\url{https://vega.github.io/vega-lite/}} to render the interactive story arcs and scatter plots, and BlockNote\footnote{\url{https://www.blocknotejs.org/}} to build the Markdown editor.
Sample prompts and outputs for inferring narrative strategies, generating story content, identifying the protagonist, and extracting emotional adjectives are provided in Appendix~\ref{appendix:technical_details}.\looseness=-1


\section{User Study}
We conducted a \revision{preliminary} within-subjects study with 12 novice story writers to evaluate the efficacy of \tool.
The study aims to answer the following questions:

\begin{enumerate}
    \item[\textbf{RQ1}] \textit{How do novice writers perceive the usability, usefulness, creativity support, and overall experience of \tool?}
    \item[\textbf{RQ2}] \textit{In what ways does \tool help novice writers learn narrative strategies from examples during story writing?}
    \item[\textbf{RQ3}] \textit{In what ways does \tool help novice writers remix narrative strategies from examples into their own stories?}  
\end{enumerate} 

\subsection{Baseline}
Given that no existing solutions are directly comparable to \tool in terms of supporting interaction with narrative strategies in examples, we implemented a baseline system that closely resembled \tool in UI but excluded the key features unique to our approach.

Both systems shared the same Markdown text editor, ensuring a consistent writing environment. However, in the baseline, we removed the interactive story arc visualization. In the Browser panel, example stories were shown as story cards; each card could be clicked to reveal the full text of the example story, but without highlighting or extracting narrative strategies.
The Remix panel in \tool was replaced by an AI chat assistant. This assistant, powered by the same underlying LLM but designed to mimic mainstream chat-based writing interfaces (such as ChatGPT Canvas): users could interact via free-form chat prompts, and the system would respond conversationally in a dedicated output panel. This baseline followed design conventions used in prior work~\cite{rezaABScribeRapidExploration2024,massonTextoshopInteractionsInspired2025,zhangFrictionDecipheringWriting2025}, which also implemented chat-based AI writing assistants as comparison conditions when evaluating novel writing tools.\looseness=-1

In summary, this baseline design (1) retains basic non-contributory features of \tool (\eg the Markdown editor) to minimize interface confounds, (2) uses the same underlying model to isolate interaction design effects, (3) mirrors real-world practices where users have access to general-purpose AI tools like ChatGPT, and (3) integrates story editing, example browsing, and AI assistance within a unified workspace to avoid unnecessary window switching.
A screenshot of the baseline interface is provided in Appendix~\ref{appendix:baseline_interface}.\looseness=-1

\subsection{Participants}
We recruited 12 participants (8 female, 4 male), ages 23--28 (\(M = 26.27\), \(SD = 1.68\)), from a large software organization in the United States via internal communication channels and word of mouth. We sought novice writers and, following prior work that recruited ESL (English as a Second Language) writers as novices for writing tasks~\cite{zhangFrictionDecipheringWriting2025,huangFeedbackOrchestrationStructuring2018}, targeted non-native English speakers during recruitment. All participants reported regularly engaging in writing and wanting to improve their creative writing skills; each had prior creative writing experience and self-rated their expertise on a 7-point scale (\(M = 2.33\), \(SD = 1.07\); 1 = beginner, 7 = professional).\looseness=-1

On a 5-point scale, participants reported actively seeking and referring to examples in writing (\(M = 4.08\), \(SD = 0.90\); 1 = never, 5 = always) and regularly using generative AI tools (\eg ChatGPT) in their writing activities (\(M = 4.67\), \(SD = 0.50\); 1 = never, 5 = daily).
Appendix~\ref{appendix:participant_information} provides detailed participant information (gender, age, creative writing proficiency, example-usage frequency, and AI-tool usage and frequency). We compensated each participant with a \$30 digital gift card.
\looseness=-1

\subsection{Tasks and Materials}

We designed two themed micro-story writing tasks.
Each task prompted participants to write a short story in response to a writing prompt.
This format draw on established microfiction practices, such as flash fiction or five-sentence fiction, which challenge writers to convey a complete story in just a few lines.
The two writing prompt were adapted from the online writing community r/WritingPrompts\footnote{\url{https://www.reddit.com/r/WritingPrompts/}}: (1) write a story that begins with a feeling of joy and ends with a sense of surprise or uncertainty; and (2) write a story that begins with a feeling of confidence and ends with a sense of doubt or realization.
These two prompts were selected because they exemplify widely used creative writing exercises that encourage the practice of core narrative skills and are commonly featured in both online writing communities and traditional writing workshops.\looseness=-1

We collected twenty widely recognized literary stories (\eg \textit{Cinderella}, \textit{Little Red Riding Hood}, and \textit{The Little Match Girl}) as examples.
By providing universally familiar and shared cultural stories as materials, we aimed to reduce extensive reading time required to comprehend unfamiliar plots and help participants focus on the writing task itself.
In addition, the choice of these stories addresses practical considerations, as they are all in the public domain and free from copyright restrictions.
The twenty stories were randomly split into two sets of ten, one for each prompt, controlling the length (1{,}448.8 vs. 1{,}461.6 words) to ensure consistency across tasks. 
We also piloted both writing prompts to ensure they were comparable in difficulty and creative potential. \looseness=-1


\subsection{Study Procedure}
The study began with informed consent and a demographics quetionnaire, followed by a brief orientation introducing the concept of learning from examples, with guidelines adapted from cognitive apprenticeship theory~\cite{collinsCognitiveApprenticeship2006,collinsCognitiveApprenticeshipMaking1991} and community writing resources.
The orientation emphasized actively analyzing example stories to surface underlying strategies, rather than copying surface features, and introduced example strategies.
Participants were instructed to focus on identifying and interpreting deeper techniques from examples, and to consider how they might adapt them in their own writing.\looseness=-1

Participants then completed two 30-minute writing sessions, one with \tool and one with the baseline; the order of systems and task materials was counterbalanced across participants. Each session began with a 3--5-minute tutorial covering key features of the assigned system. Participants were then encouraged to explore the provided example stories and compose a micro-story, using system features to revisit examples and experiment as needed.\looseness=-1

After each story, participants were asked to perform a brief recall exercise, listing narrative strategies learned from the examples and providing a short definition for each, followed by a post-task survey (see §\ref{sec:measures}).
During the recall tasks, participants were not allowed to check the examples or the system.
Both task sessions were completed in a single sitting, with a short break (approximately 5 minutes) between conditions. 
The study concluded with a 15-minute semi-structured interview to gather qualitative reflections on their experiences across the two conditions. 
The entire study lasted approximately 90 minutes per participant.\looseness=-1

\subsection{Measures}
\label{sec:measures}

We aligned our measures directly with the research questions, combining standardized survey instruments, performance in recall tasks, quantitative usage logs, and qualitative coding from interviews.

To address \textbf{RQ1}, we assessed participants' subjective experiences via post-task surveys: the short-form Usability Metric for User Experience (UMUX-LITE)~\cite{lewisUMUXLITEWhenTheres2013}, NASA Task Load Index (NASA-TLX)~\cite{hartDevelopmentNASATLXTask1988}, Creativity Support Index (CSI)~\cite{cherryQuantifyingCreativitySupport2014}, and five items adapted from Wu et al.~\cite{wuAIChainsTransparent2022} to capture experiences using an AI system. All items used a 7-point Likert scale.\looseness=-1

To address \textbf{RQ2}, we measured \textit{information retention}---the immediate internalization and recall of narrative strategies---following prior work~\cite{fowlerEffectivenessHighlightingRetention1974}. 
We assessed the quantity of information retention by counting the number of distinct strategies each participant recalled and and the quality of information retention by coding their written descriptions as vague (0), partial understanding (0.5), or full understanding (1). 
We also included a survey item on self-reported confidence in understanding these strategies, along with two items assessing whether the system helped participants identify narrative strategies in examples and understand strategies in context. 
In interviews, we further probed whether and how the system supported learning narrative strategies and participants' perceptions of its impact on longer-term writing skill development.

To address \textbf{RQ3}, we evaluated how participants remixed strategies into their own story writing.
Two survey items captured self-reported confidence and satisfaction in adapting strategies with each system, supplemented by three items assessing whether the system helped them apply strategies in their stories, reflect on their usage, and experiment with combining strategies for creative exploration. 
We analyzed usage logs to characterize remix behaviors (\eg frequency of revisions/continuations linked to strategies, number of tracks used across creative dimensions, number of distinct strategies applied). 
Finally, interview themes revealed participants' strategies on selecting and remixing narrative strategies, their practices for steering AI generation, and evolving attitudes toward example-based writing across conditions.\looseness=-1

\subsection{Analysis}

\correct{For quantitative measures (\eg Likert-scale responses, number of narrative strategies recalled, and feature-usage logs), we used the Wilcoxon signed-rank test due to the small sample size and the non-normal distribution of the data.}
For qualitative analysis of interview transcripts, we followed established open-coding protocols~\cite{braunUsingThematicAnalysis2006,scupinKJMethodTechnique1997}.
Two researchers independently coded the transcripts, then discussed, reached a consensus, and created a consolidated codebook.
This codebook was then used for thematic analysis to identify emerging topics from the interviews.
The entire research team collectively reviewed the coding outcomes to refine high-level themes.\looseness=-1
\section{Findings}
In the following sections, we will investigate our research questions in depth and present the corresponding findings.
\looseness=-1

\begin{figure*}
\centering
\includegraphics[width=\linewidth]{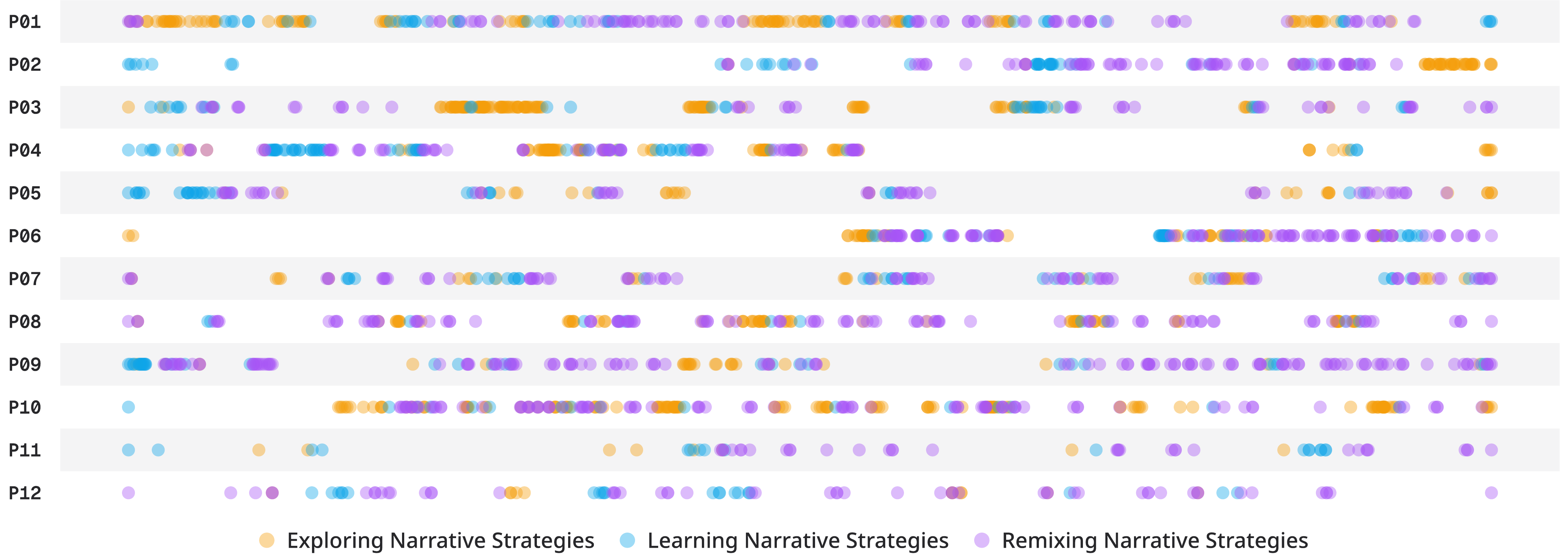}
\caption{The distribution of events related to exploring, learning, and remixing narrative strategies across the normalized time span. Exploring Narrative Strategies (yellow circles) includes actions such as interacting with the interactive story arc and applying filters. Learning Narrative Strategies (blue circles) includes actions such as checking and reading stories, strategies, explanations, and lexical hints. Remixing Narrative Strategies (purple circles) includes actions such as manipulating strategies and tracks within the Remixer, as well as revising or continuing the story.}
\Description{The distribution of events related to exploring, learning, and remixing narrative strategies across the normalized time span. Exploring Narrative Strategies (yellow circles) includes actions such as interacting with the interactive story arc and applying filters. Learning Narrative Strategies (blue circles) includes actions such as checking and reading stories, strategies, explanations, and lexical hints. Remixing Narrative Strategies (purple circles) includes actions such as manipulating strategies and tracks within the Remixer, as well as revising or continuing the story.}
\label{fig:usage_event}
\end{figure*}

\subsection{RQ1: General Usage and Perception of \tool}

\begin{table*}
\centering
\caption{Survey results of perceived experience on task workload, creativity support, and AI system experience under two conditions, where the Wilcoxon signed-rank paired t-test W-values and p-values (*: \(p<.05\), **: \(p<.01\), ***: \(p<.001\)) are reported.}
\label{tab:survey}
\begin{tabular}{llrrrrrl}
\toprule
\multicolumn{2}{c}{\multirow{2}{*}{\textbf{Scales}}} & \multicolumn{2}{c}{\textbf{\tool}} & \multicolumn{2}{c}{\textbf{Baseline}} & \multicolumn{2}{c}{\textbf{Statistics}}  
\\ \cmidrule(lr){3-3}\cmidrule(lr){4-4}\cmidrule(lr){5-5}\cmidrule(lr){6-6}\cmidrule(lr){7-7}\cmidrule(lr){8-8}
\multicolumn{1}{c}{} & \multicolumn{1}{c}{} & \multicolumn{1}{c}{\textbf{M}} & \multicolumn{1}{c}{\textbf{SD}} & \multicolumn{1}{c}{\textbf{M}} & \multicolumn{1}{c}{\textbf{SD}} & \multicolumn{1}{c}{\textbf{W}} & \multicolumn{1}{c}{\textbf{p}} \\ 
\toprule
\multirow{6}{*}{\textbf{NASA Task Load}} & Mental & 4.42 & 1.44 & 4.00 & 2.05 & 23.00 & .525 \\ & Physical & 4.00 & 1.86 & 3.00 & 1.28 & 34.50 & .170 \\ & Temporal & 3.83 & 1.80 & 4.00 & 1.86 & 25.50 & .878 \\ & Performance & 5.42 & 1.08 & 4.17 & 1.12 & 56.00 & .043* \\ & Effort & 3.67 & 1.23 & 4.33 & 1.16 & 2.00 & .089 \\ & Frustration & 2.58 & 1.24 & 3.83 & 1.34 & 2.50 & .019*\\
\midrule
\multirow{5}{*}{\textbf{Creativity Support Index}} & Enjoyment & 6.25 & .87 & 3.58 & 1.78 & 45.00 & .009** \\ & Exploration & 6.08 & 1.00 & 3.42 & 1.88 & 52.00 & .014* \\ & Expressiveness & 6.25 & .75 & 4.33 & 1.67 & 45.00 & .009** \\ & Immersion & 5.08 & 1.50 & 3.08 & 1.56 & 41.00 & .032*\\ & Results Worth Effort & 6.00 & .95 & 4.17 & 1.53 & 63.00 & .008**\\
\midrule
\multirow{5}{*}{\textbf{AI Experience}} & Match Goal & 5.92 & .90 & 4.17 & 1.40 & 43.00 & .017* \\ & Think Through & 5.67 & 1.07 & 2.92 & 1.68 & 53.00 & .010* \\ & Transparent & 6.00 & .95 & 2.67 & 1.50 & 78.00 & .002** \\ & Controllable & 6.00 & 1.13 & 3.33 & 1.56 & 63.00 & .008**\\ & Collaborative & 6.33 & .78 & 3.42 & 1.62 & 55.00 & .006**\\
\bottomrule
\end{tabular}
\end{table*}

\subsubsection{Usage Patterns}
We first examine participant usage of \tool from event logs. 
As shown in \figref{fig:usage_event}, we categorized events into three groups: 
Exploring (\eg interacting with the story arc inspector and applying filters), 
Learning (\eg checking and reading stories, strategies, explanations, and lexical hints), and 
Remixing (\eg manipulating strategies and tracks within the Remixer workspace, as well as revising or continuing the story) Narrative Strategies.\looseness=-1

Most participants (P01, P03-P05, P07-P10) frequently switched among Explore, Learn, and Remix throughout their session, \textbf{engaging in a cyclical process of trying, checking, and adjusting rather than following a linear pipeline}.
On average, Remix (702.56s, 48.56\%) was the most frequent and widely distributed activity, particularly in the mid-to-late stages of the session, serving as the backbone with exploration and learning woven around it.
Exploration (460.79s, 31.85\%) typically appeared early or in bursts to spark new ideas, often preceding or punctuating remix phases.
Learning (283.50s, 19.59\%) occurred in shorter, episodic clusters between remix phases.

\subsubsection{Perceived Usability}
We computed System Usability Scale (SUS) scores using the UMUX-LITE.
Both our system and Baseline received reasonable usability ratings, with \tool scoring 82.64 (typically considered ``good''~\cite{bangorDeterminingWhatIndividual2009}) and Baseline scoring 59.03 (typically considered ``ok'').
A Wilcoxon signed-rank test showed that the SUS score for \tool was significantly higher than that of Baseline (\(W=57.00, p=.036^{*}\)).

\subsubsection{Perceived Cognitive Load}
The overall perceived workload, calculated by averaging all six raw NASA-TLX scores (with the ``Performance'' measure inverted), did not differ significantly (\(W=32.50\), \(p=.638\)) between \tool (\(M_N=3.35\), \(SD_N=.87\)) and Baseline (\(M_B=3.67\), \(SD_B=1.23\)).
However, participants reported being significantly more satisfied with their performance (\(W=56.00\), \(p=.043^{*}\)) when using \tool (\(M_N=5.42\), \(SD_N=1.08\)) compared to Baseline (\(M_B=4.17\), \(SD_B=1.12\)).
They also perceived significantly less frustration (\(W=2.50\), \(p=.019^{*}\)) with \tool (\(M_N=2.58\), \(SD_N=1.24\)) than with Baseline (\(M_B=3.83\), \(SD_B=1.34\)).\looseness=-1

\subsubsection{Perceived Creativity Support}
Participants rated \tool significantly higher than Baseline across multiple CSI dimensions. They reported greater enjoyment (\(M_N=6.25, SD_N=.87\) vs. \(M_B=3.58, SD_B=1.78\), \(W=45.00, p=.009^{**}\)) and immersion (\(M_N=5.08, SD_N=1.50\) vs. \(M_B=3.08, SD_B=1.56\), \(W=41.00, p=.032^{*}\)). \tool also supported exploration (\(M_N=6.08, SD_N=1.00\) vs. \(M_B=3.42, SD_B=1.88\), \(W=52.00, p=.014^{*}\)) and expressiveness (\(M_N=6.25, SD_N=.75\) vs. \(M_B=4.33, SD_B=1.67\), \(W=45.00, p=.009^{**}\)) more effectively than Baseline. Participants further felt their results were more worth the effort (\(M_N=6.00, SD_N=.95\) vs. \(M_B=4.17, SD_B=1.53\), \(W=63.00, p=.008^{**}\)). Together, these results indicate that \tool not only facilitated story writing but also enriched the creative process by making it more engaging, exploratory, and expressive.

\subsubsection{Perceived AI Experience}
Participants also perceived \tool as a more effective and collaborative AI partner. They felt the system better matched their goals (\(M_N=5.92, SD_N=.90\) vs. \(M_B=4.17, SD_B=1.40\), \(W=43.00, p=.017^{*}\)) and encouraged deeper thinking (\(M_N=5.67, SD_N=1.07\) vs. \(M_B=2.92, SD_B=1.68\), \(W=53.00, p=.010^{*}\)). Ratings of transparency (\(M_N=6.00, SD_N=.95\) vs. \(M_B=2.67, SD_B=1.50\), \(W=78.00, p=.002^{**}\)) and controllability (\(M_N=6.00, SD_N=1.13\) vs. \(M_B=3.33, SD_B=1.56\), \(W=63.00, p=.008^{**}\)) were also significantly higher. Importantly, participants highlighted a stronger sense of collaboration with \tool (\(M_N=6.33, SD_N=.78\) vs. \(M_B=3.42, SD_B=1.62\), \(W=55.00, p=.006^{**}\)).
Overall, \tool was experienced as a more collaborative, steerable partner, clarifying what the AI was doing and enabling users to direct assistance toward their narrative intents.\looseness=-1

\subsection{RQ2: Learning Narrative Strategies with \tool}

\subsubsection{Quantitative Findings}
Participants perceived \tool as significantly more effective in helping them identify narrative strategies in examples \textbf{[DG-M]} (\(M_N=6.58, SD_N=.52\) vs. \(M_B=1.92, SD_B=1.51\); \(W=78.00, p=.002^{**}\)) and understand those strategies in context \textbf{[DG-C]} (\(M_N=6.08, SD_N=1.17\) vs. \(M_B=1.67, SD_B=.99\); \(W=78.00, p=.002^{**}\)).\looseness=-1

\begin{figure*}
\centering
\includegraphics[width=\linewidth]{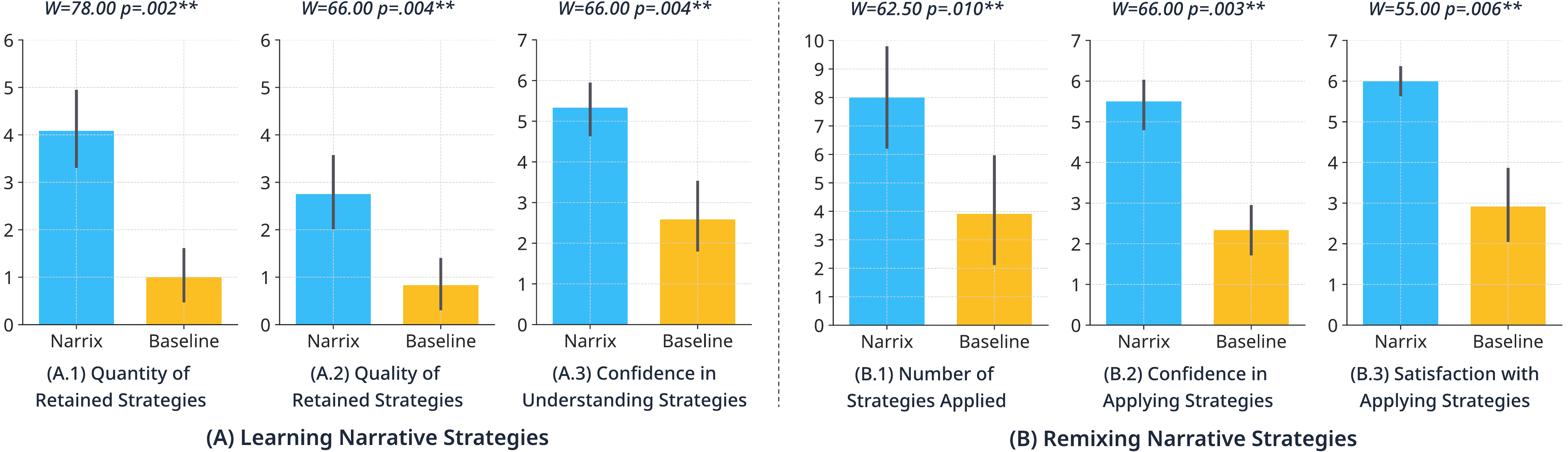}
\caption{Bar plots illustrating the statistical metrics of participant performance in (A) learning narrative strategies and (B) remixing narrative strategies, where the t-values from the Student's paired t-test, W-values from the Wilcoxon signed-rank paired test, and p-values (*: \textit{p}<.05, **: \textit{p}<.01, ***: \textit{p}<.001) are reported. Error bars represent 95\% confidence intervals (CIs).}
\Description{A grouped bar chart comparing Narrix and the baseline on two categories: (A) learning narrative strategies and (B) remixing narrative strategies. Panel A has three sub-metrics: (A.1) quantity of retained strategies, (A.2) quality of retained strategies, and (A.3) confidence in understanding strategies. Panel B has three sub-metrics: (B.1) number of strategies applied, (B.2) confidence in applying strategies, and (B.3) satisfaction with applying strategies. For every metric, Narrix shows higher mean values than the baseline (e.g., Narrix participants retained more strategies, reported higher confidence, and applied more strategies). Error bars represent 95\% confidence intervals (CIs). Statistical test results above each bar pair (t-tests or Wilcoxon tests) show significant differences in all comparisons, favoring Narrix. Overall, the figure indicates that Narrix substantially outperforms the baseline in both learning and remixing narrative strategies.}
\label{fig:performance}
\end{figure*}

\namedparagraph{How participants retained strategies after exposure from the tool}
We examined participants' performance in information retention (\figref{fig:performance}A).  
The results showed that participants recalled significantly more narrative strategies \correct{(\(W=78.00\), \(p=.002^{**}\))} after using \tool (\(M_N=4.08, SD_N=1.51\)) than after using Baseline (\(M_B=1.00, SD_B=.95\)).
When considering the quality of their descriptions of the recalled strategies (\ie their understanding of the strategies), participants using \tool (\(M_N=2.75, SD_N=1.47\)) also achieved significantly higher scores \correct{(\(W=66.00, p=.004^{**}\))} than with Baseline (\(M_B=.83, SD_B=.91\)).
They further reported feeling significantly more confident in their understanding of narrative strategies when supported by \tool (\(M_N=5.33, SD_N=1.16\)) compared to Baseline (\(M_B=2.58, SD_B=1.51\); \(W=66.00, p=.004^{**}\)).\looseness=-1

\begin{figure*}
\centering
\includegraphics[width=\linewidth]{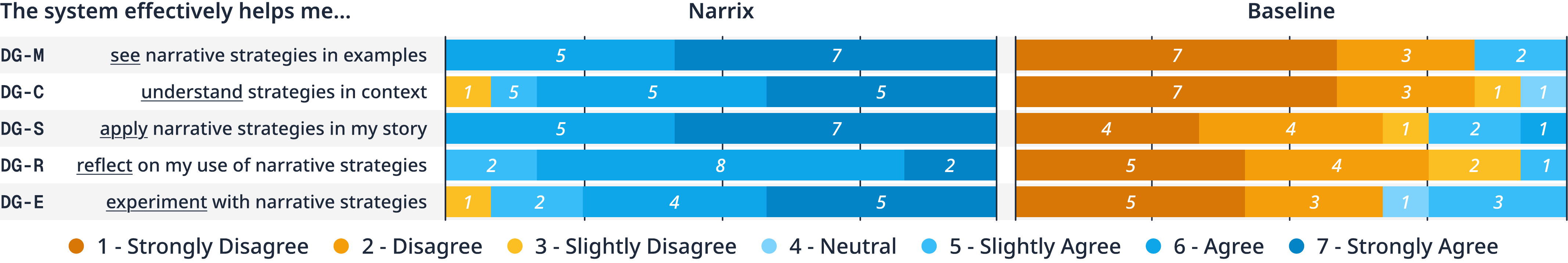}
\caption{Participants' responses to a 7-point Likert-scale questionnaire, assessing perceived support across five design goals (see \secref{sec:design_goals}) in both the Baseline condition and our system.}
\Description{A horizontal stacked bar chart comparing user survey responses for Narrix (our system) versus a baseline across five design goals. The goals are: DG-M, seeing narrative strategies in examples; DG-C, understanding strategies in context; DG-S, applying strategies in a story; DG-R, reflecting on use of strategies; and DG-E, experimenting with strategies. For each goal, responses are shown on a 7-point Likert scale from 1 (strongly disagree) to 7 (strongly agree). Bars for Narrix are dominated by higher agreement values (5–7), while baseline responses cluster more toward disagreement and neutral. Overall, the figure indicates that Narrix is perceived as more effective across all five design goals.}
\label{fig:survey_goal}
\end{figure*}

\subsubsection{Qualitative Findings}
\label{sec:rq2_qual}
Next, we report the key findings identified from our qualitative analysis of interviews regarding how \tool helped participants learn narrative strategies in story writing.\looseness=-1

\namedparagraph{Making tacit strategies explicit and nameable}
Participants consistently emphasized that \tool surfaced strategies they had previously used only ``\textit{by feel},'' making them visible and nameable. 
For example, P01 remarked, ``\textit{before I relied on intuition, but it extracts the strategies and lets me examine them more rationally.}''
Similarly, P05 explained: ``\textit{It let me learn... stories are split into small segments with strategies... I didn't know these strategies before. As a novice, without this assistance I wouldn't know where to start.}''
Naming strategies was seen as the first step toward deliberate learning and transfer.\looseness=-1

\namedparagraph{Supporting intentional and systematic learning}
Participants emphasized that explicit strategy extraction, filtering, and highlighting allowed them to study examples purposefully rather than passively. 
P02 reflected that ``\textit{selecting strategies itself was a learning process... it extracts narrative strategies and I can consciously decide which to learn and use.}''  
P04 praised the ability to click on a strategy ``\textit{to see its description and better understand its definition,}'' and P10 highlighted that strategies were ``\textit{clearly explained, with text passages showing where each strategy applied.}'' 
Others stressed that story arcs made structural choices concrete and helped them link strategies to narrative placement (\eg P03, P04, P10, P11). 
As P04 described, ``\textit{the curve was especially helpful... I could directly see which sentiment I preferred, and clicking a strategy showed the corresponding section, which deepened my understanding of how it was defined.}''\looseness=-1

\namedparagraph{Modeling and coaching strategies through structured exemplars}
Several participants described \tool as ``\textit{like a teacher}'' (P03) or ``\textit{like a mentor}'' (P09) that guided them with structured exemplars. 
Pre-segmented examples reduced the cognitive burden of reading long texts (P07), while visualizations such as story arcs provided scaffolding by making story structure and emotional flow visible (\eg P09, P12). 
This teacher-like quality reflects the principles of cognitive apprenticeship: modeling strategies explicitly and coaching learners through contextual examples. In contrast, the chat baseline was often seen as simply producing text without fostering these learning processes (P11).
P12 similarly noted that Baseline ``\textit{does many steps at once, writing directly to the end,}'' which lowered their involvement.

\namedparagraph{Fostering long-term awareness and transfer}
\revision{Our findings also show preliminary evidence of} longer-term benefits of naming and practicing strategies. 
P02 explained that ``\textit{next time I'll think to apply certain strategies again.}'' 
Similarly, P03, P06, P07, and P10 believed that repeatedly seeing strategies classified and applied would promote durable awareness and skill transfer in their own future writing.
As P03 reflected, ``\textit{With longer use my skills would improve... I'd become more aware of specific strategies I may already use but didn't know by name.}'' 
Echoing this, P07 remarked, ``\textit{...in the future, when I read novels, I'll pay more attention to which strategies the author is using.}''
\revision{Future long-term deployment studies could further quantitatively examine how writers retain and apply the strategies they learn from \tool.}

\subsection{RQ3: Remixing Narrative Strategies with \tool}
\subsubsection{Quantitative Findings}
Participants perceived \tool as significantly more effective in supporting them across three dimensions of remixing (\figref{fig:survey_goal}). 
It helped them apply narrative strategies in their stories \textbf{[DG-S]} (\(M_N=6.58, SD_N=.52\) vs. \(M_B=2.58, SD_B=1.78\); \(W=66.00, p=.003^{**}\)), reflect on their usage of strategies \textbf{[DG-R]} (\(M_N=6.00, SD_N=.60\) vs. \(M_B=2.00, SD_B=1.21\); \(W=78.00, p=.004^{**}\)), and experiment with different combinations of strategies \textbf{[DG-E]} (\(M_N=6.00, SD_N=1.21\) vs. \(M_B=2.50, SD_B=1.73\); \(W=66.00, p=.004^{**}\)).

\revision{To objectively evaluate the writing quality of stories produced using the two systems, we employed the \textsc{Lamp-P-Writing-Quality-RM} model~\cite{chakrabartyAISlopAIPolishAligning2025}, which was trained on expert preference data and has been shown to align with expert judgments.
We used the model to conduct pairwise comparisons between all stories written with \tool and those from the baseline system (12 stories in each condition, yielding 144 cross-set comparisons).
\tool won 107 out of 144 comparisons (74.3\%), and a binomial test confirmed that stories produced with \tool were significantly better than those written using the baseline system (\(p<.001^{***}\)).
}

\namedparagraph{How participants invoked the tool for remixing}
As show in \figref{fig:performance}, participants using \tool (\(M_N=8.00, SD_N=3.25\)) applied significantly more narrative strategies than those using Baseline (\(M_B=3.92, SD_B=3.53\); \correct{\(W=62.50, p=.010^{**}\)}) when steering the AI to revise or continue their stories. 
Participants also reported feeling more confident in applying strategies with \tool (\(M_N=5.33, SD_N=1.16\) vs. \(M_B=2.58, SD_B=1.51\); \(W=66.00, p=.003^{**}\)) and more satisfied with their performance in integrating strategies into their writing (\(M_N=6.00, SD_N=.60\) vs. \(M_B=2.92, SD_B=1.83\); \(W=55.00, p=.006^{**}\)).
\revision{Participants typically drafted their stories themselves and used \tool primarily to revise their own text. On average, they invoked the revision feature 5.42 times (\(SD = 3.29\)) per story, versus only 1.83 uses (\(SD = 2.41\)) of the AI continue function. 
Put differently, about 75\% of \tool interactions started from user-written text followed by AI revisions, while only 25\% were direct AI continuations. 
This pattern indicates that participants used \tool mainly to explore and apply narrative strategies to improve their drafts rather than delegating content generation to the AI.}\looseness=-1

\begin{table*}
  \centering
  \caption{Usage of tracks across creative dimensions in the Remixer workspace of \tool. Row 1 shows the number of tracks per dimension added by all participants; Row 2 shows the number of distinct strategies added within those tracks from all participants.}
  \Description{}
  \label{tab:dimension_usage}
  \begin{tabular}{lcccccccc|c}
    \toprule
    & \textbf{Plot} & \textbf{Character} & \textbf{Information} & \textbf{Emotional} & \textbf{Linguistic} & \textbf{Pacing} & \textbf{Thematic} & \textbf{Engagement} & \textbf{Total} \\
    \midrule
    \textbf{\# Tracks} & 9 (23.68\%) & 6 (15.79\%) & 3 (7.89\%) & 10 (26.32\%) & 3 (7.89\%) & 3 (7.89\%) & 1 (2.63\%) & 3 (7.89\%) & 38 \\
    \textbf{\# Strategies} & 29 (30.21\%) & 19 (19.79\%) & 5 (5.21\%) & 18 (18.75\%) & 10 (10.42\%) & 5 (5.21\%) & 0 (0.00\%) & 10 (10.42\%) & 96 \\
    \bottomrule
  \end{tabular}
\end{table*}

\namedparagraph{What dimensions were remixed in the workspace}
As summarized in Table~\ref{tab:dimension_usage}, participants created 38 total tracks in the Remixer (3.17 per participant) and added 96 distinct strategies overall (8.00 per participant). Track usage was concentrated in the Emotional (26.32\%) and Plot (23.68\%) dimensions, followed by Character (15.79\%), with smaller but notable use of Information, Linguistic, Pacing, and Engagement (each 7.89\%). Strategy additions were led by Plot (30.21\%), followed by Character (19.79\%) and Emotional (18.75\%), with Linguistic and Engagement contributing 10.42\% each. Information and Pacing were less frequently used (5.21\% each), and Thematic showed minimal uptake (1 track, 0 strategies). Overall, participants focused their remixing on plot turns and emotional shaping, while thematic scaffolding was rarely used.\looseness=-1

\subsubsection{Qualitative Findings}
\label{sec:rq3_qual}
Next, we present key findings from our qualitative analysis of interviews regarding how \tool supported participants in remixing narrative strategies during story writing.

\namedparagraph{Encouraging directed yet open-ended creative exploration}
Participants described \tool as expanding idea space and enabling more expressive outcomes. Curated, strategy-tagged examples ``\textit{remind[ed] me of techniques and spark[ed] ideas}'' (P01), while simplifying text into selectable strategies yielded ``\textit{faster, higher-signal ideation}'' than reading long generations (P02). Several participants reported concrete gains in expressiveness---\eg automatically calling back earlier motifs for foreshadowing at the ending (P03)---and cited the arc visualizations as a prompt for directions they would not have considered otherwise (P09, P12). Overall, \tool made creative exploration feel directed yet open-ended.\looseness=-1

\revision{
\namedparagraph{Connecting high-level story goals with concrete writing techniques} 
Participants selected narrative strategies largely based on their storytelling goals.
They typically followed a three-step process, starting from (1) using the story-arc visualization to identify examples that pursued similar goals (\eg turning points, emotional shifts), (2) reading those examples to understand which narrative strategies were used and how, and (3) applying the same strategies in their own drafts to test whether they achieved the desired effect.
For example, P08 described planning the overall style first and then selecting strategies to realize it: ``\textit{I first think about the style and whether the plot should stay steady or rise and fall... then I look for something that matches that plan and put it into my story.}'' 
P12 used the story arc as feedback to evaluate whether applying a certain strategy shifted the narrative in the intended direction: ``\textit{The visualization was rising, but after I rewrote it, it turned downward... the chart made that visible.}'' 
P10 studied how exemplar blocks deploy strategies at specific structural moments (\eg a happy opening) to identify similar techniques to use: ``\textit{If the opening is happy, I find the point closest to 1.0 on the arc, then look for strategies there to understand how they're used.}''\looseness=-1
}


\namedparagraph{Scaffolding targeted, controllable remixing by tracks and block-based edits}
Participants used tracks to map priorities (\eg plot, character, emotion) and to ``\textit{spec}'' tactics per scene, then invoked block-scoped generation for targeted local changes (\eg P02, P04, P05, P06, P08, P12). 
They described dragging strategies tied to the current scene intent, performing segment-level rewrites, and immediately inspecting effects (often via the story arc). 
For example, P06 emphasized the value of block-scoped rewrites for fine-grained control: ``\textit{I would pick a strategy and apply it only to a segment, then rewrite that part... this gave me more control than letting the AI regenerate everything.}'' 
Several highlighted the immediate feedback of the emotion curve, as P12 explained: ``\textit{I could see how the curve changed after a rewrite, which made it clear whether the strategy worked.}''\looseness=-1 

\namedparagraph{Enhancing controllability with block-level, strategy-steered generation}
Participants contrasted the chat baseline's \emph{summarize-generate} workflow with \tool's block-level, strategy-based steering. 
They noted that chat was good for quick pre-writing or one-shot follow-ups but offered low strategy salience and limited fine-grained control (\eg P02, P04, P07, P08, P09, P11, P12). 
With chat, they often fed an outline or whole draft and then pruned unintended additions; with \tool, they targeted specific passages and effects using named strategies (\eg P04, P06). 
Several emphasized that \tool reduced prompt guesswork and clean-up while increasing ownership of the revision process; chat felt more like ``\textit{AI writes for you}'' (P08, P12).








\section{Discussion}
In this paper, we propose and evaluate \tool, an interactive system that supports novice writers in discovering, interpreting, and remixing narrative strategies from examples into their own stories. Based on findings from system design and user evaluation, we discuss key design implications for future writing interfaces and creativity support tools.\looseness=-1


\subsection{Reifying Tacit Knowledge into Visible, Reusable Units}
Our results show that making tacit techniques explicit, nameable, and placeable turns ``\textit{writing by feel}'' into deliberate practice (\secref{sec:rq2_qual}) and targeted remix (\secref{sec:rq3_qual}).
This design pattern generalizes to other creative domains (\eg UI design). 
For instance, Misty~\cite{luMistyUIPrototyping2025} supports aspect-level blending (\eg color, layout, content) from exemplars into work-in-progress designs; while it does not explicitly name strategies, our findings suggest that exposing techniques as explicit units would further strengthen learning, control, and reuse.
To this end, future interfaces could:
(i) extract and label tacit techniques directly in examples and user work;
(ii) bind each technique to where it occurs (\eg section/beat, UI element) and why it works (intended effect);
(iii) represent techniques as lightweight, manipulatable tokens (\eg chips on tracks) that can be inserted, reordered, or removed within a work; and
(iv) preview local impact before committing (\eg emotion/arc deltas in writing; simulated user interactions in UI).\looseness=-1

\subsection{Building Personal ``Strategy Libraries'' for Transfer and Long-Term Reuse}
\label{sec:strategy_libraries}
Participants reported growing awareness and reuse intentions after naming and practicing strategies (\secref{sec:rq2_qual}). 
Future systems could therefore support ``personal strategy libraries'': save any strategy the user tries (or edits) along with micro-examples, before/after diffs, and tags for goal, tone, and placement. 
Beyond a single writing sessions, future tools could also let users collect strategies during everyday reading, Pinterest-style~\cite{linderEverydayIdeationAll2014}. 
Concretely: provide a lightweight reader/extension to ``pin'' snippets from the web, PDFs, or e-books; run the paper's techniques (see \secref{sec:technial_pipeline}) in the capture flow (\eg automatic strategy extraction, sentence-level highlights, and affect/arc estimation) to store each pin as a strategy token with provenance (source, author), context (where in the narrative it occurs), intended effect, and representative text.
Over time, surface analytics to reveal patterns (e.g., what techniques the author tends to use/avoid) and suggest complementary or under-used strategies. This will support long-term skill development, enabling users to build repertoire across sessions and genres. \looseness=-1 

\subsection{Steering Generation with Abstract Strategies and Concrete Intent}
\secref{sec:rq3_qual} shows users prefer to state an effect (goal-driven) or browse and try (exploratory), then place strategies at specific locations.
Future work could further support steering content generation by combining abstract strategies and concrete intent specifications.
For example, 
tools could (i) accept intent specifications (desired effect, narrative position, constraints) and compile them into strategy bundles that guide localized generation; 
(ii) expose both abstract knobs (like in~\cite{chungPatchviewLLMpoweredWorldbuilding2024a} and~\cite{luWhatELSEShapingNarrative2025}) and concrete moves (\eg ``\textit{echo earlier image},'' ``\textit{interiority: sensory detail + self-talk}'') so users can steer at the right level; 
(iii) enable fast retrieval by intent (\eg ``\textit{raise stakes at midpoint},'' ``\textit{amplify interiority}'') and recommend previously successful moves when similar contexts recur, helping writers reuse and adapt strategies effectively across different drafts.\looseness=-1

\revision{
\subsection{Linking Local Strategies to Global Story Structure}
\tool primarily emphasizes surfacing strategy applied at the paragraph or block level, which is an intentional first step that gives novice writers more manageable techniques to experiment with before tackling higher-level story structure. While the story-arc visualization provides some connection between local and global reasoning, helping writers reflect on emotional trajectories and structural turning points (\secref{sec:rq3_qual}), participants sometimes noticed small inconsistencies when strategies were combined across sections.
Because broader narrative coherence depends on linking local moves to global story logic (\eg character arcs, causal progression, and structures like the Hero's Journey~\cite{mirowskiCoWritingScreenplaysTheatre2023}), a natural next step is to extend our block-scoped, controllable design toward multi-level coherence.
Concretely, we envision future extensions on hierarchical tracks (like in~\cite{mirowskiCoWritingScreenplaysTheatre2023}) that tie block-level strategies to overarching trajectories, cross-block coherence checks that flag drift or redundancy, and visual summaries of global patterns (such as cumulative emotional arcs or character development) to help writers see how local revisions accumulate into story-wide structure.\looseness=-1
}

\subsection{Blending Chat for Pre-Writing with Controllable Remix for Refinement}
Participants saw value in both chat-style interfaces (for fast ideation and upfront drafting) and \tool's strategy-steered editing (for precise and fine-grained control) (\secref{sec:rq3_qual}). 
Future systems could \emph{hybridize} the strengths: (i) a Pre-write/Chat space for rapid ideation and scaffolding (summaries, beat lists), followed by (ii) a Remix/Edit space where strategy tokens are placed on tracks and applied locally; 
(iii) provide a seamless handoff (import outline → auto-seed tracks and candidate strategies), and a reversible ``promote to draft / demote to sketch'' mechanism so users can fluidly move between broad strokes and fine control;
and (iv) default to local-first edits to reduce unintended global changes in chat-only one-shot rewrites.\looseness=-1

\subsection{\revision{Supporting Cognitive Apprenticeship with AI for Long-Term Learning}}
\tool operationalizes the five components of cognitive apprenticeship ~\cite{collinsCognitiveApprenticeship2006}: modeling, coaching, scaffolding, reflection, and exploration (\secref{sec:design_goals} and \figref{fig:survey_goal}). 
\revision{Previous work has shown that cognitive apprenticeship can foster long-term skill development and knowledge retention~\cite{collinsCognitiveApprenticeshipMaking1991,wangInvestigatingImpactStratified2024} and shows promise in computer-based learning environments~\cite{hennessySituatedCognitionCognitive1993}. 
Building on this, we can expect that future writing tools like \tool can position the AI not merely as a co-author that generates text, but as a long-term ``cognitive mentor'' for users that externalizes reasoning and augments long-term writing skill development.
For example,} each suggestion should carry a rationale (modeling), surface provenance from examples (coaching), and support quick ``apply this effect'' (scaffolding) or ``show alternatives'' (exploration) actions. 
\revision{Lightweight prompts to encourage reflection, like in \cite{zhangFrictionDecipheringWriting2025}, would further help users develop strategy awareness rather than deferring entirely to AI outputs.} 
\revision{Finally, logging applied strategies into a personal library (\secref{sec:strategy_libraries}) could transform momentary co-creation into cumulative learning resources and gains, aligning with the apprenticeship cycle of moving from guided observation to independent, strategic practice.\looseness=-1}

\section{Limitations}
Our system design has several limitations.
First, \revision{despite our mitigation strategies, LLM errors cannot be fully eliminated from our technical pipeline due to the statistical nature of these models.
We encourage future work to enhance our methods for extracting narrative strategies, such as by establishing benchmarks, developing specialized models, and post-training LLMs to better align with expert narrative analysts.}
\revision{Second}, while \tool makes individual strategies explicit and manipulatable, it offers limited support for global-level strategies, such as maintaining coherence across multiple sections or harmonizing style across blocks.
Some participants noted inconsistencies when combining strategies across tracks or blocks, suggesting the need for better mechanisms to manage cross-cutting narrative patterns. 
\revision{Third}, although the current interface allows users to apply strategies iteratively, it offers limited functionality for comparing different combinations in parallel.
Future iterations could address this gap by enabling side-by-side prototyping and evaluation of multiple strategy combinations.
Finally, writing excessively long stories (such as a 50-page novel) remains a challenge; future iterations could incorporate hierarchical tracks (from chapter to scene to paragraph), batched apply/rollback of strategies, and cross-block coherence checks (style, voice, references) to keep long-form revision tractable.\looseness=-1

Our study methodology also has limitations. 
\revision{First, w}e recruited a relatively small sample of 12 participants, \revision{all non-native English speakers recruited from a single software organization.}
Additionally, to manage potential confounds and keep sessions within time limits, we controlled task materials with preset across conditions.
\revision{The fairy-tale examples used are culturally homogeneous, which may not represent the diversity of narratives that writers engage with, limiting the generalizability of our findings to broader global contexts.
While we computed statistical significance, we do not claim the results to be conclusive but should be interpreted as promising but preliminary.
The specific demographics of our participants may also have influenced how they interacted with the system. 
As non-native English speakers, who are typically recruited as novice writers in prior work~\cite{zhangFrictionDecipheringWriting2025,huangFeedbackOrchestrationStructuring2018}, they might have benefited more from the system's explicit scaffolding of narrative strategies and lexical cues, compared to native speakers who could draw more intuitively on their linguistic and cultural repertoire.
Similarly, the use of well-known fairy tales, while helpful for controlling familiarity, may have biased participants toward recognizing conventional Western narrative structures and emotional arcs.
Future work should examine how \tool performs with a more diverse population of writers and story corpora, including different genres, languages, and cultural traditions, to better understand how narrative strategy learning and remixing generalize across contexts.
}

\revision{Finally, our study only provides initial evidence of users' short-term learning gains in awareness and retention.
To understand the longer-term impact of \tool on writing skill development (\eg long-term retention of strategies, transfer to independent writing), we plan to conduct a longitudinal study that follows participants over multiple weeks or months, tracking how their strategy use evolves across repeated writing tasks and whether they continue to apply and adapt learned strategies in their independent practice.}\looseness=-1


\section{Conclusion}
We presented \tool, a writing interface that makes narrative strategies in examples visible, understandable, and ready to remix. 
Our \revision{preliminary user} study with 12 novice writers suggests that \tool shows promise in shifting AI-assisted writing away from one-shot content generation toward a more deliberate, transparent, and controllable practice, where examples become teachable moments and edits become opportunities to learn the craft.
Looking forward, we envision future writing tools that not only help authors produce text, but also cultivate their narrative repertoire, supporting long-term creative growth, transferable skills, and more collaborative forms of human-AI co-creation.
\begin{acks}
We sincerely thank our participants for sharing their feedback on our system, and we are grateful to all reviewers for their valuable insights and suggestions. This research was conducted during an internship at Adobe Research. We thank the Adobe Research EEL group for their guidance and support throughout this work.
\end{acks}

\balance
\bibliographystyle{ACM-Reference-Format}
\bibliography{reference}

@String(CHI  = {Proceedings of the SIGCHI Conference on Human Factors in Computing Systems})

@String(DIS  = {Proceedings of the SIGCHI Conference on Designing Interactive Systems})

@String(CSCW  = {Proceedings of the ACM Conference on Computer Supported Cooperative Work and Social Computing})

@article{charneyLearningWriteGenre1995,
  title = {Learning to Write in a Genre: What Student Writers Take from Model Texts},
  shorttitle = {Learning to Write in a Genre},
  author = {Charney, Davida H. and Carls, Richard A.},
  year = {1995},
  month = feb,
  journal = {Research in the Teaching of English},
  volume = {29},
  number = {1},
  pages = {88--125},
  issn = {0034-527X, 1943-2348},
  doi = {10.58680/rte199515358},
  urldate = {2025-06-16},
  langid = {english}
}

@inproceedings{dangCorpusStudioSurfacingEmergent2025,
  title = {CorpusStudio: Surfacing Emergent Patterns In A Corpus Of Prior Work While Writing},
  shorttitle = {CorpusStudio},
  booktitle = {Proceedings of the 2025 CHI Conference on Human Factors in Computing Systems},
  author = {Dang, Hai and Swoopes, Chelse and Buschek, Daniel and Glassman, Elena L.},
  year = {2025},
  month = apr,
  series = {CHI '25},
  pages = {1--19},
  publisher = {Association for Computing Machinery},
  address = {New York, NY, USA},
  doi = {10.1145/3706598.3713974},
  urldate = {2025-05-23},
  isbn = {979-8-4007-1394-1},
  langid = {american}
}

@article{driscollGenreKnowledgeWriting2020,
  title = {Genre Knowledge and Writing Development: Results From the Writing Transfer Project},
  shorttitle = {Genre Knowledge and Writing Development},
  author = {Driscoll, Dana Lynn and Paszek, Joseph and Gorzelsky, Gwen and Hayes, Carol L. and Jones, Edmund},
  year = {2020},
  month = jan,
  journal = {Written Communication},
  volume = {37},
  number = {1},
  pages = {69--103},
  issn = {0741-0883, 1552-8472},
  doi = {10.1177/0741088319882313},
  urldate = {2025-06-16},
  langid = {english}
}

@inproceedings{huiIntroAssistToolSupport2018,
  title = {IntroAssist: A Tool to Support Writing Introductory Help Requests},
  shorttitle = {IntroAssist},
  booktitle = {Proceedings of the 2018 CHI Conference on Human Factors in Computing Systems},
  author = {Hui, Julie S. and Gergle, Darren and Gerber, Elizabeth M.},
  year = {2018},
  month = apr,
  series = {CHI '18},
  pages = {1--13},
  publisher = {Association for Computing Machinery},
  address = {New York, NY, USA},
  doi = {10.1145/3173574.3173596},
  urldate = {2025-05-21},
  isbn = {978-1-4503-5620-6},
  langid = {american}
}

@inproceedings{huiLettersmithScaffoldingWritten2023,
  title = {Lettersmith: Scaffolding Written Professional Communication Among College Students},
  shorttitle = {Lettersmith},
  booktitle = {Proceedings of the 2023 CHI Conference on Human Factors in Computing Systems},
  author = {Hui, Julie and Sprouse, Michelle L.},
  year = {2023},
  month = apr,
  series = {CHI '23},
  pages = {1--17},
  publisher = {Association for Computing Machinery},
  address = {New York, NY, USA},
  doi = {10.1145/3544548.3581029},
  urldate = {2025-05-21},
  isbn = {978-1-4503-9421-5},
  langid = {american}
}

@misc{itoLangsmithInteractiveAcademic2020,
  title = {Langsmith: An Interactive Academic Text Revision System},
  shorttitle = {Langsmith},
  author = {Ito, Takumi and Kuribayashi, Tatsuki and Hidaka, Masatoshi and Suzuki, Jun and Inui, Kentaro},
  year = {2020},
  month = oct,
  number = {arXiv:2010.04332},
  eprint = {2010.04332},
  primaryclass = {cs},
  publisher = {arXiv},
  urldate = {2024-04-05},
  archiveprefix = {arXiv},
  langid = {american}
}

@article{naeemSmartEmailClient2018,
  title = {A Smart Email Client Prototype for Effective Reuse of Past Replies},
  author = {Naeem, M. Asif and Linggawa, I. Wayan S. and Mughal, Aftab A. and Lutteroth, Christof and Weber, Gerald},
  year = {2018},
  journal = {IEEE Access},
  volume = {6},
  pages = {69453--69471},
  issn = {2169-3536},
  doi = {10.1109/ACCESS.2018.2878523},
  urldate = {2025-06-17}
}

@book{tardyBuildingGenreKnowledge2009,
  title = {Building Genre Knowledge},
  author = {Tardy, Christine},
  year = {2009},
  publisher = {Parlor Press LLC},
  urldate = {2025-06-16},
  address = {Anderson, South Carolina, USA}
}

@article{tardyTeachingResearchingGenre2020,
  title = {Teaching and Researching Genre Knowledge: Toward an Enhanced Theoretical Framework},
  shorttitle = {Teaching and Researching Genre Knowledge},
  author = {Tardy, Christine M. and {Sommer-Farias}, Bruna and Gevers, Jeroen},
  year = {2020},
  month = jul,
  journal = {Written Communication},
  volume = {37},
  number = {3},
  pages = {287--321},
  issn = {0741-0883, 1552-8472},
  doi = {10.1177/0741088320916554},
  urldate = {2025-06-16},
  langid = {english}
}

@inproceedings{zhangMathemythsLeveragingLarge2024,
  title = {Mathemyths: Leveraging Large Language Models to Teach Mathematical Language through Child-AI Co-Creative Storytelling},
  shorttitle = {Mathemyths},
  booktitle = {Proceedings of the 2024 CHI Conference on Human Factors in Computing Systems},
  author = {Zhang, Chao and Liu, Xuechen and Ziska, Katherine and Jeon, Soobin and Yu, Chi-Lin and Xu, Ying},
  year = {2024},
  month = may,
  series = {CHI '24},
  pages = {1--23},
  publisher = {Association for Computing Machinery},
  address = {New York, NY, USA},
  doi = {10.1145/3613904.3642647},
  urldate = {2024-11-14},
  isbn = {979-8-4007-0330-0},
  langid = {american}
}

@inproceedings{changWriteAhead2MiningLexical2015,
  title = {WriteAhead2: Mining Lexical Grammar Patterns for Assisted Writing},
  shorttitle = {WriteAhead2},
  booktitle = {Proceedings of the 2015 Conference of the North American Chapter of the Association for Computational Linguistics: Demonstrations},
  author = {Chang, Jim and Chang, Jason},
  year = {2015},
  pages = {106--110},
  publisher = {Association for Computational Linguistics},
  address = {Denver, Colorado},
  doi = {10.3115/v1/N15-3022},
  urldate = {2025-06-16},
  langid = {english}
}

@inproceedings{kangParagonOnlineGallery2018,
  title = {Paragon: An Online Gallery for Enhancing Design Feedback with Visual Examples},
  shorttitle = {Paragon},
  booktitle = {Proceedings of the 2018 CHI Conference on Human Factors in Computing Systems},
  author = {Kang, Hyeonsu B. and Amoako, Gabriel and Sengupta, Neil and Dow, Steven P.},
  year = {2018},
  month = apr,
  pages = {1--13},
  publisher = {ACM},
  address = {Montreal QC Canada},
  doi = {10.1145/3173574.3174180},
  urldate = {2024-01-27},
  isbn = {978-1-4503-5620-6},
  langid = {english}
}

@inproceedings{leeDesigningInteractiveExample2010,
  title = {Designing with Interactive Example Galleries},
  booktitle = {Proceedings of the SIGCHI Conference on Human Factors in Computing Systems},
  author = {Lee, Brian and Srivastava, Savil and Kumar, Ranjitha and Brafman, Ronen and Klemmer, Scott R.},
  year = {2010},
  month = apr,
  series = {CHI '10},
  pages = {2257--2266},
  publisher = {Association for Computing Machinery},
  address = {New York, NY, USA},
  doi = {10.1145/1753326.1753667},
  urldate = {2022-06-22},
  isbn = {978-1-60558-929-9},
  langid = {american}
}

@inproceedings{luMistyUIPrototyping2025,
  title = {Misty: UI Prototyping Through Interactive Conceptual Blending},
  shorttitle = {Misty},
  booktitle = {Proceedings of the 2025 CHI Conference on Human Factors in Computing Systems},
  author = {Lu, Yuwen and Leung, Alan and Swearngin, Amanda and Nichols, Jeffrey and Barik, Titus},
  year = {2025},
  month = apr,
  series = {CHI '25},
  eprint = {2409.13900},
  primaryclass = {cs},
  pages = {1--17},
  publisher = {Association for Computing Machinery},
  address = {New York, NY, USA},
  doi = {10.1145/3706598.3713924},
  urldate = {2025-07-17},
  archiveprefix = {arXiv},
  isbn = {979-8-4007-1394-1}
}

@inproceedings{mansourUsingCOCAFoster2017,
  title = {Using COCA to Foster Students' Use of English Collocations in Academic Writing},
  booktitle = {Proceedings of the 3rd International Conference on Higher Education Advances},
  author = {Mansour, Deena Mohammad},
  year = {2017},
  month = jun,
  publisher = {Universitat Polit{\`e}cnica Val{\`e}ncia},
  address = {Valencia, Spain},
  doi = {10.4995/HEAD17.2017.5301},
  urldate = {2025-06-16},
  isbn = {978-84-9048-590-3},
  pages = {600--607}
}

@inproceedings{ngoonShownAdaptiveConceptual2021,
  title = {Sh{\"o}wn: Adaptive Conceptual Guidance Aids Example Use in Creative Tasks},
  shorttitle = {Sh{\"o}wn},
  booktitle = {Designing Interactive Systems Conference 2021},
  author = {Ngoon, Tricia J. and Kim, Joy O and Klemmer, Scott},
  year = {2021},
  month = jun,
  series = {DIS '21},
  pages = {1834--1845},
  publisher = {Association for Computing Machinery},
  address = {New York, NY, USA},
  doi = {10.1145/3461778.3462072},
  urldate = {2022-11-09},
  isbn = {978-1-4503-8476-6},
  langid = {american}
}

@inproceedings{ritchieDtourStylebasedExploration2011,
  title = {D.Tour: Style-Based Exploration of Design Example Galleries},
  shorttitle = {D.Tour},
  booktitle = {Proceedings of the 24th Annual ACM Symposium on User Interface Software and Technology - UIST '11},
  author = {Ritchie, Daniel and Kejriwal, Ankita Arvind and Klemmer, Scott R.},
  year = {2011},
  pages = {165},
  publisher = {ACM Press},
  address = {Santa Barbara, California, USA},
  doi = {10.1145/2047196.2047216},
  urldate = {2021-04-22},
  isbn = {978-1-4503-0716-1},
  langid = {english}
}

@misc{wangSchemexInteractiveStructural2025,
  title = {Schemex: Interactive Structural Abstraction from Examples with Contrastive Refinement},
  shorttitle = {Schemex},
  author = {Wang, Sitong and Menon, Samia and Li, Dingzeyu and Ma, Xiaojuan and Zemel, Richard and Chilton, Lydia B.},
  year = {2025},
  month = apr,
  number = {arXiv:2504.11795},
  eprint = {2504.11795},
  primaryclass = {cs},
  publisher = {arXiv},
  doi = {10.48550/arXiv.2504.11795},
  urldate = {2025-06-09},
  archiveprefix = {arXiv},
  langid = {american}
}

@article{wardStructuredImaginationRole1994,
  title = {Structured Imagination: The Role of Category Structure in Exemplar Generation},
  shorttitle = {Structured Imagination},
  author = {Ward, T. B.},
  year = {1994},
  month = aug,
  journal = {Cognitive Psychology},
  volume = {27},
  number = {1},
  pages = {1--40},
  issn = {0010-0285},
  doi = {10.1006/cogp.1994.1010},
  urldate = {2025-06-13}
}

@article{xuIdeateRelateExamplesGallery2021,
  title = {IdeateRelate: An Examples Gallery That Helps Creators Explore Ideas in Relation to Their Own},
  shorttitle = {IdeateRelate},
  author = {Xu, Xiaotong (Tone) and Xiong, Rosaleen and Wang, Boyang and Min, David and Dow, Steven P.},
  year = {2021},
  month = oct,
  journal = {Proc. ACM Hum.-Comput. Interact.},
  volume = {5},
  number = {CSCW2},
  pages = {352:1--352:18},
  doi = {10.1145/3479496},
  urldate = {2024-01-23},
  langid = {american}
}

@misc{tianAreLargeLanguage2024,
  title = {Are Large Language Models Capable of Generating Human-Level Narratives?},
  author = {Tian, Yufei and Huang, Tenghao and Liu, Miri and Jiang, Derek and Spangher, Alexander and Chen, Muhao and May, Jonathan and Peng, Nanyun},
  year = {2024},
  month = oct,
  number = {arXiv:2407.13248},
  eprint = {2407.13248},
  primaryclass = {cs},
  publisher = {arXiv},
  doi = {10.48550/arXiv.2407.13248},
  urldate = {2025-05-21},
  archiveprefix = {arXiv},
  langid = {american}
}

@misc{ammanabroluAutomatedStorytellingCausal2020,
  title = {Automated Storytelling via Causal, Commonsense Plot Ordering},
  author = {Ammanabrolu, Prithviraj and Cheung, Wesley and Broniec, William and Riedl, Mark O.},
  year = {2020},
  month = dec,
  number = {arXiv:2009.00829},
  eprint = {2009.00829},
  primaryclass = {cs},
  publisher = {arXiv},
  doi = {10.48550/arXiv.2009.00829},
  urldate = {2025-06-18},
  archiveprefix = {arXiv}
}

@inproceedings{cavazzaCharactersSearchAuthor2001,
  title = {Characters in Search of an Author: AI-Based Virtual Storytelling},
  shorttitle = {Characters in Search of an Author},
  booktitle = {Virtual Storytelling Using Virtual Reality Technologies for Storytelling},
  author = {Cavazza, Marc and Charles, Fred and Mead, Steven J.},
  editor = {Balet, Olivier and Subsol, G{\'e}rard and Torguet, Patrice},
  year = {2001},
  pages = {145--154},
  publisher = {Springer},
  address = {Berlin, Heidelberg},
  doi = {10.1007/3-540-45420-9_16},
  isbn = {978-3-540-45420-5},
  langid = {english}
}

@inproceedings{chouTaleStreamSupportingStory2023,
  title = {TaleStream: Supporting Story Ideation with Trope Knowledge},
  shorttitle = {TaleStream},
  booktitle = {Proceedings of the 36th Annual ACM Symposium on User Interface Software and Technology},
  author = {Chou, Jean-Pe{\"i}c and Siu, Alexa Fay and Lipka, Nedim and Rossi, Ryan and Dernoncourt, Franck and Agrawala, Maneesh},
  year = {2023},
  month = oct,
  series = {UIST '23},
  pages = {1--12},
  publisher = {ACM},
  address = {San Francisco CA USA},
  doi = {10.1145/3586183.3606807},
  urldate = {2024-09-25},
  isbn = {979-8-4007-0132-0},
  langid = {english}
}

@inproceedings{chungPatchviewLLMpoweredWorldbuilding2024a,
  title = {Patchview: LLM-Powered Worldbuilding with Generative Dust and Magnet Visualization},
  shorttitle = {Patchview},
  booktitle = {Proceedings of the 37th Annual ACM Symposium on User Interface Software and Technology},
  author = {Chung, John Joon Young and Kreminski, Max},
  year = {2024},
  month = oct,
  series = {UIST '24},
  pages = {1--19},
  publisher = {Association for Computing Machinery},
  address = {New York, NY, USA},
  doi = {10.1145/3654777.3676352},
  urldate = {2025-06-17},
  isbn = {979-8-4007-0628-8}
}

@inproceedings{chungTaleBrushSketchingStories2022,
  title = {TaleBrush: Sketching Stories with Generative Pretrained Language Models},
  shorttitle = {TaleBrush},
  booktitle = {Proceedings of the 2022 CHI Conference on Human Factors in Computing Systems},
  author = {Chung, John Joon Young and Kim, Wooseok and Yoo, Kang Min and Lee, Hwaran and Adar, Eytan and Chang, Minsuk},
  year = {2022},
  month = apr,
  series = {CHI '22},
  pages = {1--19},
  publisher = {Association for Computing Machinery},
  address = {New York, NY, USA},
  doi = {10.1145/3491102.3501819},
  urldate = {2024-01-19},
  isbn = {978-1-4503-9157-3},
  langid = {american}
}

@inproceedings{duvalBreakingWritersBlock2021,
  title = {Breaking Writer's Block: Low-Cost Fine-Tuning of Natural Language Generation Models},
  shorttitle = {Breaking Writer's Block},
  booktitle = {Proceedings of the 16th Conference of the European Chapter of the Association for Computational Linguistics: System Demonstrations},
  author = {Duval, Alexandre and Lamson, Thomas and {de L{\'e}s{\'e}leuc de K{\'e}rouara}, Ga{\"e}l and Gall{\'e}, Matthias},
  editor = {Gkatzia, Dimitra and Seddah, Djam{\'e}},
  year = {2021},
  month = apr,
  pages = {278--287},
  publisher = {Association for Computational Linguistics},
  address = {Online},
  doi = {10.18653/v1/2021.eacl-demos.33},
  urldate = {2025-06-18}
}

@misc{fanHierarchicalNeuralStory2018,
  title = {Hierarchical Neural Story Generation},
  author = {Fan, Angela and Lewis, Mike and Dauphin, Yann},
  year = {2018},
  month = may,
  number = {arXiv:1805.04833},
  eprint = {1805.04833},
  primaryclass = {cs},
  publisher = {arXiv},
  doi = {10.48550/arXiv.1805.04833},
  urldate = {2025-06-18},
  archiveprefix = {arXiv}
}

@article{gervasStoryPlotGeneration2005,
  title = {Story Plot Generation Based on CBR},
  author = {Gerv{\'a}s, Pablo and {D{\'i}az-Agudo}, Bel{\'e}n and Peinado, Federico and Herv{\'a}s, Raquel},
  year = {2005},
  month = aug,
  journal = {Knowledge-Based Systems},
  series = {AI-2004, Cambridge, England, 13th-15th December 2004},
  volume = {18},
  number = {4},
  pages = {235--242},
  issn = {0950-7051},
  doi = {10.1016/j.knosys.2004.10.011},
  urldate = {2025-06-18}
}

@inproceedings{huangHeteroglossiaInSituStory2020,
  title = {Heteroglossia: In-Situ Story Ideation with the Crowd},
  shorttitle = {Heteroglossia},
  booktitle = {Proceedings of the 2020 CHI Conference on Human Factors in Computing Systems},
  author = {Huang, Chieh-Yang and Huang, Shih-Hong and Huang, Ting-Hao Kenneth},
  year = {2020},
  month = apr,
  series = {CHI '20},
  pages = {1--12},
  publisher = {Association for Computing Machinery},
  address = {New York, NY, USA},
  doi = {10.1145/3313831.3376715},
  urldate = {2025-06-09},
  isbn = {978-1-4503-6708-0},
  langid = {american}
}

@inproceedings{huangINSETSentenceInfilling2020,
  title = {INSET: Sentence Infilling with INter-SEntential Transformer},
  shorttitle = {INSET},
  booktitle = {Proceedings of the 58th Annual Meeting of the Association for Computational Linguistics},
  author = {Huang, Yichen and Zhang, Yizhe and Elachqar, Oussama and Cheng, Yu},
  editor = {Jurafsky, Dan and Chai, Joyce and Schluter, Natalie and Tetreault, Joel},
  year = {2020},
  month = jul,
  pages = {2502--2515},
  publisher = {Association for Computational Linguistics},
  address = {Online},
  doi = {10.18653/v1/2020.acl-main.226},
  urldate = {2025-06-18}
}

@inproceedings{ippolitoUnsupervisedHierarchicalStory2019,
  title = {Unsupervised Hierarchical Story Infilling},
  booktitle = {Proceedings of the First Workshop on Narrative Understanding},
  author = {Ippolito, Daphne and Grangier, David and {Callison-Burch}, Chris and Eck, Douglas},
  editor = {Bamman, David and Chaturvedi, Snigdha and Clark, Elizabeth and Fiterau, Madalina and Iyyer, Mohit},
  year = {2019},
  month = jun,
  pages = {37--43},
  publisher = {Association for Computational Linguistics},
  address = {Minneapolis, Minnesota},
  doi = {10.18653/v1/W19-2405},
  urldate = {2025-06-18}
}

@inproceedings{ippolitoUnsupervisedHierarchicalStory2019a,
  title = {Unsupervised Hierarchical Story Infilling},
  booktitle = {Proceedings of the First Workshop on Narrative Understanding},
  author = {Ippolito, Daphne and Grangier, David and {Callison-Burch}, Chris and Eck, Douglas},
  editor = {Bamman, David and Chaturvedi, Snigdha and Clark, Elizabeth and Fiterau, Madalina and Iyyer, Mohit},
  year = {2019},
  month = jun,
  pages = {37--43},
  publisher = {Association for Computational Linguistics},
  address = {Minneapolis, Minnesota},
  doi = {10.18653/v1/W19-2405},
  urldate = {2025-06-18}
}

@article{lebowitzCreatingCharactersStorytelling1984,
  title = {Creating Characters in a Story-Telling Universe},
  author = {Lebowitz, Michael},
  year = {1984},
  journal = {Poetics},
  volume = {13},
  number = {3},
  pages = {171--194},
  publisher = {Elsevier},
  urldate = {2025-06-18}
}

@inproceedings{meehanTALESPINInteractiveProgram1977,
  title = {TALE-SPIN, an Interactive Program That Writes Stories},
  booktitle = {Proceedings of the 5th International Joint Conference on Artificial Intelligence - Volume 1},
  author = {Meehan, James R.},
  year = {1977},
  month = aug,
  series = {IJCAI'77},
  pages = {91--98},
  publisher = {Morgan Kaufmann Publishers Inc.},
  address = {San Francisco, CA, USA},
  urldate = {2025-07-17}
}

@article{perezMEXICAComputerModel2001,
  title = {MEXICA: A Computer Model of a Cognitive Account of Creative Writing},
  shorttitle = {MEXICA},
  author = {P{\'E}rez, Rafael P{\'E}rez {\'Y} and {and Sharples}, Mike},
  year = {2001},
  month = apr,
  journal = {Journal of Experimental \& Theoretical Artificial Intelligence},
  volume = {13},
  number = {2},
  pages = {119--139},
  publisher = {Taylor \& Francis},
  issn = {0952-813X},
  doi = {10.1080/09528130010029820},
  urldate = {2025-06-18}
}

@inproceedings{raoScriptVizVisualizationTool2024a,
  title = {ScriptViz: A Visualization Tool to Aid Scriptwriting Based on a Large Movie Database},
  shorttitle = {ScriptViz},
  booktitle = {Proceedings of the 37th Annual ACM Symposium on User Interface Software and Technology},
  author = {Rao, Anyi and Chou, Jean-Pe{\"i}c and Agrawala, Maneesh},
  year = {2024},
  month = oct,
  series = {UIST '24},
  pages = {1--13},
  publisher = {Association for Computing Machinery},
  address = {New York, NY, USA},
  doi = {10.1145/3654777.3676402},
  urldate = {2025-06-17},
  isbn = {979-8-4007-0628-8}
}

@article{riedlNarrativePlanningBalancing2010,
  title = {Narrative Planning: Balancing Plot and Character},
  shorttitle = {Narrative Planning},
  author = {Riedl, Mark O. and Young, Robert Michael},
  year = {2010},
  journal = {Journal of Artificial Intelligence Research},
  volume = {39},
  pages = {217--268},
  urldate = {2025-06-18}
}

@inproceedings{riedlVignettebasedStoryPlanning2008,
  title = {Vignette-Based Story Planning: Creativity through Exploration and Retrieval},
  shorttitle = {Vignette-Based Story Planning},
  booktitle = {Proceedings of the 5th International Joint Workshop on Computational Creativity},
  author = {Riedl, Mark O.},
  year = {2008},
  pages = {41--50},
  urldate = {2025-06-18},
  publisher = {Association for Computational Creativity},
  address = {Madrid, Spain}
}

@inproceedings{sunIGAIntentGuidedAuthoring2021,
  title = {IGA: An Intent-Guided Authoring Assistant},
  shorttitle = {IGA},
  booktitle = {Proceedings of the 2021 Conference on Empirical Methods in Natural Language Processing},
  author = {Sun, Simeng and Zhao, Wenlong and Manjunatha, Varun and Jain, Rajiv and Morariu, Vlad and Dernoncourt, Franck and Srinivasan, Balaji Vasan and Iyyer, Mohit},
  editor = {Moens, Marie-Francine and Huang, Xuanjing and Specia, Lucia and Yih, Scott Wen-tau},
  year = {2021},
  month = nov,
  pages = {5972--5985},
  publisher = {Association for Computational Linguistics},
  address = {Online and Punta Cana, Dominican Republic},
  doi = {10.18653/v1/2021.emnlp-main.483},
  urldate = {2025-06-18}
}

@book{turnerMinstrelComputerModel1993,
  title = {Minstrel: A Computer Model of Creativity and Storytelling},
  shorttitle = {Minstrel},
  author = {Turner, Scott R.},
  year = {1993},
  publisher = {University of California, Los Angeles},
  urldate = {2025-06-18},
  address = {Los Angeles, California, USA}
}

@misc{wangNarrativeInterpolationGenerating2020,
  title = {Narrative Interpolation for Generating and Understanding Stories},
  author = {Wang, Su and Durrett, Greg and Erk, Katrin},
  year = {2020},
  month = aug,
  number = {arXiv:2008.07466},
  eprint = {2008.07466},
  primaryclass = {cs},
  publisher = {arXiv},
  doi = {10.48550/arXiv.2008.07466},
  urldate = {2025-06-18},
  archiveprefix = {arXiv}
}

@misc{wanPolymindParallelVisual2025a,
  title = {Polymind: Parallel Visual Diagramming with Large Language Models to Support Prewriting Through Microtasks},
  shorttitle = {Polymind},
  author = {Wan, Qian and Li, Jiannan and Wang, Huanchen and Lu, Zhicong},
  year = {2025},
  month = feb,
  number = {arXiv:2502.09577},
  eprint = {2502.09577},
  primaryclass = {cs},
  publisher = {arXiv},
  doi = {10.48550/arXiv.2502.09577},
  urldate = {2025-06-17},
  archiveprefix = {arXiv},
  langid = {american}
}

@inproceedings{xuMEGATRONCNTRLControllableStory2020,
  title = {MEGATRON-CNTRL: Controllable Story Generation with External Knowledge Using Large-Scale Language Models},
  shorttitle = {MEGATRON-CNTRL},
  booktitle = {Proceedings of the 2020 Conference on Empirical Methods in Natural Language Processing (EMNLP)},
  author = {Xu, Peng and Patwary, Mostofa and Shoeybi, Mohammad and Puri, Raul and Fung, Pascale and Anandkumar, Anima and Catanzaro, Bryan},
  editor = {Webber, Bonnie and Cohn, Trevor and He, Yulan and Liu, Yang},
  year = {2020},
  month = nov,
  pages = {2831--2845},
  publisher = {Association for Computational Linguistics},
  address = {Online},
  doi = {10.18653/v1/2020.emnlp-main.226},
  urldate = {2025-06-18}
}

@article{collinsCognitiveApprenticeshipMaking1991,
  title = {Cognitive Apprenticeship: Making Thinking Visible},
  shorttitle = {Cognitive Apprenticeship},
  author = {Collins, Allan and Brown, John Seely and Holum, Ann},
  year = {1991},
  journal = {American educator},
  volume = {15},
  number = {3},
  pages = {6--11},
  urldate = {2024-11-26},
  langid = {american}
}

@inproceedings{gatysImageStyleTransfer2016a,
  title = {Image Style Transfer Using Convolutional Neural Networks},
  booktitle = {Proceedings of the IEEE Conference on Computer Vision and Pattern Recognition},
  author = {Gatys, Leon A. and Ecker, Alexander S. and Bethge, Matthias},
  year = {2016},
  address = {Las Vegas, Nevada, USA},
  publisher = {IEEE},
  pages = {2414--2423},
  urldate = {2025-06-17}
}

@inproceedings{kimMechanicalNovelCrowdsourcing2017,
  title = {Mechanical Novel: Crowdsourcing Complex Work through Reflection and Revision},
  shorttitle = {Mechanical Novel},
  booktitle = {Proceedings of the 2017 ACM Conference on Computer Supported Cooperative Work and Social Computing},
  author = {Kim, Joy and Sterman, Sarah and Cohen, Allegra Argent Beal and Bernstein, Michael S.},
  year = {2017},
  month = feb,
  series = {CSCW '17},
  pages = {233--245},
  publisher = {Association for Computing Machinery},
  address = {New York, NY, USA},
  doi = {10.1145/2998181.2998196},
  urldate = {2024-02-03},
  isbn = {978-1-4503-4335-0}
}

@inproceedings{papalampidiMoviePlotAnalysis2019,
  title = {Movie Plot Analysis via Turning Point Identification},
  booktitle = {Proceedings of the 2019 Conference on Empirical Methods in Natural Language Processing and the 9th International Joint Conference on Natural Language Processing (EMNLP-IJCNLP)},
  author = {Papalampidi, Pinelopi and Keller, Frank and Lapata, Mirella},
  editor = {Inui, Kentaro and Jiang, Jing and Ng, Vincent and Wan, Xiaojun},
  year = {2019},
  month = nov,
  pages = {1707--1717},
  publisher = {Association for Computational Linguistics},
  address = {Hong Kong, China},
  doi = {10.18653/v1/D19-1180},
  urldate = {2025-06-20}
}

@book{balNarratologyIntroductionTheory2004,
  title = {Narratology: Introduction to the Theory of Narrative},
  shorttitle = {Narratology},
  author = {Bal, Mieke},
  year = {2004},
  publisher = {University of Toronto Press, Scholarly Publishing Division},
  address = {Toronto},
  isbn = {978-0-8020-7806-3},
  langid = {english}
}

@book{kennedyLiteratureIntroductionFiction2016,
  title = {Literature: An Introduction to Fiction, Poetry, Drama, and Writing},
  shorttitle = {Literature},
  author = {Kennedy, X. J. and Gioia, Dana},
  year = {2016},
  publisher = {Pearson},
  address = {Boston},
  isbn = {978-0-321-97166-1},
  langid = {english}
}

@book{mckeeStoryStyleStructure1997,
  title = {Story: Style, Structure, Substance, and the Principles of Screenwriting},
  shorttitle = {Story},
  author = {McKee, Robert},
  year = {1997},
  publisher = {Dey Street Books},
  address = {New York},
  isbn = {978-0-06-203982-8},
  langid = {english}
}

@book{princeNarratologyFormFunctioning2012,
  title = {Narratology: The Form and Functioning of Narrative},
  shorttitle = {Narratology},
  author = {Prince, Gerald},
  year = {2012},
  volume = {108},
  publisher = {Walter de Gruyter},
  urldate = {2025-06-20},
  address = {Berlin, Germany}
}

@article{cicchettiGuidelinesCriteriaRules1994,
  title = {Guidelines, Criteria, and Rules of Thumb for Evaluating Normed and Standardized Assessment Instruments in Psychology},
  author = {Cicchetti, Domenic V.},
  year = {1994},
  journal = {Psychological Assessment},
  volume = {6},
  number = {4},
  pages = {284--290},
  publisher = {American Psychological Association},
  address = {US},
  issn = {1939-134X},
  doi = {10.1037/1040-3590.6.4.284},
  langid = {american}
}

@article{jiSurveyHallucinationNatural2023,
  title = {Survey of Hallucination in Natural Language Generation},
  author = {Ji, Ziwei and Lee, Nayeon and Frieske, Rita and Yu, Tiezheng and Su, Dan and Xu, Yan and Ishii, Etsuko and Bang, Ye Jin and Madotto, Andrea and Fung, Pascale},
  year = {2023},
  month = mar,
  journal = {ACM Comput. Surv.},
  volume = {55},
  number = {12},
  pages = {248:1--248:38},
  issn = {0360-0300},
  doi = {10.1145/3571730},
  urldate = {2023-10-17}
}

@misc{mohammadNRCVADLexicon2025,
  title = {NRC VAD Lexicon v2: Norms for Valence, Arousal, and Dominance for over 55k English Terms},
  shorttitle = {NRC VAD Lexicon V2},
  author = {Mohammad, Saif M.},
  year = {2025},
  month = mar,
  number = {arXiv:2503.23547},
  eprint = {2503.23547},
  primaryclass = {cs},
  publisher = {arXiv},
  doi = {10.48550/arXiv.2503.23547},
  urldate = {2025-07-16},
  archiveprefix = {arXiv}
}

@inproceedings{zhangFrictionDecipheringWriting2025,
  title = {Friction: Deciphering Writing Feedback into Writing Revisions through LLM-Assisted Reflection},
  shorttitle = {Friction},
  booktitle = {Proceedings of the 2025 CHI Conference on Human Factors in Computing Systems},
  author = {Zhang, Chao and Ju, Kexin and Bidoshi, Peter and Yen, Yu-Chun Grace and Rzeszotarski, Jeffrey M.},
  year = {2025},
  month = apr,
  series = {CHI '25},
  pages = {1--27},
  publisher = {Association for Computing Machinery},
  address = {New York, NY, USA},
  doi = {10.1145/3706598.3714316},
  urldate = {2025-04-29},
  isbn = {979-8-4007-1394-1},
  langid = {american}
}

@inproceedings{huangFeedbackOrchestrationStructuring2018,
  title = {Feedback Orchestration: Structuring Feedback for Facilitating Reflection and Revision in Writing},
  shorttitle = {Feedback Orchestration},
  booktitle = {Companion of the 2018 ACM Conference on Computer Supported Cooperative Work and Social Computing},
  author = {Huang, Yi-Ching and Wang, Hao-Chuan and Hsu, Jane Yung-jen},
  year = {2018},
  month = oct,
  series = {CSCW '18 Companion},
  pages = {257--260},
  publisher = {Association for Computing Machinery},
  address = {New York, NY, USA},
  doi = {10.1145/3272973.3274069},
  urldate = {2024-01-18},
  isbn = {978-1-4503-6018-0},
  langid = {american}
}

@inproceedings{wuAIChainsTransparent2022,
  title = {AI Chains: Transparent and Controllable Human-AI Interaction by Chaining Large Language Model Prompts},
  shorttitle = {AI Chains},
  booktitle = {CHI Conference on Human Factors in Computing Systems},
  author = {Wu, Tongshuang and Terry, Michael and Cai, Carrie Jun},
  year = {2022},
  month = apr,
  pages = {1--22},
  publisher = {ACM},
  address = {New Orleans LA USA},
  doi = {10.1145/3491102.3517582},
  urldate = {2024-01-24},
  isbn = {978-1-4503-9157-3},
  langid = {english}
}

@article{braunUsingThematicAnalysis2006,
  title = {Using Thematic Analysis in Psychology},
  author = {Braun, Virginia and Clarke, Victoria},
  year = {2006},
  month = jan,
  journal = {Qualitative Research in Psychology},
  volume = {3},
  number = {2},
  pages = {77--101},
  issn = {1478-0887, 1478-0895},
  doi = {10.1191/1478088706qp063oa},
  urldate = {2021-12-02},
  langid = {english}
}

@article{fowlerEffectivenessHighlightingRetention1974,
  title = {Effectiveness of Highlighting for Retention of Text Material},
  author = {Fowler, Robert L. and Barker, Anne S.},
  year = {1974},
  journal = {Journal of Applied Psychology},
  volume = {59},
  number = {3},
  pages = {358--364},
  publisher = {American Psychological Association},
  address = {US},
  issn = {1939-1854},
  doi = {10.1037/h0036750}
}

@inproceedings{berndtUsingDynamicTime1994,
  title = {Using Dynamic Time Warping to Find Patterns in Time Series},
  booktitle = {Proceedings of the 3rd International Conference on Knowledge Discovery and Data Mining},
  author = {Berndt, Donald J. and Clifford, James},
  year = {1994},
  month = jul,
  series = {AAAIWS'94},
  pages = {359--370},
  publisher = {AAAI Press},
  address = {Seattle, WA},
  urldate = {2025-07-24}
}

@inproceedings{massonTextoshopInteractionsInspired2025,
  title = {Textoshop: Interactions Inspired by Drawing Software to Facilitate Text Editing},
  shorttitle = {Textoshop},
  booktitle = {Proceedings of the 2025 CHI Conference on Human Factors in Computing Systems},
  author = {Masson, Damien and Kim, Young-Ho and Chevalier, Fanny},
  year = {2025},
  month = apr,
  pages = {1--14},
  publisher = {ACM},
  address = {Yokohama Japan},
  doi = {10.1145/3706598.3713862},
  urldate = {2025-06-19},
  isbn = {979-8-4007-1394-1},
  langid = {english}
}

@inproceedings{rezaABScribeRapidExploration2024,
  title = {ABScribe: Rapid Exploration \& Organization of Multiple Writing Variations in Human-AI Co-Writing Tasks Using Large Language Models},
  shorttitle = {ABScribe},
  booktitle = {Proceedings of the 2024 CHI Conference on Human Factors in Computing Systems},
  author = {Reza, Mohi and Laundry, Nathan M and Musabirov, Ilya and Dushniku, Peter and Yu, Zhi Yuan ``Michael'' and Mittal, Kashish and Grossman, Tovi and Liut, Michael and Kuzminykh, Anastasia and Williams, Joseph Jay},
  year = {2024},
  month = may,
  series = {CHI '24},
  pages = {1--18},
  publisher = {Association for Computing Machinery},
  address = {New York, NY, USA},
  doi = {10.1145/3613904.3641899},
  urldate = {2025-07-31},
  isbn = {979-8-4007-0330-0}
}

@article{reaganEmotionalArcsStories2016,
  title = {The Emotional Arcs of Stories Are Dominated by Six Basic Shapes},
  author = {Reagan, Andrew J and Mitchell, Lewis and Kiley, Dilan and Danforth, Christopher M and Dodds, Peter Sheridan},
  year = {2016},
  month = dec,
  journal = {EPJ Data Sci.},
  volume = {5},
  number = {1},
  pages = {31},
  issn = {2193-1127},
  doi = {10.1140/epjds/s13688-016-0093-1},
  urldate = {2025-06-03},
  langid = {english}
}

@inproceedings{whitePromptPatternCatalog2023b,
  title = {A Prompt Pattern Catalog to Enhance Prompt Engineering with ChatGPT},
  booktitle = {Proceedings of the 30th Conference on Pattern Languages of Programs},
  author = {White, Jules and Fu, Quchen and Hays, Sam and Sandborn, Michael and Olea, Carlos and Gilbert, Henry and Elnashar, Ashraf and {Spencer-Smith}, Jesse and Schmidt, Douglas C.},
  year = {2023},
  month = oct,
  series = {PLoP '23},
  pages = {1--31},
  publisher = {The Hillside Group},
  address = {USA},
  urldate = {2025-08-10},
  isbn = {978-1-941652-19-0}
}

@book{grahamBestPracticesWriting2007,
  title = {Best Practices in Writing Instruction},
  editor = {Graham, Steve and MaCarthur, Charles A. and Fitzgerald, Jill},
  year = {2007},
  series = {Best Practices in Writing Instruction},
  pages = {xii, 340},
  publisher = {The Guilford Press},
  address = {New York, NY, US},
  isbn = {978-1-59385-432-4 978-1-59385-433-1}
}

@book{grahamWritingNextEffective2007,
  title = {Writing Next: Effective Strategies to Improve Writing of Adolescents in Middle and High Schools},
  shorttitle = {Writing Next},
  author = {Graham, Steve and Perin, Dolores},
  year = {2007},
  publisher = {Alliance for Excellent Education},
  googlebooks = {c7ZpHAAACAAJ},
  langid = {english},
  address = {Washington, D.C., US},
}

@book{murrayWriterTeachesWriting2004,
  title = {A Writer Teaches Writing Revised},
  author = {Murray, Donald M.},
  year = {2004},
  publisher = {Cengage Learning},
  address = {Boston, MA},
  isbn = {978-0-7593-9829-0},
  langid = {english}
}

@article{bangorDeterminingWhatIndividual2009,
  title = {Determining What Individual SUS Scores Mean: Adding an Adjective Rating Scale},
  shorttitle = {Determining What Individual SUS Scores Mean},
  author = {Bangor, Aaron and Kortum, Philip and Miller, James},
  year = {2009},
  month = may,
  journal = {J. Usability Studies},
  volume = {4},
  number = {3},
  pages = {114--123},
  issn = {1931-3357},
  langid = {american}
}

@inproceedings{linderEverydayIdeationAll2014,
  title = {Everyday Ideation: All of My Ideas Are on Pinterest},
  shorttitle = {Everyday Ideation},
  booktitle = {Proceedings of the SIGCHI Conference on Human Factors in Computing Systems},
  author = {Linder, Rhema and Snodgrass, Clair and Kerne, Andruid},
  year = {2014},
  month = apr,
  series = {CHI '14},
  pages = {2411--2420},
  publisher = {Association for Computing Machinery},
  address = {New York, NY, USA},
  doi = {10.1145/2556288.2557273},
  urldate = {2025-08-30},
  isbn = {978-1-4503-2473-1}
}

@inproceedings{luWhatELSEShapingNarrative2025,
  title = {WhatELSE: Shaping Narrative Spaces at Configurable Level of Abstraction for AI-Bridged Interactive Storytelling},
  shorttitle = {WhatELSE},
  booktitle = {Proceedings of the 2025 CHI Conference on Human Factors in Computing Systems},
  author = {Lu, Zhuoran and Zhou, Qian and Wang, Yi},
  year = {2025},
  month = apr,
  series = {CHI '25},
  pages = {1--18},
  publisher = {Association for Computing Machinery},
  address = {New York, NY, USA},
  doi = {10.1145/3706598.3713363},
  urldate = {2025-08-30},
  isbn = {979-8-4007-1394-1}
}

@misc{leeDesignSpaceIntelligent2024,
  title = {A Design Space for Intelligent and Interactive Writing Assistants},
  author = {Lee, Mina and Gero, Katy Ilonka and Chung, John Joon Young and Shum, Simon Buckingham and Raheja, Vipul and Shen, Hua and Venugopalan, Subhashini and Wambsganss, Thiemo and Zhou, David and Alghamdi, Emad A. and August, Tal and Bhat, Avinash and Choksi, Madiha Zahrah and Dutta, Senjuti and Guo, Jin L. C. and Hoque, Md Naimul and Kim, Yewon and Knight, Simon and Neshaei, Seyed Parsa and Sergeyuk, Agnia and Shibani, Antonette and Shrivastava, Disha and Shroff, Lila and Stark, Jessi and Sterman, Sarah and Wang, Sitong and Bosselut, Antoine and Buschek, Daniel and Chang, Joseph Chee and Chen, Sherol and Kreminski, Max and Park, Joonsuk and Pea, Roy and Rho, Eugenia H. and Shen, Shannon Zejiang and Siangliulue, Pao},
  year = {2024},
  month = mar,
  eprint = {2403.14117},
  primaryclass = {cs},
  doi = {10.1145/3613904.3642697},
  urldate = {2024-04-03},
  archiveprefix = {arxiv}
}

@inproceedings{geroSparksInspirationScience2022,
  title = {Sparks: Inspiration for Science Writing Using Language Models},
  shorttitle = {Sparks},
  booktitle = {Proceedings of the 2022 ACM Designing Interactive Systems Conference},
  author = {Gero, Katy Ilonka and Liu, Vivian and Chilton, Lydia},
  year = {2022},
  month = jun,
  series = {DIS '22},
  pages = {1002--1019},
  publisher = {Association for Computing Machinery},
  address = {New York, NY, USA},
  doi = {10.1145/3532106.3533533},
  urldate = {2023-10-31},
  isbn = {978-1-4503-9358-4},
  langid = {english}
}

@inproceedings{schmittCharacterChatSupportingCreation2021,
  title = {CharacterChat: Supporting the Creation of Fictional Characters through Conversation and Progressive Manifestation with a Chatbot},
  shorttitle = {CharacterChat},
  booktitle = {Creativity and Cognition},
  author = {Schmitt, Oliver and Buschek, Daniel},
  year = {2021},
  month = jun,
  pages = {1--10},
  publisher = {ACM},
  address = {Virtual Event Italy},
  doi = {10.1145/3450741.3465253},
  urldate = {2024-04-03},
  isbn = {978-1-4503-8376-9},
  langid = {english}
}

@inproceedings{zhangVISARHumanAIArgumentative2023,
  title = {VISAR: A Human-AI Argumentative Writing Assistant with Visual Programming and Rapid Draft Prototyping},
  shorttitle = {VISAR},
  booktitle = {Proceedings of the 36th Annual ACM Symposium on User Interface Software and Technology},
  author = {Zhang, Zheng and Gao, Jie and Dhaliwal, Ranjodh Singh and Li, Toby Jia-Jun},
  year = {2023},
  month = oct,
  series = {UIST '23},
  pages = {1--30},
  publisher = {Association for Computing Machinery},
  address = {New York, NY, USA},
  doi = {10.1145/3586183.3606800},
  urldate = {2023-12-10},
  isbn = {9798400701320},
  langid = {english}
}

@inproceedings{chungTaleBrushSketchingStories2022a,
  title = {TaleBrush: Sketching Stories with Generative Pretrained Language Models},
  shorttitle = {TaleBrush},
  booktitle = {Proceedings of the 2022 CHI Conference on Human Factors in Computing Systems},
  author = {Chung, John Joon Young and Kim, Wooseok and Yoo, Kang Min and Lee, Hwaran and Adar, Eytan and Chang, Minsuk},
  year = {2022},
  month = apr,
  series = {CHI '22},
  pages = {1--19},
  publisher = {Association for Computing Machinery},
  address = {New York, NY, USA},
  doi = {10.1145/3491102.3501819},
  urldate = {2024-01-19},
  isbn = {978-1-4503-9157-3}
}

@inproceedings{jakeschCoWritingOpinionatedLanguage2023,
  title = {Co-Writing with Opinionated Language Models Affects Users' Views},
  booktitle = {Proceedings of the 2023 CHI Conference on Human Factors in Computing Systems},
  author = {Jakesch, Maurice and Bhat, Advait and Buschek, Daniel and Zalmanson, Lior and Naaman, Mor},
  year = {2023},
  month = apr,
  pages = {1--15},
  publisher = {ACM},
  address = {Hamburg Germany},
  doi = {10.1145/3544548.3581196},
  urldate = {2023-10-04},
  isbn = {978-1-4503-9421-5},
  langid = {english}
}

@misc{kimAuthorsValuesAttitudes2024,
  title = {Authors' Values and Attitudes Towards AI-Bridged Scalable Personalization of Creative Language Arts},
  author = {Kim, Taewook and Han, Hyomin and Adar, Eytan and Kay, Matthew and Chung, John Joon Young},
  year = {2024},
  month = mar,
  eprint = {2403.00439},
  primaryclass = {cs},
  doi = {10.1145/3613904.3642529},
  urldate = {2024-03-04},
  archiveprefix = {arxiv}
}

@inproceedings{yuanWordcraftStoryWriting2022,
  title = {Wordcraft: Story Writing With Large Language Models},
  shorttitle = {Wordcraft},
  booktitle = {27th International Conference on Intelligent User Interfaces},
  author = {Yuan, Ann and Coenen, Andy and Reif, Emily and Ippolito, Daphne},
  year = {2022},
  month = mar,
  series = {IUI '22},
  pages = {841--852},
  publisher = {Association for Computing Machinery},
  address = {New York, NY, USA},
  doi = {10.1145/3490099.3511105},
  urldate = {2023-02-13},
  isbn = {978-1-4503-9144-3}
}

@inproceedings{afrinEffectiveInterfacesStudentDriven2021a,
  title = {Effective Interfaces for Student-Driven Revision Sessions for Argumentative Writing},
  booktitle = {Proceedings of the 2021 CHI Conference on Human Factors in Computing Systems},
  author = {Afrin, Tazin and Kashefi, Omid and Olshefski, Christopher and Litman, Diane and Hwa, Rebecca and Godley, Amanda},
  year = {2021},
  month = may,
  series = {CHI '21},
  pages = {1--13},
  publisher = {Association for Computing Machinery},
  address = {New York, NY, USA},
  doi = {10.1145/3411764.3445683},
  urldate = {2024-01-18},
  isbn = {978-1-4503-8096-6}
}

@inproceedings{itoUseAIpoweredRewriting2023,
  title = {Use of an AI-Powered Rewriting Support Software in Context with Other Tools: A Study of Non-Native English Speakers},
  shorttitle = {Use of an AI-Powered Rewriting Support Software in Context with Other Tools},
  booktitle = {Proceedings of the 36th Annual ACM Symposium on User Interface Software and Technology},
  author = {Ito, Takumi and Yamashita, Naomi and Kuribayashi, Tatsuki and Hidaka, Masatoshi and Suzuki, Jun and Gao, Ge and Jamieson, Jack and Inui, Kentaro},
  year = {2023},
  month = oct,
  series = {UIST '23},
  pages = {1--13},
  publisher = {Association for Computing Machinery},
  address = {New York, NY, USA},
  doi = {10.1145/3586183.3606810},
  urldate = {2024-02-24},
  isbn = {9798400701320},
  langid = {english}
}

@inproceedings{leeInteractiveChildrenStory2022,
  title = {Interactive Children's Story Rewriting Through Parent-Children Interaction},
  booktitle = {Proceedings of the First Workshop on Intelligent and Interactive Writing Assistants (In2Writing 2022)},
  author = {Lee, Yoonjoo and Kim, Tae Soo and Chang, Minsuk and Kim, Juho},
  year = {2022},
  pages = {62--71},
  publisher = {Association for Computational Linguistics},
  address = {Dublin, Ireland},
  doi = {10.18653/v1/2022.in2writing-1.9},
  urldate = {2023-02-06},
  langid = {english}
}

@misc{rezaABScribeRapidExploration2023,
  title = {ABScribe: Rapid Exploration of Multiple Writing Variations in Human-AI Co-Writing Tasks Using Large Language Models},
  shorttitle = {ABScribe},
  author = {Reza, Mohi and Laundry, Nathan and Musabirov, Ilya and Dushniku, Peter and Yu, Zhi Yuan "Michael" and Mittal, Kashish and Grossman, Tovi and Liut, Michael and Kuzminykh, Anastasia and Williams, Joseph Jay},
  year = {2023},
  month = oct,
  number = {arXiv:2310.00117},
  eprint = {2310.00117},
  primaryclass = {cs},
  publisher = {arXiv},
  doi = {10.48550/arXiv.2310.00117},
  urldate = {2024-02-24},
  archiveprefix = {arxiv},
  langid = {english}
}

@inproceedings{turkayIteroRevisionHistory2018,
  title = {Itero: A Revision History Analytics Tool for Exploring Writing Behavior and Reflection},
  shorttitle = {Itero},
  booktitle = {Extended Abstracts of the 2018 CHI Conference on Human Factors in Computing Systems},
  author = {T{\"u}rkay, Selen and Seaton, Daniel and Ang, Andrew M.},
  year = {2018},
  month = apr,
  series = {CHI EA '18},
  pages = {1--6},
  publisher = {Association for Computing Machinery},
  address = {New York, NY, USA},
  doi = {10.1145/3170427.3188474},
  urldate = {2024-01-18},
  isbn = {978-1-4503-5621-3}
}

@article{cherryQuantifyingCreativitySupport2014,
  title = {Quantifying the Creativity Support of Digital Tools through the Creativity Support Index},
  author = {Cherry, Erin and Latulipe, Celine},
  year = {2014},
  month = aug,
  journal = {ACM Trans. Comput.-Hum. Interact.},
  volume = {21},
  number = {4},
  pages = {1--25},
  issn = {1073-0516, 1557-7325},
  doi = {10.1145/2617588},
  urldate = {2020-12-03},
  langid = {english}
}

@incollection{hartDevelopmentNASATLXTask1988,
  title = {Development of NASA-TLX (Task Load Index): Results of Empirical and Theoretical Research},
  shorttitle = {Development of NASA-TLX (Task Load Index)},
  booktitle = {Advances in Psychology},
  author = {Hart, Sandra G. and Staveland, Lowell E.},
  editor = {Hancock, Peter A. and Meshkati, Najmedin},
  year = {1988},
  month = jan,
  series = {Human Mental Workload},
  volume = {52},
  pages = {139--183},
  publisher = {North-Holland},
  address = {Amsterdam, The Netherlands},
  doi = {10.1016/S0166-4115(08)62386-9},
  urldate = {2024-08-31}
}

@misc{dhillonShapingHumanAICollaboration2024,
  title = {Shaping Human-AI Collaboration: Varied Scaffolding Levels in Co-Writing with Language Models},
  shorttitle = {Shaping Human-AI Collaboration},
  author = {Dhillon, Paramveer S. and Molaei, Somayeh and Li, Jiaqi and Golub, Maximilian and Zheng, Shaochun and Robert, Lionel P.},
  year = {2024},
  month = feb,
  number = {arXiv:2402.11723},
  eprint = {2402.11723},
  primaryclass = {cs},
  publisher = {arXiv},
  urldate = {2024-02-24},
  archiveprefix = {arxiv},
  langid = {english}
}

@article{scupinKJMethodTechnique1997,
  title = {The KJ Method: A Technique for Analyzing Data Derived from Japanese Ethnology},
  shorttitle = {The KJ Method},
  author = {Scupin, Raymond},
  year = {1997},
  journal = {Human Organization},
  volume = {56},
  number = {2},
  eprint = {44126786},
  eprinttype = {jstor},
  pages = {233--237},
  publisher = {Society for Applied Anthropology},
  issn = {0018-7259},
  urldate = {2021-12-14}
}

@inproceedings{hoqueHaLLMarkEffectSupporting2024,
  title = {The HaLLMark Effect: Supporting Provenance and Transparent Use of Large Language Models in Writing with Interactive Visualization},
  shorttitle = {The HaLLMark Effect},
  booktitle = {Proceedings of the CHI Conference on Human Factors in Computing Systems},
  author = {Hoque, Md Naimul and Mashiat, Tasfia and Ghai, Bhavya and Shelton, Cecilia D. and Chevalier, Fanny and Kraus, Kari and Elmqvist, Niklas},
  year = {2024},
  month = may,
  series = {CHI '24},
  pages = {1--15},
  publisher = {Association for Computing Machinery},
  address = {New York, NY, USA},
  doi = {10.1145/3613904.3641895},
  urldate = {2024-08-19},
  isbn = {979-8-4007-0330-0},
  langid = {american}
}

@inproceedings{lewisUMUXLITEWhenTheres2013,
  title = {UMUX-LITE: When There's No Time for the SUS},
  shorttitle = {UMUX-LITE},
  booktitle = {Proceedings of the SIGCHI Conference on Human Factors in Computing Systems},
  author = {Lewis, James R. and Utesch, Brian S. and Maher, Deborah E.},
  year = {2013},
  month = apr,
  series = {CHI '13},
  pages = {2099--2102},
  publisher = {Association for Computing Machinery},
  address = {New York, NY, USA},
  doi = {10.1145/2470654.2481287},
  urldate = {2025-03-04},
  isbn = {978-1-4503-1899-0}
}

@book{collinsCognitiveApprenticeship2006,
  title = {Cognitive Apprenticeship},
  author = {Collins, Allan and Kapur, Manu},
  year = 2006,
  series = {The Cambridge Handbook of the Learning Sciences},
  volume = {291},
  publisher = {Cambridge University Press},
  address = {Cambridge, UK},
  urldate = {2024-11-26}
}

@inproceedings{zamfirescu-pereiraWhyJohnnyCant2023,
  title = {Why Johnny Can't Prompt: How Non-AI Experts Try (and Fail) to Design LLM Prompts},
  shorttitle = {Why Johnny Can't Prompt},
  booktitle = {Proceedings of the 2023 CHI Conference on Human Factors in Computing Systems},
  author = {{Zamfirescu-Pereira}, J.D. and Wong, Richmond Y. and Hartmann, Bjoern and Yang, Qian},
  year = 2023,
  month = apr,
  series = {CHI '23},
  pages = {1--21},
  publisher = {Association for Computing Machinery},
  address = {New York, NY, USA},
  doi = {10.1145/3544548.3581388},
  urldate = {2025-11-07},
  isbn = {978-1-4503-9421-5}
}

@inproceedings{mirowskiCoWritingScreenplaysTheatre2023,
  title = {Co-Writing Screenplays and Theatre Scripts with Language Models: Evaluation by Industry Professionals},
  shorttitle = {Co-Writing Screenplays and Theatre Scripts with Language Models},
  booktitle = {Proceedings of the 2023 CHI Conference on Human Factors in Computing Systems},
  author = {Mirowski, Piotr and Mathewson, Kory W. and Pittman, Jaylen and Evans, Richard},
  year = 2023,
  month = apr,
  series = {CHI '23},
  pages = {1--34},
  publisher = {Association for Computing Machinery},
  address = {New York, NY, USA},
  doi = {10.1145/3544548.3581225},
  urldate = {2025-09-30},
  isbn = {978-1-4503-9421-5},
  langid = {american}
}

@inproceedings{shenConvXAIDeliveringHeterogeneous2023a,
  title = {ConvXAI : Delivering Heterogeneous AI Explanations via Conversations to Support Human-AI Scientific Writing},
  shorttitle = {ConvXAI},
  booktitle = {Companion Publication of the 2023 Conference on Computer Supported Cooperative Work and Social Computing},
  author = {Shen, Hua and Huang, Chieh-Yang and Wu, Tongshuang and Huang, Ting-Hao Kenneth},
  year = 2023,
  month = oct,
  series = {CSCW '23 Companion},
  pages = {384--387},
  publisher = {Association for Computing Machinery},
  address = {New York, NY, USA},
  doi = {10.1145/3584931.3607492},
  urldate = {2024-01-20},
  isbn = {979-8-4007-0129-0}
}

@article{sunMetaWriterExploringPotential2024,
  title = {MetaWriter: Exploring the Potential and Perils of AI Writing Support in Scientific Peer Review},
  shorttitle = {MetaWriter},
  author = {Sun, Lu and Tao, Stone and Hu, Junjie and Dow, Steven P.},
  year = 2024,
  month = apr,
  journal = {Proc. ACM Hum.-Comput. Interact.},
  volume = {8},
  number = {CSCW1},
  pages = {94:1--94:32},
  doi = {10.1145/3637371},
  urldate = {2024-09-11}
}

@inproceedings{zhangSynthiaVisuallyInterpreting2025,
  title = {Synthia: Visually Interpreting and Synthesizing Feedback for Writing Revision},
  shorttitle = {Synthia},
  booktitle = {Proceedings of the 38th Annual ACM Symposium on User Interface Software and Technology},
  author = {Zhang, Chao and Ju, Kexin and Han, Zhuolun and Yen, Yu-Chun Grace and Rzeszotarski, Jeffrey M.},
  year = 2025,
  month = sep,
  series = {UIST '25},
  pages = {1--16},
  publisher = {Association for Computing Machinery},
  address = {New York, NY, USA},
  doi = {10.1145/3746059.3747703},
  urldate = {2025-09-27},
  isbn = {979-8-4007-2037-6},
  langid = {american}
}

@inproceedings{dhuliawalaChainofVerificationReducesHallucination2024,
  title = {Chain-of-Verification Reduces Hallucination in Large Language Models},
  booktitle = {Findings of the Association for Computational Linguistics: ACL 2024},
  author = {Dhuliawala, Shehzaad and Komeili, Mojtaba and Xu, Jing and Raileanu, Roberta and Li, Xian and Celikyilmaz, Asli and Weston, Jason},
  editor = {Ku, Lun-Wei and Martins, Andre and Srikumar, Vivek},
  year = 2024,
  month = aug,
  pages = {3563--3578},
  publisher = {Association for Computational Linguistics},
  address = {Bangkok, Thailand},
  doi = {10.18653/v1/2024.findings-acl.212},
  urldate = {2025-11-08}
}

@inproceedings{leiRationalizingNeuralPredictions2016a,
  title = {Rationalizing Neural Predictions},
  booktitle = {Proceedings of the 2016 Conference on Empirical Methods in Natural Language Processing},
  author = {Lei, Tao and Barzilay, Regina and Jaakkola, Tommi},
  editor = {Su, Jian and Duh, Kevin and Carreras, Xavier},
  year = 2016,
  month = nov,
  pages = {107--117},
  publisher = {Association for Computational Linguistics},
  address = {Austin, Texas},
  doi = {10.18653/v1/D16-1011},
  urldate = {2025-11-08}
}

@misc{chakrabartyAISlopAIPolishAligning2025,
  title = {AI-Slop to AI-Polish? Aligning Language Models through Edit-Based Writing Rewards and Test-Time Computation},
  shorttitle = {AI-Slop to AI-Polish?},
  author = {Chakrabarty, Tuhin and Laban, Philippe and Wu, Chien-Sheng},
  year = 2025,
  month = apr,
  number = {arXiv:2504.07532},
  eprint = {2504.07532},
  primaryclass = {cs},
  publisher = {arXiv},
  doi = {10.48550/arXiv.2504.07532},
  urldate = {2025-11-06},
  archiveprefix = {arXiv},
  langid = {american}
}

@article{hennessySituatedCognitionCognitive1993,
  title = {Situated Cognition and Cognitive Apprenticeship: Implications for Classroom Learning},
  shorttitle = {Situated Cognition and Cognitive Apprenticeship},
  author = {Hennessy, Sara},
  year = 1993,
  month = jan,
  journal = {Studies in Science Education},
  volume = {22},
  number = {1},
  pages = {1--41},
  issn = {0305-7267, 1940-8412},
  doi = {10.1080/03057269308560019},
  urldate = {2025-12-05},
  langid = {english}
}

@article{wangInvestigatingImpactStratified2024,
  title = {Investigating the Impact of the Stratified Cognitive Apprenticeship Model on High School Students' Math Performance},
  author = {Wang, Ruimei and Zulkifli, Nurul Nadwa and Mohd Ayub, Ahmad Fauzi},
  year = 2024,
  month = aug,
  journal = {Education Sciences},
  volume = {14},
  number = {8},
  pages = {898},
  publisher = {Multidisciplinary Digital Publishing Institute},
  issn = {2227-7102},
  doi = {10.3390/educsci14080898},
  urldate = {2025-12-05},
  copyright = {http://creativecommons.org/licenses/by/3.0/},
  langid = {english}
}

@inproceedings{zhangNavigatingFogHow2025,
  title = {Navigating the Fog: How University Students Recalibrate Sensemaking Practices to Address Plausible Falsehoods in LLM Outputs},
  shorttitle = {Navigating the Fog},
  booktitle = {Proceedings of the 7th ACM Conference on Conversational User Interfaces},
  author = {Zhang, Chao and Zhu, Shengqi and Yang, Xinyu and Tseng, Yu-Chia and Jiang, Shenrong and Rzeszotarski, Jeffrey M.},
  year = 2025,
  month = jul,
  series = {CUI '25},
  pages = {1--15},
  publisher = {Association for Computing Machinery},
  address = {New York, NY, USA},
  doi = {10.1145/3719160.3736618},
  urldate = {2025-09-05},
  isbn = {979-8-4007-1527-3},
  langid = {american}
}

\clearpage
\appendix
\section{Definition of Narrative Concepts}

\subsection{Turning Points}
\label{appendix:turning_points}
A turning point is an event (or plot moment) that significantly influences a plot progression~\cite{papalampidiMoviePlotAnalysis2019}. 
These turning points are generally in sequential order in a narrative (\ie Opportunity happens first; Climax happens last).
The definitions are shown in \tabref{tab:turning_points}.

\begin{table}[H]
  \centering
  \caption{Definition of turning points~\cite{tianAreLargeLanguage2024}.}
  \Description{}
  \label{tab:turning_points}
  \small
  \begin{tabular}{p{0.26\linewidth} p{0.66\linewidth}}
    \toprule
    \textbf{Turning Point} & \textbf{Description} \\
    \midrule
    Opportunity & The introductory event that sets the stage for the narrative. \\
    Change of Plans & A pivotal moment where the main goal of the narrative is defined or altered. \\
    Point of No Return & The commitment point beyond which the protagonists are invested in goals. \\
    Major Setback & A critical juncture where the protagonists face significant challenges or failures. \\
    Climax & The peak of the narrative arc, encompassing the resolution of the central conflict. \\
    \bottomrule
  \end{tabular}
\end{table}

\subsection{Creative Dimensions}
\label{appendix:creative_dimensions}
We derived a taxonomy of eight creative dimensions from theories in narrative and writing studies (\tabref{tab:creative_dimensions}). 
These dimensions capture distinct aspects of narrative construction emphasized across narratology, creative writing pedagogy, and rhetorical narrative theory~\cite{balNarratologyIntroductionTheory2004,kennedyLiteratureIntroductionFiction2016,mckeeStoryStyleStructure1997,princeNarratologyFormFunctioning2012}.
This taxonomy is not exhaustive; rather, it offers an interpretive lens for examining the diverse narrative strategies employed in story writing.

\begin{table}[H]
  \centering
  \caption{Definition of creative dimensions~\cite{balNarratologyIntroductionTheory2004,kennedyLiteratureIntroductionFiction2016,mckeeStoryStyleStructure1997,princeNarratologyFormFunctioning2012}.}
  \Description{}
  \label{tab:creative_dimensions}
  \small
  \begin{tabular}{p{0.2\linewidth} p{0.72\linewidth}}
    \toprule
    \textbf{Dimension} & \textbf{Description} \\
    \midrule
    Plot &
    Strategies for plot construction and story progression, \eg causation, escalation, conflict setup and resolution, reversals and twists, and act and beat frameworks. \\
    Character &
    Strategies for character development and portrayal, \eg growth, traits, relationships, and archetypal roles. \\
    Information &
    Strategies for information control and perspective, \eg revelation, concealment, misdirection, foreshadowing, and point-of-view manipulation. \\
    Emotional &
    Strategies for emotional effect, \eg tension, empathy, surprise, catharsis, and atmosphere. \\
    Linguistic &
    Strategies for language style, \eg voice, imagery, syntax and rhythm, dialogue, and rhetorical devices. \\
    Pacing &
    Strategies for pacing at the moment and segment levels, \eg scene versus summary, time compression and expansion, beat density, sentence and paragraph cadence, cutaways and cross-cutting, time skips, and arrive-late, leave-early trims. \\
    Thematic &
    Strategies for theme and meaning, \eg symbolism, allegory, and philosophical exploration. \\
    Engagement &
    Strategies for reader engagement, \eg hooks, immersion techniques, curiosity creation, suspense management, and narrative payoffs. \\
    \bottomrule
  \end{tabular}
\end{table}

\section{LLM Prompts Used in \tool}
\label{appendix:technical_details}

In this section, we present the prompts used to instruct GPTs to segment stories, infer narrative strategies, categorize strategies, infer protagonist, and infer emotional adjectives.

\subsection{Segmenting Stories into Blocks}
We prompt GPT-4o to segment stories into blocks (\figref{fig:technical_pipeline}A).

\begin{myfancybox}
\footnotesize

\noindent\textbf{System Prompt:}\\
\texttt{You are an expert story analyst. Your task is to segment a given story into coherent plot segments that represent distinct narrative beats.\\ \\
**Segment Criteria:**\\
- Each segment should be self-contained enough to understand independently\\
- Aim for 5-10 segments depending on story length and complexity\\
- Each segment should represent a meaningful story progression\\
- Avoid overly short segments (less than 50 words) or overly long segments (more than 300 words)\\ \\
**Output Format:**\\
Return your response as a JSON object with this exact structure:\\
\{"plots": [{"title": "Brief, descriptive title (3-8 words)", "plot": "Original text content from this story segment (extracted verbatim from the source)", "summary": "Concise summary of what happens in this segment"}]\}.}

\noindent\dotfill

\par\medskip
\noindent\textbf{Context Prompt:}\\
\texttt{
The story title is \colorbox{gray-bg}{<insert story title>} and the story is: \colorbox{gray-bg}{<insert full story content>}
}

\end{myfancybox}

\subsection{Inferring Story Protagonist}
We instruct GPT-4o to identify the main character (\figref{fig:technical_pipeline}D).

\begin{myfancybox}
\footnotesize

\noindent\textbf{System Prompt:}\\
\texttt{Who is the main character of the this story?\\The output should just be a name or a short phrase. Do not include any other information or context.}

\noindent\dotfill

\par\medskip
\noindent\textbf{Context Prompt:}\\
\texttt{
The story is: \colorbox{gray-bg}{<insert full story content>}
}

\end{myfancybox}

\subsection{Inferring Emotional Adjectives}
For each content block in a narrative, we ask GPT-4o to infer three adjectives that describe the protagonist's emotions as the plot progresses (\eg \textit{amused}, \textit{relaxed}, \textit{anxious}) (\figref{fig:technical_pipeline}E).

\begin{myfancybox}
\footnotesize

\noindent\textbf{System Prompt:}\\
\texttt{Use three different words to describe the character's feeling in a given story plot.\\The output should be a list of words. For example, [happy, sad, joyful]. Do not include other outputs.}

\noindent\dotfill

\par\medskip
\noindent\textbf{Context Prompt:}\\
\texttt{
How does \colorbox{gray-bg}{<insert protagonist>} feel in this plot? 
\colorbox{gray-bg}{<insert plot content>}
}

\end{myfancybox}

\subsection{Inferring Narrative Strategies}
We prompt GPT-4.1 to infer narrative strategies (\figref{fig:technical_pipeline}B).

\begin{myfancybox}
\footnotesize

\noindent\textbf{System Prompt:}\\
\texttt{You are an expert literary analyst tasked with identifying creative strategies used in story plots. Creative strategies are any storytelling techniques, narrative devices, plot mechanisms, stylistic choices, or structural elements that authors use to create compelling narratives, engage readers, or achieve specific artistic effects.\\ \\
**Your task:**\\Analyze the given plot and identify all creative strategies employed. For each strategy, provide:\\
1. A concise phrase describing the strategy (2-6 words)\\
2. A detailed explanation of how and why this strategy is used effectively\\
3. Specific lexical features (words, phrases, linguistic patterns) that contribute to or signal this strategy\\ \\
**What Constitutes a Creative Strategy:**\\
Any deliberate creative choice that serves a narrative purpose, including but not limited to:\\
- How information is revealed or withheld\\
- Character development and interaction patterns\\
- Structural and pacing decisions\\
- Language and tone choices\\
- Conflict creation and resolution approaches\\
- Thematic development techniques\\
- Reader engagement mechanisms\\
- Innovative or unexpected narrative elements\\ \\
**Lexical Features to Identify:**\\
For each strategy, extract EXACT verbatim text from the original plot that contributes to or signals the strategy:\\
- Copy exact words, phrases, or sentences as they appear in the plot\\
- Include direct quotes from dialogue exactly as written\\
- Extract precise descriptive language or imagery\\
- Identify specific repeated words or phrases\\
- Copy transitional phrases or structural markers verbatim\\
- Quote any language that contributes to the strategy's effectiveness\\
- Do NOT paraphrase, interpret, or modify the original text - use exact quotations only
Return output\\ \\
**Output Format:**\\
Return your response as a JSON object with this exact structure:\\
\{"strategies":[\{"strategy":"Brief strategy name (2--6 words)","reasoning":"1--3 sentence explanation of how this strategy functions and why it's effective.","lexicon":["word1","phrase2","linguistic pattern3"]\}]\}.}

\noindent\dotfill

\par\medskip
\noindent\textbf{Context Prompt:}\\
\texttt{The story plot is: \colorbox{gray-bg}{<insert plot content>}\\ \\
Please analyze this plot and identify all creative strategies employed. Look for any storytelling techniques, narrative choices, or creative elements that serve a purpose in the story - don't limit yourself to traditional categories.\\ \\
For each strategy you identify:\\
1. Name the strategy clearly and concisely\\
2. Explain how it functions in the plot and why it's effective\\
3. Extract EXACT verbatim words, phrases, or sentences from the plot text above that contribute to or signal this strategy - use precise quotations only, do not paraphrase or modify the original text\\ \\
Be thorough and creative in your analysis. Consider both obvious techniques and subtle creative choices that make this plot work.}

\end{myfancybox}

\subsection{Categorizing Narrative Strategies}
We prompt GPT-4.1 to categorize narrative strategies into the eight creative dimensions in \tabref{tab:creative_dimensions}.

\begin{myfancybox}
\footnotesize

\noindent\textbf{System Prompt:}\\
\texttt{You are an expert literary analyst tasked with categorizing creative strategies according to a comprehensive taxonomy. Each strategy should be assigned to one or two primary categories based on its main functions and effects.\\ \\
**Your task:**\\For each creative strategy provided, determine the PRIMARY CATEGORY (or two categories if the strategy serves multiple major functions) that best describes the strategy's main function(s).\\ \\
**Taxonomy Categories:**\\
\colorbox{gray-bg}{<insert definitions of the eight dimensions>}\\ \\
**Guidelines:**\\
- Choose 1-2 categories that represent the strategy's primary functions\\
- If a strategy clearly serves two major narrative purposes, assign both categories\\
- Focus on what the strategy DOES rather than what it contains\\
- Consider the strategy's main purpose(s) in the narrative context\\
- Only use two categories if the strategy genuinely has dual primary functions - avoid over-categorizing \\ \\
**Output Format:**\\
Return your response as a JSON object with only the category assignment(s):\\
\{"category": ["PRIMARY\_CATEGORY"]\} or \{"category": ["CATEGORY\_1","CATEGORY\_2"]\}.}

\noindent\dotfill

\par\medskip
\noindent\textbf{Context Prompt:}\\
\texttt{Please categorize the following creative strategy: \colorbox{gray-bg}{<insert strategy name>} used in the plot: \colorbox{gray-bg}{<insert plot content>}.\\ \\
The explanation for this strategy is: \colorbox{gray-bg}{<insert strategy explanation>}.\\ \\
Assign it to one or two primary categories based on its main function(s) and narrative purpose(s). Use two categories only when the strategy genuinely serves dual major functions.}

\end{myfancybox}

\begin{table*}
  \centering
  \caption{\revision{Failure cases in LLM-extracted strategies and explanations flagged by raters.}}
  \Description{}
  \label{tab:strategy_failure_examples}
  \small
  \begin{tabular}{p{0.48\textwidth} p{0.48\textwidth}}
    \toprule
    \textbf{Story Excerpts} & \textbf{Extracted Strategies \& Explanations} \\
    \midrule
    ``\textit{...Maya and Willy went to the glow worms, but they took Maya and Willy up to wreck the hive and the precious sunstone, much to the displeasure of the Queen...}'' 
      & \textbf{Use of MacGuffin}: The ``precious sunstone'' functions as a MacGuffin, an object that propels the plot and is central to the characters' actions and the conflict. \\
    ``\textit{...Will reaches a lower road and encounters a man who turns out to be a Frank's guard, but talks him into letting him pass. He finds Frank's camp in the woods, a derelict farm, the main building of which is an extensive but rudimentary drug lab. Most of the inhabitants are unconcerned at his presence...}'' 
      & \textbf{Unexpected Lack of Conflict}: Rather than a tense or violent encounter, Will persuades the guard and the camp's inhabitants are unconcerned, subverting expectations for dramatic confrontation and promoting intrigue about their motives or broader circumstances. \\
    ``\textit{...Reconstructing K.C.'s consciousness, the Moon's operating system appears to him as his cat, Fuzz Aldrin, and his mother, remarking that they must now `get started'...}''
      & \textbf{Abrupt Narrative Start}: The action begins in medias res, with K.C.'s consciousness being reconstructed and immediate engagement with the central mystery. This technique hooks the reader and leaves important details to be filled in later. \\
    ``\textit{...After Lydell is arrested in a DEA sting, a desperate Liza asks Neel for help paying for the surgery but he exhorts her to use Phoebe's situation as `fire' for inspiration, like he did when his wife was dying from cancer and he came up with Lonafen. When a friend's husband dies from a Lonafen overdose and her condolences are rejected, Liza agrees to testify to the U.S. Attorney's office that's investigating Zanna. She admits her involvement in Zanna's speaker and bribery programs...}''
      & \textbf{Chain Reaction Causality}: The plot unfolds through a tight sequence of cause and effect, where each character's action directly provokes consequential reactions, propelling the narrative forward and intensifying momentum. \\
    ``\textit{...Meanwhile at the hive, Crawley tries to fix the sunstone, but to no avail, much to his dismay. Miss Cassandra calls out that Maya and Willy has left the meadow...}''
      & \textbf{Direct Character Dialogue}: Involving Miss Cassandra's speech adds immediacy and a sense of presence to the scene, making the action feel ongoing and interactive while also revealing plot developments through character voices. \\
    ``\textit{...Alone at last, Shirley allows Stanley to read her work on Hangsaman. He declares it to be a work of genius. Shirley acknowledges his praise. It's obviously a familiar pattern in their peculiar partnership. The two celebrate by drinking and dancing — together and yet alone in their cluttered house...}''
      & \textbf{Symbolic Setting}: The 'cluttered house' setting at the conclusion operates symbolically, reflecting the characters' internal lives and the complexity of their relationship. \\
    ``\textit{...The nun then shows the brothers that Eloise really is an actual plane built by Flavio to pass down to his sons. Now having a way of getting home, Renato gets ready to part ways with Asher and the goat; Renatito. However, he changes his mind and invites them both to the wedding. They fly away together and return to Mexico as they pass above the Sumidero Canyon...}''
      & \textbf{Symbolic Setting Transition}: Flying above the Sumidero Canyon serves as a visual marker of transition—both geographic and emotional—underscoring the theme of crossing thresholds into new phases of life. \\
    ``\textit{...Ava Faulkner is a recovering addict and former soldier turned assassin...}''
      & \textbf{Economical Exposition}: The plot delivers significant backstory and premise in a single sentence, using minimal language to maximum effect. This directness creates instant engagement without over-explanation, trusting the audience to infer dramatic stakes. \\
    ``\textit{...Among the guests are Rowena's housekeeper Olga Seminoff, Drake family doctor Leslie Ferrier and his son Leopold, and Joyce's Romani assistant Desdemona Holland; they are joined by Maxime right before the séance, and during it Poirot reveals Desdemona's half-brother Nicholas—and Joyce's second assistant—hiding in the chimney...}''
      & \textbf{Romani Identity Signification}: Distinctly identifying Desdemona as 'Romani' marks cultural and social difference, potentially invoking theme, bias, or stereotype as part of the plot's societal commentary. \\
    ``\textit{...In small-town Florida, Chris is working at a car wash when a woman, Maria, arrives. Chris tells the man whose car he's washing that she is his high school crush. He begins to vacuum the car, but the overpowered vacuum sucks his clothes off, leaving him naked. He panics and hides in the car. The man has a conversation with Maria, attempting to get her number for Chris, but she declines and leaves...}''
      & \textbf{Setting as Character}: Locating the plot in 'small-town Florida' uses regional specificity to evoke a sense of place, potential small-town gossip, and intimacy, affecting both character behavior and plot plausibility. \\
    ``\textit{...In Mexico City, Sebastián Silva is a depressed, ketamine-addicted artist and filmmaker contemplating suicide by taking pentobarbital...}''
      & \textbf{Setting Establishment}: By explicitly setting the story in Mexico City, the narrative immediately grounds the plot in a specific cultural and urban environment, providing context and atmosphere that can shape the reader’s expectations and the protagonist’s experiences. \\
    \bottomrule
  \end{tabular}
\end{table*}

\begin{table*}
  \centering
  \caption{\revision{Failure cases in LLM-extracted lexical cues flagged by raters.}}
  \Description{}
  \label{tab:cue_failure_examples}
  \small
  \begin{tabular}{p{0.48\textwidth} p{0.48\textwidth}}
    \toprule
    \textbf{Story Excerpts} & \textbf{Extracted Strategies \& Explanations} \\
    \midrule
    ``\textit{...Maya and Willy went to the glow worms, but they took Maya and Willy up to wreck the hive and the precious sunstone, much to the displeasure of the Queen...}'' 
      & \textbf{Use of MacGuffin}: The ``precious sunstone'' functions as a MacGuffin, an object that propels the plot and is central to the characters' actions and the conflict.\newline \textbf{Lexical Cues}: ``\textit{the precious sunstone}'' \\
    ``\textit{...Reconstructing K.C.'s consciousness, the Moon's operating system appears to him as his cat, Fuzz Aldrin, and his mother, remarking that they must now `get started'...}'' 
      & \textbf{Abrupt Narrative Start}: The action begins in medias res, with K.C.'s consciousness being reconstructed and immediate engagement with the central mystery. This technique hooks the reader and leaves important details to be filled in later.\newline \textbf{Lexical Cues}: ``\textit{Reconstructing K.C.'s consciousness}'' \\
    ``\textit{...After Lydell is arrested in a DEA sting, a desperate Liza asks Neel for help paying for the surgery but he exhorts her to use Phoebe's situation as `fire' for inspiration, like he did when his wife was dying from cancer and he came up with Lonafen. When a friend's husband dies from a Lonafen overdose and her condolences are rejected, Liza agrees to testify to the U.S. Attorney's office that's investigating Zanna. She admits her involvement in Zanna's speaker and bribery programs...}''
      & \textbf{Chain Reaction Causality}: The plot unfolds through a tight sequence of cause and effect, where each character's action directly provokes consequential reactions, propelling the narrative forward and intensifying momentum.\newline \textbf{Lexical Cues}: ``\textit{After Lydell is arrested in a DEA sting, a desperate Liza asks Neel for help}'', ``\textit{When a friend's husband dies from a Lonafen overdose and her condolences are rejected, Liza agrees to testify}'' \\
    ``\textit{...Hutch Mansell sits inside of an interrogation room, heavily bruised and injured. Two FBI agents interrogate him about his identity, before Hutch responds that he's simply "nobody." In a flashback to a few days earlier, Hutch leads an ordinary, mundane life as an office worker with his emotionally estranged wife Becca, teenager Blake and a much younger daughter, Abby...}''
      & \textbf{Emotionally Distant Familial Relationships}: By describing Hutch's home life as 'emotionally estranged,' the story hints at inner conflicts and character motivations, adding depth and potential for emotional development.\newline \textbf{Lexical Cues}: ``\textit{with his emotionally estranged wife Becca}'' \\
    ``\textit{...Meanwhile at the hive, Crawley tries to fix the sunstone, but to no avail, much to his dismay. Miss Cassandra calls out that Maya and Willy has left the meadow...}''
      & \textbf{Direct Character Dialogue}: Involving Miss Cassandra's speech adds immediacy and a sense of presence to the scene, making the action feel ongoing and interactive while also revealing plot developments through character voices.\newline \textbf{Lexical Cues}: ``\textit{Miss Cassandra calls out that Maya and Willy has left the meadow}'' \\
    ``\textit{...Her brother Kevin is caught dancing in the restroom by some older students who record him and bully him, but Clancy's and Kevin's mother Margot, the lunch monitor, scares them away. The bullies upload a remix video of Kevin dancing and Margot admonishing them to YouTube, where it quickly goes viral, gaining over two million views...}''
      & \textbf{Authority Figure Intervention}: Margot, as a lunch monitor and mother, intervenes to protect her son, adding a layer of authority and adult presence that both disrupts the bullying and sets up later plot developments. This juxtaposition of child vulnerability and adult protection intensifies reader investment in the outcome.\newline \textbf{Lexical Cues}: ``\textit{Margot, the lunch monitor, scares them away}'' \\
    ``\textit{...Alone at last, Shirley allows Stanley to read her work on Hangsaman. He declares it to be a work of genius. Shirley acknowledges his praise. It's obviously a familiar pattern in their peculiar partnership. The two celebrate by drinking and dancing — together and yet alone in their cluttered house...}''
      & \textbf{Symbolic Setting}: The `cluttered house' setting at the conclusion operates symbolically, reflecting the characters' internal lives and the complexity of their relationship.\newline \textbf{Lexical Cues}: ``\textit{in their cluttered house}'' \\
    ``\textit{...The next morning, Kath awakens to find everyone gone. Venturing into the surrounding woods, she finds Al, crying. He explains that he found Greta and Max having sex and that the two left together...}''
      & \textbf{Withholding Information}: Key details about what happened the night before are initially omitted, heightening suspense and encouraging curiosity. This deliberate omission motivates the reader to follow Kath as she seeks answers.\newline \textbf{Lexical Cues}: ``\textit{The next morning, Kath awakens to find everyone gone.}'' \\
    \bottomrule
  \end{tabular}
\end{table*}

\revision{
\section{Failure Cases from Human Evaluation of  LLM-Extracted Strategies}
\label{appendix:failure_case}
Across the 11 extraction/explanation cases and 8 lexical-cue cases that at least one evaluator flagged as incorrect, raters highlighted points where the LLM analyses failed.
\tabref{tab:strategy_failure_examples} and \tabref{tab:cue_failure_examples} shows typical failure cases.
For instance, describing the ``precious sunstone'' as a MacGuffin gestures in a sensible direction, but the explanation would benefit from pointing more explicitly to how characters’ choices turn on the object. 
Likewise, the ``chain reaction'' reading of one plot summarizes a rich sequence, but could be made more informative by isolating the immediate trigger within the plot that escalates consequences. 
We also saw a few cases where noticeable surface features, such as an in-medias-res opening (``Reconstructing K.C.’s consciousness…''), a reported line of speech (``Miss Cassandra calls out…''), or setting details (``small-town Florida,'' ``Mexico City,'' ``Sumidero Canyon''), were highlighted without fully spelling out their local function (\eg how the opener withholds specifics to invite inference, what the utterance changes in the scene, or how the locale shapes stakes and behavior). 
On evidence selection, raters were concerned when quoted cues were very general (\eg quoting only the noun ``the precious sunstone''), leaned on temporal or expository phrases (``The next morning...'', ``with his emotionally estranged wife'') rather than the textual signals of the claimed device, or lifted long recap-style spans that restate events instead of pinpointing the terms that create causal or thematic pressure.\looseness=-1
}

\begin{table*}
  \centering
  \small
\caption{Demographic information of participants and their usage of examples and AI writing tools in daily writing practice. *Writing proficiency; **English proficiency (both on Likert scales from 1/low to 7/high); ***Frequency of referring to others' work in writing (from Never to Always).}
  \label{table:demographics}
  \begin{tabular}{lllcclll}
    \toprule
    \textbf{ID} & \textbf{Gender} & \textbf{Age} & \textbf{Writing*} & \textbf{English**} & \textbf{Example***} & \textbf{AI Writing Tools} & \textbf{AI Usage Freq.} \\ 
    \midrule
    P01  & Female & 27 & 1 & 4 & Always & ChatGPT, Grammarly, DeepL & Weekly \\
    P02  & Male & 28 & 2 & 5 & Always & ChatGPT, Grammarly, Cursor & Daily \\
    P03  & Male & 25 & 3 & 5 & Sometimes & ChatGPT, Cursor & Daily \\
    P04  & Female & 25 & 2 & 6 & Often & ChatGPT, Grammarly & Daily \\
    P05  & Female & 28 & 1 & 5 & Always & ChatGPT & Weekly \\
    P06  & Male & 25 & 2 & 5 & Always & ChatGPT, Grok, Gemini, Grammarly & Daily \\
    P07  & Female & 26 & 3 & 6 & Always & ChatGPT, NotebookLM, Grammarly & Weekly \\
    P08  & Female & --- & 2 & 3 & Sometimes & ChatGPT, Grammarly & Daily \\
    P09  & Female & 28 & 3 & 7 & Sometimes & ChatGPT, Grammarly & Daily \\
    P10 & Male & 28 & 4 & 4 & Often & ChatGPT, Notion AI, Grammarly & Daily \\
    P11 & Female & 23 & 4 & 5 & Sometimes & ChatGPT, Grammarly & Weekly \\
    P12 & Female & 26 & 1 & 5 & Often & ChatGPT & Daily \\
    \bottomrule
\end{tabular}
\end{table*}

\section{User Study}

\subsection{Baseline System Interface}
\label{appendix:baseline_interface}

\begin{figure}[H]
\centering
\includegraphics[width=\linewidth]{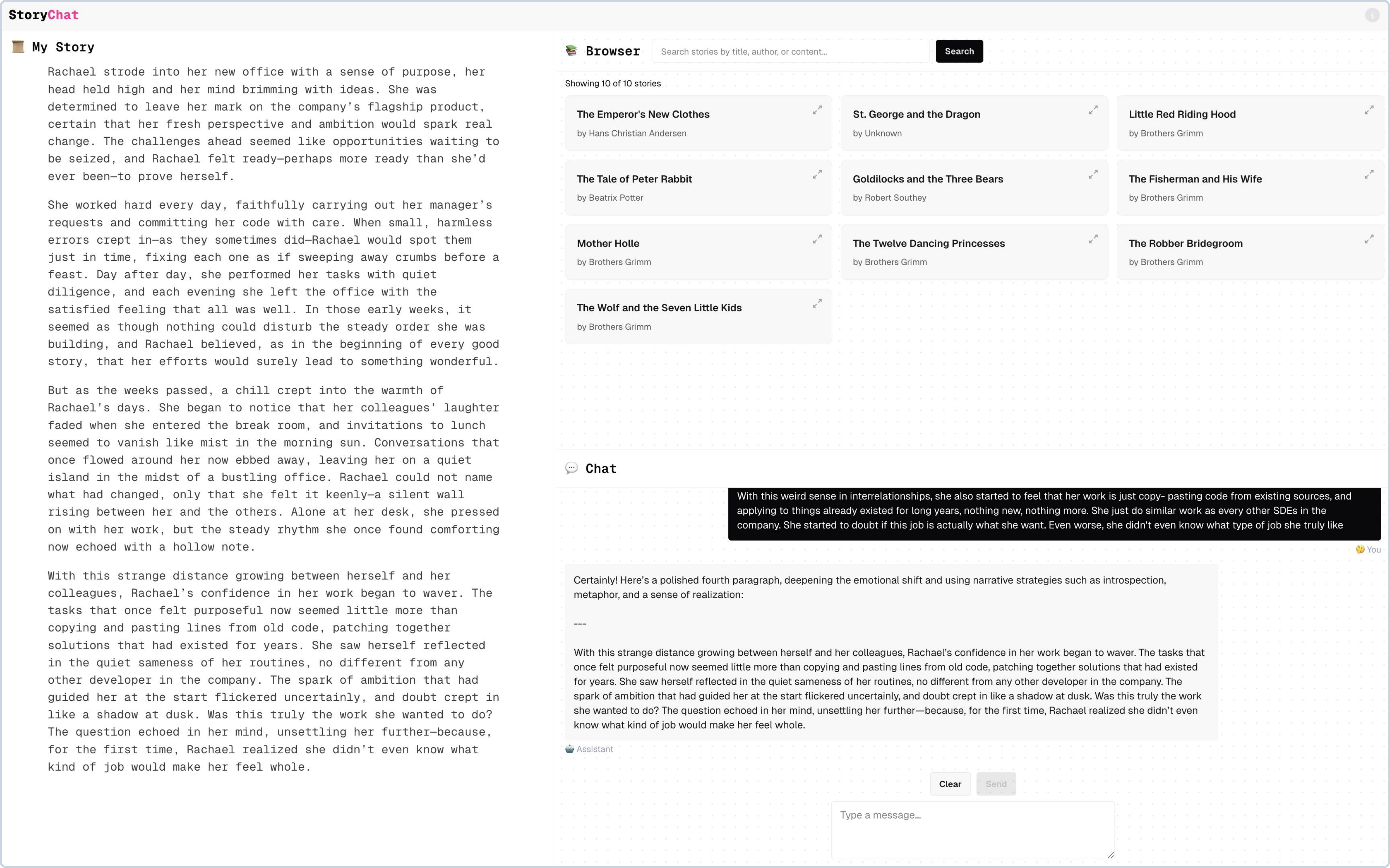}
\caption{Baseline system interface, which includes the same Markdown editor as \tool, example stories presented as whole-text cards without surfacing narrative strategies, and a chat-based AI assistant, similar to interfaces such as ChatGPT Canvas.}
\Description{Baseline system interface, which retains the same Markdown editor as Narrix but removes the interactive story arc visualization. The Browser panel displays example stories as whole-text cards without surfacing narrative strategies. The Remix panel is replaced with a chat-based AI assistant, allowing users to send free-form prompts and receive conversational responses, similar to interfaces like ChatGPT Canvas.}
\label{fig:baseline_interface}
\end{figure}

Given that no existing solutions are directly comparable to \tool in terms of supporting interaction with narrative strategies in examples, we implemented a baseline system (\figref{fig:baseline_interface}) that closely resembled \tool in UI but excluded the key features unique to our approach.
Both systems shared the same Markdown text editor, ensuring a consistent writing environment. However, in the baseline, we removed the interactive story arc visualization. In the Browser panel, example stories were shown as story cards; each card could be clicked to reveal the full text of the example story, but without highlighting or extracting narrative strategies.
The Remix panel in \tool was replaced by an AI chat assistant. This assistant, powered by the same underlying LLM but designed to mimic mainstream chat-based writing interfaces (such as ChatGPT Canvas): users could interact via free-form chat prompts, and the system would respond conversationally in a dedicated output panel. This baseline followed design conventions used in prior work~\cite{rezaABScribeRapidExploration2024,massonTextoshopInteractionsInspired2025,zhangFrictionDecipheringWriting2025}, which also implemented chat-based AI writing assistants as comparison conditions when evaluating novel writing tools.\looseness=-1

In summary, this baseline design (1) retains basic non-contributory features of \tool (\eg the Markdown editor) to minimize interface confounds, (2) uses the same underlying model to isolate interaction design effects, (3) mirrors real-world practices where users have access to general-purpose AI tools like ChatGPT, and (3) integrates story editing, example browsing, and AI assistance within a unified workspace to avoid unnecessary window switching.

\subsection{Participant Information}
\label{appendix:participant_information}

We recruited 12 participants (Table~\ref{table:demographics}; 8 female, 4 male), ages 23--28 (\(M = 26.27\), \(SD = 1.68\)), from a large software organization in the United States via internal communication channels and word of mouth. We sought novice writers and, following prior work that recruited ESL writers as novices for writing tasks~\cite{zhangFrictionDecipheringWriting2025,huangFeedbackOrchestrationStructuring2018}, targeted non-native English speakers during recruitment. All participants reported regularly engaging in writing and wanting to improve their creative writing skills; each had prior creative writing experience and self-rated their expertise on a 7-point scale (\(M = 2.33\), \(SD = 1.07\); 1 = beginner, 7 = professional).
On a 5-point scale, participants reported actively seeking and referring to examples in writing (\(M = 4.08\), \(SD = 0.90\); 1 = never, 5 = always) and regularly using generative AI tools (\eg ChatGPT) in their writing activities (\(M = 4.67\), \(SD = 0.50\); 1 = never, 5 = daily).\looseness=-1

\subsection{Interview Questions}
\label{appendix:interview_protocol}


\begin{enumerate}
    \item What was your overall experience with the two tools? Was there anything that excited or frustrated you?
    \item Did you have any ``aha'' moments or memorable experiences while working with either tool?
    \item Which features did you find most beneficial in each tool, and in what scenarios were they especially useful?
    \item How did your strategies for working with or using AI differ between the two systems?
    \item Before this study, how did you typically use examples in writing? After using the tool, did your approach to using examples in writing change? If so, how?
    \item Can you describe any specific narrative strategies or techniques you learned from the examples and how you tried to use them in your own story?
    \item What was your approach to selecting narrative strategies that best fit your goals?
    \item Do you think the tool will contribute to your long-term writing skill development? If so, how?
    \item Do you have any suggestions or ideas to improve the tool?
\end{enumerate}

\end{document}